\documentclass[
	nofootinbib,
         amsmath,
         amssymb,
         superscriptaddress,
        groupedaddress,
         aps,
         prb,
         longbibliography,
         floatfix,
         twocolumn
]{revtex4-2}
\usepackage[english]{babel}
\usepackage[utf8]{inputenc} 
\usepackage{dsfont}
\usepackage{SIunits}
\usepackage[caption=false,label font=bf,labelformat=simple]{subfig}
\usepackage{graphicx}     
\usepackage{color}
\usepackage[normalem]{ulem}
\usepackage{soul}
\usepackage{cancel}
\usepackage{hyperref}

\begin{document}
\title{Periodic source of energy-entangled electrons in helical states coupled to a BCS superconductor}
\author{Flavio Ronetti}
\author{Bruno Bertin-Johannet}
\author{Jérôme Rech}
\author{Thibaut Jonckheere}
\author{Benoît Grémaud}
\author{Laurent Raymond}
\author{Thierry Martin}
\affiliation{Aix Marseille Univ, Universit\'e de Toulon, CNRS, CPT, Marseille, France}

\begin{abstract}
We propose a source of purely electronic energy-entangled states implemented in a solid-state system with potential applications in quantum information protocols based on electron flying qubits. The proposed device relies on the standard tools of Electron Quantum Optics (EQO) and exploits entanglement of the Cooper pairs of a BCS superconductor. The latter is coupled via an adjustable quantum point contact to two opposite spin polarized electron wave-guides, which are driven by trains of Lorentzian pulses. This specific choice for the drive is crucial to inject purely electronic entangled-states devoid of spurious electron-hole pairs. In the Andreev regime, a perturbative calculation in the tunnel coupling confirms that entangled electrons states (EES) are generated at the output of the normal side. We introduce a quantity related to charge current cross-correlations which allows one to verify experimentally the entangled nature of the emitted state.
\end{abstract}
\let\endtitlepage\relax
\maketitle
\section{Introduction}
Quantum electronics is a fascinating research field addressing the fundamental properties of matter by accessing its quantum properties, such as interference and entanglement~\cite{Josephson1962,Liu1998,Laflorencie2016,Nature2016}. Nowadays on-chip devices have been realized and tested in ground-breaking experiments to probe the quantum nature of electronic states down to the single-electron level~\cite{Feve07,Bocquillon2013,Pekola2013,Bauerle2018,Edlbauer2022}. These major achievements have been enabled by the implementation of a variety of single-electron emission protocols, such as acoustic surface waves~\cite{Hermelin2011,Takada2019}, periodically-driven quantum dots~\cite{Mahe2010,Grenier2011,Parmentier2012}, or Lorentzian-shaped voltage drives~\cite{Glattli2016}. The latter source is based on the proposal of Levitov \textit{et al.} that pulses with a Lorentzian profile can be tuned to excite purely electronic excitations, i.e. devoid of any additional electron hole pairs, on top of the Fermi sea~\cite{Levitov1996,Ivanov1997,Keeling2006}. These quasi-particles have then been termed \textit{Levitons}~\cite{Flindt2013}. By properly setting the drive parameters, the total charge emitted in one period can be varied, thus allowing for the simultaneous injection of multiple Levitons~\cite{Glattli2017,Moskalets2018}. Compared to other sources of isolated electrons, the excitations emitted by quantized Lorentzian-shaped pulses possess an intriguing property for individual quantum systems: they are not entangled with their environment~\cite{Ferraro2014}. 

This initial theoretical proposal originated several experimental breakthroughs using Levitonic quasi-particles~\cite{Dubois2013,Assouline2023}. For instance, the measurement of quantum transport properties of systems driven by Lorentzian-shaped pulses allowed to observe fermion anti-bunching~\cite{Dubois2013,Glattli2016,Glattli2017} and to perform electron tomography and time-resolved reconstruction of the Leviton wave-function~\cite{Grenier2011,Jullien2014,Bisognin2019,Roussel2021}. Many theoretical proposals are still increasing the interest for Levitons, which include the excitations of half-charge zero-energy quasi-particles~\cite{Moskalets2017} or the generation of electron-hole entanglement in Mach-Zender interferometers~\cite{Dasenbrook2015,Hofer2016}. An appealing research direction is to investigate the effects of electron correlations, which are ubiquitous in mesoscopic physics, on the generation and dynamics of single-electron excitations~\cite{Ronetti2024}. The properties of Levitons are currently assessed theoretically in strongly-correlated systems, such as the fractional quantum Hall effect~\cite{Stormer1999}, where their stability with respect to the interaction has been proven~\cite{Rech2017}, even for thermal transport~\cite{Vannucci2017,Ronetti2019,Bertin2024}, and peculiar effects, such as an analog of Wigner crystallization, have been proposed~\cite{Ronetti2018,Ferraro2018,Vannucci2018}. Moreover, the effects of superconducting correlations have also been taken into account for Levitons in the presence of tunneling junctions between two superconductors or between a normal system and a superconductor~\cite{Acciai2019,Ronetti2020,Bertin2022,Bertin2023,Bertin2024b}. 

This vast research framework, built on the close cooperation between theory and experiments, is called \textit{electron quantum optics}~\cite{Grenier2011b,Bocquillon2012,Bocquillon2014,Roussel2017,Marguerite2016}. Indeed, the starting motivation for this research field was to reproduce quantum optics experiments by replacing photons by electrons. Nevertheless, the fact that single-electron excitations are easily exposed to interactions is a marking difference in comparison with photon quantum states~\cite{Wang2023,Ubbelohde2023,Fletcher2023}. In this sense, the electron quantum optics scenarios have been increasingly attracting a speculative interest due to possible applications for quantum computation schemes based on \textit{electron flying qubits}~\cite{Roussely2018,Bauerle2018,Edlbauer2022,Aluffi2023,Wang2024}. In this framework, one-dimensional channels existing in mesoscopic systems, as a consequence of quantum confinement or topological properties, are exploited as wave-guides for these electronic states. These channels are termed quantum rails and quantum information is encoded by accounting for the presence or the absence of the flying electron in each of them. Combining together several quantum rails and single-electron sources would allow for a purely electronic quantum computation scheme, which possesses a great potential for scalability~\cite{Takeda2017,Wang2022Rev}. 

One of the main ingredients for the success of this scenario is to find realistic implementations of two-qubit gates, which requires the generation of entanglement between electron flying qubits~\cite{Edlbauer2022,Zhang2024}. The main existing proposals address this issue by considering the entanglement induced in Mach-Zender interferometers where quantum rails, over finite-length regions, are brought sufficiently close to induce a phase shift due to the Coulomb interaction between electrons on neighboring channels~\cite{Vyshnevyy2013}. While the latter proposal is certainly of great interest, other sources of entanglement deserve to be explored in order to offer multiple paths to quantum information processing with electrons. 

In this paper we investigate a mechanism to generate entangled pairs of Levitons by resorting to the entanglement naturally existing in the Cooper pair condensate of a BCS superconductor~\cite{Chtchelkatchev2002,Recher2001,Samuelsson2003,Sauret2005}. More specifically, we will elaborate on the proposal for an on-demand source of energy-entangled electron states (EES), based on a hybrid superconducting system~\cite{Martin1996,Anantram1996,Torres2001,Sauret2004,Taddei2005,Bayandin06,Chevallier2011,Rech2012,Jacquet2020,Benito2022}. The ballistic channels of a single edge of a two-dimensional topological insulator (2DTI)~\cite{Bernevig2006,Konig2007,Qi2011} are coupled via an adjustable quantum point contact (QPC) to the BCS superconductor. These one-dimensional topological edge states, which are termed \textit{helical}, propagate with opposite chiralities and, according to spin-momentum locking, they possess an opposite spin-polarization axis~\cite{Kane2005a,Kane2005b,Dolcetto2016}. Levitons can be injected into the system by an AC periodic voltage or by optical generation with radio-frequency frequency combs~\cite{Aluffi2023}. We decide to focus on the first case for our calculations in order to exploit the theoretical framework of the \textit{photo-assisted formalism}\cite{Vannucci2018,Ronetti2024}, but our results are valid for any type of source of Lorentzian-shaped pulses.

We derive a perturbative expression for the quantum state emitted in this configuration. Since two simultaneous tunneling events are required for one Cooper pair to be created, we employ perturbation theory up to second order in the tunneling amplitude. Our focus is on the Andreev regime where the superconducting gap is the largest energy scale. In this limit, the BCS ground state is unperturbed and BCS excitations are excluded, thus also validating the mean field approach here considered.

The emitted state propagates non-locally by spreading over the two spin-polarized edge channels. In order to assess its properties, we compute analytically the charge locally backscattered at the QPC by a quantum average over the emitted state. Importantly, we show that the entangled nature of the quantum state can be tested in a multiple-QPC setup with five terminals, i.e. one superconducting lead, two sources and two detectors. In this situation, we compute a quantity related to the detector's cross-correlations~\cite{Martin2005} and show that it is always monotonous for separable states, while it can change sign as a function of the system parameters for entangled states.

The paper is organized as follows.
In Section~\ref{ModelSec} we introduce the Hamiltonian model we use for a time-dependent perturbation theory in the tunnel coupling.
Then in Sec.~\ref{sec:pert-calc} this formalism is used to compute the quantum state analytically, when the junction is driven by a periodic bias, in the Andreev regime of frequencies.
This allows us to compute transport properties, such as the charge transmitted, in Sec.~\ref{sec:charge}. 
Finally, in Sec.~\ref{sec:noise}, a way to measure the entanglement of the output state is also proposed, based on a multiple-QPC scheme.
In the following, units where $\hbar=1$ and $k_B=1$ are employed.

\section{Model}\label{ModelSec}
\begin{figure}
    \includegraphics[width=\linewidth]{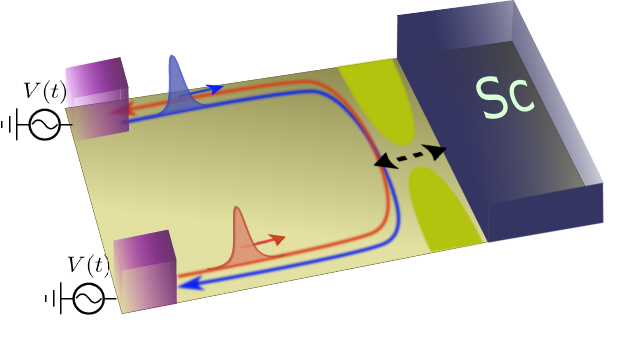}
    \caption{The setup: an adjustable QPC connects a superconductor (right, in blue) and the helical edge states of a Quantum spin Hall bar. Along the two channels quantized Lorentzian-shaped pulses are injected. The interplay between spin-momentum locking and BCS superconductivity allows to emit entangled Leviton pairs in the normal part.}
    \label{fig1}
\end{figure}
We focus on the three terminal device represented in Fig.~\ref{fig1}. {The spin $\uparrow$ ($\downarrow$) edge state is exiting an electron reservoir in the top (bottom) part of the bar and propagating with a fixed chirality to the second electron reservoir in the bottom (top) part, after having interacted with the superconducting terminals. The 2DTI edge states and the superconducting lead are described in equilibrium by the Hamiltonian
}
\begin{align}
H_{\text{N}} &= \sum_{\substack{k_N\\\sigma=\uparrow,\downarrow}} E^{\sigma}_N(k_N)c^{\dagger}_{k_N,\sigma}c^{\phantom{\dagger}}_{k_N,\sigma}\,, \nonumber\\
H_{\text{S}} &=\sum_{\substack{k_S\\\sigma=\uparrow,\downarrow}} \epsilon_{k_S}c^{\dagger}_{k_S,S,\sigma}c^{\phantom{\dagger}}_{k_S,S,\sigma}\nonumber\\
&+\Delta\sum_{k_S}\left(c_{k_S,\text{S},\downarrow}^\dagger c_{-k_S,\text{S},\uparrow}^\dagger +c_{-k_S,\text{S},\uparrow}^{\phantom{\dagger}}c_{k_S,\text{S},\downarrow}^{\phantom{\dagger}}\right),
\end{align}
where $H_{N}$ ($H_S$) is the kinetic part of the Hamiltonian of the normal (superconducting) lead, $\Delta$ is the superconducting gap and $c_{k,\sigma}^{\phantom{\dagger}}$ is the annihilation operator for electrons with momentum $k$ and spin $\sigma$ in the normal lead and $c_{k,S,\sigma}^{\phantom{\dagger}}$ in the superconducting one. 
The chemical potential is set to zero everywhere. The energy dispersion for the helical edge states is~\cite{Qi2011,Dolcetto2016}
\begin{equation}
E_N^{\uparrow/\downarrow} (k) = v_{\uparrow/\downarrow}\left(k-k_{\uparrow/\downarrow}\right),
\end{equation}
where spin-momentum locking imposes $v_{\uparrow} = -v_{\downarrow}$ and $k_{\uparrow}=-k_{\downarrow}$.

The superconducting Hamiltonian can be diagonalized by resorting to the Bogoliubov-Valatin transformation. The latter introduces new fermions in the superconductor~\cite{Schrieffer1983}:
\begin{equation}\label{eq:bvtrasfo}
  \gamma^{\phantom{\dagger}}_{k,\sigma}= u_k c_{k,S,\sigma}^\dagger + \text{sign}\left(\overline{\sigma}\right)v_k c_{-k,S,\overline{\sigma}}\, ,
\end{equation}
where $\text{sign}(\uparrow / \downarrow)=\pm $ and $u_k$ and $v_k$ are the superconductor coherence factors, defined up to a complex phase factor as 
\begin{equation}
  \begin{aligned}
    |u_k|&=\frac{1}{\sqrt{2}}\sqrt{1+\frac{\varepsilon_k}{E_S(k)}},\\
    |v_k|&=\frac{1}{\sqrt{2}}\sqrt{1-\frac{\varepsilon_k}{E_S(k)}},
  \end{aligned}
  \label{eq:ukvk}
\end{equation}
where $E_S (k)=\left(\varepsilon_k^2+\Delta^2\right)^{1/2}$ is the energy required to create an excitation of momentum $k$ in the superconductor and $\epsilon_k$ is the kinetic energy of the said excitation. We see that in the Andreev limit, i.e. for $\varepsilon_k\ll\Delta$, $E_S(k) \approx\Delta$ is independent of $k$ and $\lvert u_k\rvert=\lvert v_k\rvert=1/\sqrt{2}$ for any $k$. 

The fermion operators $\gamma_{k,\sigma}$ describe quasi-particles excited at energies above the BCS ground state, which represents the vacuum state for these fermions. Therefore, their action on the BCS ground state $\left\lvert\Psi_{\text{BCS}}\right\rangle$ is
\begin{equation}\label{eq:bvonbcs}
  \begin{aligned}
    \gamma^{\phantom{\dagger}}_{k,\sigma}\left\lvert\Psi_{\text{BCS}}\right\rangle &= 0,\\
    \gamma^{\phantom{\dagger}}_{k',\sigma'}\gamma_{k\sigma}^\dagger\left\lvert\Psi_{\text{BCS}}\right\rangle &= \delta_{k,k'}\delta_{\sigma,\sigma'}\left\lvert\Psi_{\text{BCS}}\right\rangle\, .
  \end{aligned}
\end{equation}

\subsection{Tunneling matrix and periodic drive}
In our proposal, we connect the helical edge of the 2DTI to the superconducting lead by means of a QPC. 
The formal description of the tunneling processes can be carried out in the Nambu-Keldysh formalism by identifying the edge states of the 2DTI as the left lead and the superconductor as the right one. 
Following this line, one  defines the Nambu spinors
\begin{equation}
	\psi_L^\dagger=\begin{pmatrix}
		\tilde{c}_{0,\uparrow}^\dagger & \tilde{c}_{0,\downarrow}
	\end{pmatrix}\, ,\qquad \psi_R^\dagger=\begin{pmatrix}
		\tilde{c}_{0,S,\uparrow}^\dagger & \tilde{c}_{0,S,\downarrow}
	\end{pmatrix}\, ,
\end{equation}
where we introduce the fermion operators in real space $\tilde{c}_{j,\sigma}$ and $\tilde{c}_{j,S,\sigma}$ for the normal and superconducting parts, respectively. The tunneling occurs at the site $j=0$. The corresponding tunneling Hamiltonian between the leads is
\begin{equation}\label{CurrentHamApp}
	H_\text{T}= \psi_L^\dagger W_{LR} \psi_{R}+\text{H.c.}  \, .
\end{equation}
The tunnel matrix between the coupled sites of the left and right leads of the junction is Hermitian, i.e., $W_{LR}^\dagger=W_{RL}$, and it is defined as
\begin{equation}
	W_{LR}=\lambda\sigma_z e^{i \sigma_z \phi(t)}\, ,
\end{equation}
with $\sigma_z$ the Pauli matrix in Nambu space, and $\lambda$ the tunneling amplitude. The phase
$\phi(t) = e\int_{-\infty}^{t} \mathrm{d}t'\, V(t')$ is the time-dependent phase difference between the leads which accounts for the drive $V(t)$, applied to the normal leads. We focus on periodical drives with with period $\mathcal{T}$ and frequency $\Omega /2\pi$, such that $\mathcal{T}=2\pi /\Omega$.

For later calculations, it is convenient to introduce the photo-assisted coefficients $p_l$ by means of which we can express the electron phase in terms of the following Fourier series
\begin{equation}
\label{eq:pl}
e^{i\phi(t)} = \sum_{l}p_l e^{i (l+q)\Omega t}\, .
\end{equation}

In the following we will consider a specific form of the voltage drive:
\begin{equation}
	\label{eq:Levitons}
V(t)=\sum\limits_{k=-\infty}^{+\infty}\frac{V_0}{\pi}\frac{\gamma^2}{\gamma^2+(t-k \mathcal{T})^2}\,.
\end{equation}
The above equation represents a periodic train of Lorentzian-shaped pulses with amplitude $V_0$ and width $\gamma$. Each pulse carries a charge $-q e = e/(2\pi)\int_0^{\mathcal{T}}\mathrm{d}t\, V(t)$. The main characteristic of Lorentzian-shaped pulses is that, for integer values of $q$, the corresponding photo-assisted coefficients satisfy $p_l = 0 $ for $l<-q$. 
In this case, the injected pulses are termed \textit{Levitons}. The general expression for these photo-assisted coefficients in the case of Lorentzian-shaped pulses is 
\begin{equation}
p_l = q e^{-2\pi \eta l} \sum_{s=0}^\infty (-1)^{s} \frac{\Gamma(l+s+q)}{\Gamma(1+q-s)} \frac{e^{-4\pi \eta s}}{s! (s+l)!},
\label{eq:PhotoAssistedCoefficients}
\end{equation}
where we introduced $\eta = \gamma /\mathcal{T}$.

\subsection{Time evolution and interaction picture}
Below, we introduce the time-evolution operator in the interaction picture with respect to the tunneling Hamiltonian as a preliminary step to the perturbative calculation of the electron state created in the system by the voltage drive.

As a starting point, we separate the total Hamiltonian between the contribution of the two separated leads as $H_0 \equiv H_N + H_S$ and the tunneling term, thus finding for the total Hamiltonian
\begin{equation}
	H=H_{0}+H_{\text{T}}\, .
\end{equation}

It is useful to provide an explicit expression of the tunneling Hamiltonian 
\begin{equation}
  H_T=\lambda \sum_{\substack{k_{N},k_{S},\\\sigma=\uparrow\downarrow,\epsilon=\pm}}\left(e^{-i\phi(t)}c_{k_{N},\sigma}^\dagger c_{k_S,S,\sigma}\right)^{\epsilon}\, ,
\end{equation}
where $\epsilon=-$ indicates the Hermitian conjugate. 
For the perturbative calculations, we resort to an interaction picture with $H_{\text{T}}$ as a perturbation. As a result, we can introduce the time-evolution operator which is written as
\begin{equation}
  S(t,-\infty)=T\exp\left[-i\int_{-\infty}^t\mathrm{d}t'H_{TI}(t')\right]\, ,
\end{equation}
where $T$ is the time ordering operator, and where $H_{TI}(t)$ denotes the tunnel Hamiltonian in the interaction picture as
\begin{equation}
  H_{TI}(t) = e^{iH_0t}H_Te^{-iH_0t}\,,
\end{equation}
 As anticipated, Cooper pair emission requires two tunneling events and it is therefore dominated by the $\lambda^2$ terms. 
As a result, for the following calculation of the quantum state, it is useful to present the expansion of the time-evolution operator up to the second order in the tunneling amplitude. 
The latter reads
\begin{equation}
S(t,-\infty)= \mathbf{1} + S^{(1)}(t,-\infty) + S^{(2)}(t,-\infty) + \mathcal{O}(\lambda^3),
\end{equation}
where
\begin{equation}
\label{eq:evop0}
\begin{aligned}
  S^{(1)}(t,-\infty)&=-i\int_{-\infty}^t\mathrm{d}t_1H_{TI}(t_1),\\
  S^{(2)}(t,-\infty)&=-\frac{1}{2}\int_{-\infty}^t\mathrm{d}t_1\int_{-\infty}^{t_1}\mathrm{d}t_2\, T\left[H_{TI}(t_1)H_{TI}(t_2)\right]\, .
\end{aligned}
\end{equation}

\section{Perturbative calculation of the quantum state}\label{sec:pert-calc}
In this section, we compute the quantum state induced in our system by the presence of the driving voltage $V(t)$. In order to perform this calculation, we employ perturbation theory and, therefore, we will make use of the perturbative expansion of the time-evolution operator presented in Eq.~\eqref{eq:evop0}. 
The calculation is performed by evaluating the action of the time-evolution operator on the ground state of the system, which we identify as the tensor products of the ground states of each constituent part of the system.
\begin{equation}
\left\lvert 0\right\rangle \equiv \left\lvert F_\uparrow\right\rangle\otimes\left\lvert F_\downarrow\right\rangle\otimes\left\lvert\Psi_\text{BCS}\right\rangle\, ,
\end{equation}
where $\left\lvert F_{\sigma}\right\rangle$ is the Fermi sea of spin channel $\sigma$. After the different parts of the system are coupled by the QPC and the voltage drive is switched on, the resulting quantum state is obtained by the action of the time-evolution operator $S$ on the ground state:
\begin{equation}\label{eq:fs1}
  \left\lvert\tilde{\mathcal{F}}(t)\right\rangle=S(t,-\infty)\left\lvert 0\right\rangle .
\end{equation}
Our focus is on the Andreev regime for which the gap is the largest energy scale and no BCS quasi-particle can be excited at the outcome of second order perturbation theory. As a result, in the time-evolution of the system we exclude all the tunneling processes that alter the BCS ground state of the superconductor. This superselection rule implies that only terms involving products of the pair $\gamma^{\phantom{\dagger}}_{k_1,\sigma_1}\gamma_{k_2,\sigma_2}^\dagger$ can contribute to the evolution of the quantum state. As a consequence, all the odd terms in the perturbative expansion are excluded: in particular, we can neglect the term $S^{(1)}(t,-\infty)$. 

We will focus on the second-order term in the perturbative expansion, such that
\begin{equation}\label{eq:fs2}
  \left\lvert\tilde{\mathcal{F}}^{(2)}(t)\right\rangle= S^{(2)}(t,-\infty)\left\lvert 0\right\rangle .
\end{equation}
The latter state can be formally presented in terms of two contributions as
\begin{equation}
\left\lvert\tilde{\mathcal{F}}^{(2)}(t)\right\rangle = \left\lvert\mathcal{F}_{\rm 1p}(t)\right\rangle + \left\lvert\mathcal{F}_{\rm 2p}(t)\right\rangle .
\label{eq:FormalQuantumState}
\end{equation}
In the above expression, the first term corresponds to the generation of electron-hole pairs above the Fermi ground state of the normal part. The second term describes the creation of pairs of two electrons or two holes. By excluding the creation of quasi-particle excitations above the BCS ground state, the first contribution will be neglected in the rest of this paper. In the Appendix~\ref{app:eh}, we show rigorously that the term $\left\lvert\mathcal{F}_{\rm 1p}(t)\right\rangle$ is independent of the voltage drive in the Andreev limit and can be therefore left out of our discussion. This is in agreement with the physical picture for which only Cooper pairs can be transferred in the limit of large gap. 

For these reasons, we will focus only on the two-particle terms for the rest of this work and set 
\begin{equation}
\left\lvert\tilde{\mathcal{F}}^{(2)}(t)\right\rangle \simeq \left\lvert\mathcal{F}_{\rm 2p}(t)\right\rangle.\label{eq:Identification}
\end{equation}
The above approximation is valid in the Andreev limit $\Delta / \Omega \gg 1$.

\subsection{Time-dependent quantum state}
We compute the two-particle terms in the quantum state generated at the QPC by the presence of the driving voltage. We start by replacing the Fourier coefficients in Eq.~\eqref{eq:pl} and by using Eq.~\eqref{eq:ukvk} allowing us to write $u_kv_k=\pm\Delta/[2 E_S(k)]$, so that one has 
\begin{equation}\label{eq:fs3}
    \begin{aligned}
        \left\lvert\tilde{\mathcal{F}}^{(2)}(t)\right\rangle =\frac{\lambda^2\Delta}{2}\sum_{\substack{k,k',\sigma}}\xi_{\sigma}\Big[&\mathcal{U}^{\sigma}_{kk'}(t)c_{k,\sigma}^\dagger c_{k',\overline{\sigma}}^\dagger\\
        &+ \mathcal{V}^{\sigma}_{kk'}(t)c^{\phantom{\dagger}}_{k,\sigma} c^{\phantom{\dagger}}_{k',\overline{\sigma}}\Big]\left\lvert 0\right\rangle,
    \end{aligned}
\end{equation}
where $\xi_{\uparrow/\downarrow} = \pm $ and
\begin{align}
\mathcal{U}^{\sigma}_{kk'}(t) &= -\sum_{l,m,k_S}\int_{-\infty}^{t}\mathrm{d}t_1\, \int_{-\infty}^{t_1}\mathrm{d}t_2\, \frac{p^*_lp^*_m}{E_{S}(k_S)}e^{\alpha t_2} \nonumber\\&\times e^{i\left[E^{\sigma}_{l+q}(k)-E_S(k_S)\right]t_1}e^{i\left[E^{\bar{\sigma}}_{m+q}(k')+E_S(k_S)\right]t_2},\label{eq:U}\\
\mathcal{V}^{\sigma}_{kk'}(t)&= \sum_{l,m,k_S}\int_{-\infty}^{t}\mathrm{d}t_1\, \int_{-\infty}^{t_1}\mathrm{d}t_2\, \frac{p_lp_m}{E_{S}(k_S)} e^{\alpha t_2}\nonumber\\&\times e^{i\left[-E^{\sigma}_{l+q}(k)-E_S(-k_S)\right]t_1}e^{i\left[-E^{\bar{\sigma}}_{m+q}(k')+E_S(-k_S)\right]t_2},\label{eq:V}
\end{align}
where we defined for notation convenience $E^{\sigma}_{l+q}(k)\equiv E^{\sigma}_N(k)-(l+q)\Omega$ and we introduced the parameter $\alpha \rightarrow 0^+$ to ensure the convergence of integrals. By using the anti-commutation properties of fermion operators, the sum over $\sigma$ in Eq.~\eqref{eq:fs3} can be computed and the state can be recast as
\begin{equation}\label{eq:fs_spin}
    \left\lvert\tilde{\mathcal{F}}^{(2)}(t)\right\rangle =\frac{\lambda^2}{2}\sum_{\substack{k,k'}}\left[\Upsilon^{+}_{kk'}(t)c_{k,\uparrow}^\dagger c_{k',\downarrow}^\dagger+\Upsilon^{-}_{kk'}(t)c^{\phantom{\dagger}}_{k,\uparrow} c^{\phantom{\dagger}}_{k',\downarrow}\right]\left\lvert 0\right\rangle,
\end{equation}
where we defined
\begin{align}
\Upsilon^{+}_{kk'}(t) &= \Delta\left[ \mathcal{U}^{\uparrow}_{kk'}(t) + \mathcal{U}^{\downarrow}_{k'k}(t)\right],\label{eq:yplus}\\
\Upsilon^{-}_{kk'}(t) &= \Delta\left[ \mathcal{V}^{\uparrow}_{kk'}(t) + \mathcal{V}^{\downarrow}_{k'k}(t)\right].\label{eq:yminus}
\end{align}

In the following, we will focus on the calculation of the integrals in $\mathcal{U}^{\sigma}_{kk'} $ and $\mathcal{V}^{\sigma}_{kk'} $: the derivation will be carried out in parallel.
Here, it is useful to express an intermediate result: let $I$ be the following integral
\begin{equation}\label{eq:regu}
  \begin{aligned}
    I &= \lim_{\alpha\to 0}\int_{-\infty}^{t_1}\mathrm{d}t_2 \exp\left[i(E - i\alpha)t_2\right]\\
    &= \lim_{\alpha\to 0}\frac{-ie^{iEt_1}}{E - i\alpha}\, ,
  \end{aligned}
\end{equation}
This allows us to perform the time integrals in Eqs.~\eqref{eq:U} and~\eqref{eq:V} successively.
First carrying out the integral over $t_2$, one has
\begin{align}
\mathcal{U}^{\sigma}_{kk'}(t) &= -\sum_{\substack{l,m,k_S}}\int_{-\infty}^{t}\mathrm{d}t_1\, \frac{p^*_lp^*_m e^{\alpha t_1}}{E_{S}(k_S)} \nonumber\times\\&\frac{e^{i\left[E^{\sigma}_{l+q}(k)+E^{\bar{\sigma}}_{m+q}(k')\right]t_1}}{i\left[E^{\bar{\sigma}}_{m+q}(k')+E_S(k_S)-i\alpha\right]},\label{eq:U2}\\
\mathcal{V}^{\sigma}_{kk'}(t) &= \sum_{l,m,k_S}\int_{-\infty}^{t}\mathrm{d}t_1\,\frac{p_lp_m e^{\alpha t_1}}{E_{S}(k_S)}\nonumber\times\\& \frac{e^{-i\left[E^{\sigma}_{l+q}(k)+E^{\bar{\sigma}}_{m+q}(k')\right]t_1}}{i\left[-E^{\bar{\sigma}}_{m+q}(k')+E_S(-k_S)-i\alpha\right]}.\label{eq:V2}
\end{align}
Then, by using again the result of Eq.~\eqref{eq:regu}, we compute the integral over $t_1$, yielding
\begin{align}
&\mathcal{U}^{\sigma}_{kk'}(t)= \sum_{l,m,k_S}\frac{p^*_lp^*_m}{E_{S}(k_S)} \nonumber\\ &\times \frac{e^{i\left[E^{\sigma}_{l+q}(k)+E^{\bar{\sigma}}_{m+q}(k')\right]t}e^{\alpha t}}{\left[E^{\bar{\sigma}}_{m+q}(k')+E_S(k_S)-i\alpha\right]\left[E^{\sigma}_{l+q}(k)+E^{\bar{\sigma}}_{m+q}(k')-i\alpha\right]},\label{eq:U3}\\
&\mathcal{V}^{\sigma}_{kk'}(t) = \sum_{l,m,k_S}\frac{p_lp_m}{E_{S}(k_S)}\nonumber\\ &\times \frac{e^{-i\left[E^{\sigma}_{l+q}(k)+E^{\bar{\sigma}}_{m+q}(k')\right]t}e^{\alpha t}}{\left[-E^{\bar{\sigma}}_{m+q}(k')+E_S(-k_S)-i\alpha\right]\left[E^{\sigma}_{l+q}(k)+E^{\bar{\sigma}}_{m+q}(k')+i\alpha\right]}.\label{eq:V3}
\end{align}
\subsection{Large-gap limit}
In the large-gap limit, defined as the case for which the superconducting gap $\Delta$ is the largest energy scale, the above expressions can be further simplified. In this regime, one can write
\begin{equation}
\frac{1}{\pm E^{\bar{\sigma}}_{m+q}(k') + E_S(\pm k_S)}\sim \frac{1}{E_S(\pm k_S)}
\end{equation}
as long as the terms in the sum over $k$ and $k'$ appearing in the quantum state can be neglected when $\left|E^{\bar{\sigma}}_{m+q}(k')\right|\sim \Delta$. Here, we will show that this is indeed the case. First of all, the energies appearing in $\mathcal{U}_{kk'}^{\uparrow}$ ($\mathcal{V}_{kk'}^{\uparrow}$) must all be positive (negative) because these coefficients multiply two creation (annihilation) operators acting on the vacuum state. As a consequence the expressions $E^{\bar{\sigma}}_{m+q}(k')\pm E_S(\pm k_S)$ never vanish. This means that, when $\left|E^{\bar{\sigma}}_{m+q}(k')\right|\sim \Delta$, the coefficients $\mathcal{U}^{\sigma}_{kk'}(t)$ and $\mathcal{V}^{\sigma}_{kk'}(t)$ behave as $1/\Delta ^3$, while, when $\left|E^{\bar{\sigma}}_{m+q}(k')\right|\ll \Delta$, they behave as $1/\Delta ^2$. Therefore, we can safely set in the Andreev regime:
\begin{align}
&\mathcal{U}^{\sigma}_{kk'}(t)= \frac{A_S}{\Delta} \sum_{l,m}\frac{p^*_lp^*_m e^{i\left[E^{\sigma}_{l+q}(k)+E^{\bar{\sigma}}_{m+q}(k')\right]t}e^{\alpha t}}{\left[E^{\sigma}_{l+q}(k)+E^{\bar{\sigma}}_{m+q}(k')-i\alpha\right]},\label{eq:U4}\\
&\mathcal{V}^{\sigma}_{kk'}(t) = \frac{A_S}{\Delta}\sum_{l,m}\frac{p_lp_m e^{-i\left[E^{\sigma}_{l+q}(k)+E^{\bar{\sigma}}_{m+q}(k')\right]t}e^{\alpha t}}{\left[E^{\sigma}_{l+q}(k)+E^{\bar{\sigma}}_{m+q}(k')+i\alpha\right]},\label{eq:V4}
\end{align}
where, by using $E_S(k_S) = E_S(-k_S)$, we defined the common pre-factor
\begin{equation}
A_S = \Delta \sum_{k_S}\,\frac{1}{E^2_{S}(k_S)},
\end{equation}
which includes all the information about the sum over the superconducting momenta. Interestingly, metallic and superconducting degrees of freedom are decoupled. We can further simplify these coefficients by substituting $l+m \rightarrow l$ and perform the sum over $m$
\begin{equation}
\sum_{m}p^*_{l-m}(q)p_m^*(q) = p_l^*(2q),
\end{equation}
thus obtaining
\begin{align}
&\mathcal{U}^{\sigma}_{kk'}(t)= \frac{A_S}{\Delta} \sum_{l}\frac{p^*_l(2q) e^{i\left[E^{\sigma}_{N}(k)+E^{\bar{\sigma}}_{N}(k')-\left(l+2q\right)\Omega\right]t}e^{\alpha t}}{\left[E^{\sigma}_{N}(k)+E^{\bar{\sigma}}_{N}(k')-\left(l+2q\right)\Omega-i\alpha\right]},\label{eq:U6}\\
&\mathcal{V}^{\sigma}_{kk'}(t) = \frac{A_S}{\Delta} \sum_{l}\frac{p_l(2q) e^{-i\left[E^{\sigma}_{N}(k)+E^{\bar{\sigma}}_{N}(k')-\left(l+2q\right)\Omega\right]t}e^{\alpha t}}{\left[E^{\sigma}_{N}(k)+E^{\bar{\sigma}}_{N}(k')-\left(l+2q\right)\Omega+i\alpha\right]},\label{eq:V6}
\end{align}
We observe that these two coefficients are both invariant if we simultaneously exchange $\sigma \leftrightarrow \bar{\sigma}$ and $k \leftrightarrow k'$. We exploit this invariance to simplify the coefficients defined in Eqs.~\eqref{eq:yplus} and~\eqref{eq:yminus} as
\begin{align}
\Upsilon^{+}_{kk'} &= 2\Delta \mathcal{U}_{kk'},\label{eq:yplus2}\\
\Upsilon^{-}_{kk'} &= 2\Delta \mathcal{V}_{kk'}.\label{eq:yminus2}
\end{align}
where we defined $\mathcal{U}_{kk'}\equiv  \mathcal{U}^{\uparrow}_{kk'} =  \mathcal{U}^{\downarrow}_{k'k}$ and $\mathcal{V}_{kk'} \equiv \mathcal{V}^{\uparrow}_{kk'}= \mathcal{V}^{\downarrow}_{k'k}$.

One readily sees at this stage that the state in Eq.~\eqref{eq:fs_spin} is entangled since it cannot be written as a product of two states acting separately on the two Fermi sea of electrons with spin $\uparrow$ and $\downarrow$. Indeed, as it can be seen from Eqs.~\eqref{eq:yplus} and~\eqref{eq:yminus}, the coefficients $\Upsilon^{\pm}_{kk'}$ cannot be recast as a product of two separate functions of $k$ and $k'$.

For later calculation of the transport properties, it is useful to anticipate that for the squared norms of Eqs.~\eqref{eq:yplus2} and~\eqref{eq:yminus2} one has 
\begin{equation}
\left|\Upsilon^{\pm}_{kk'}(t)\right|^2= \left|\sum_{l}\frac{2A_S ~p_l(2q)e^{i\left(l+2q\right)\Omega t}}{\left[E^{\uparrow}_{N}(k)+E^{\downarrow}_{N}(k')-\left(l+2q\right)\Omega + i\alpha\right]}\right|^2,\label{eq:yplusfinal}
\end{equation}
We comment that the above expressions are periodic functions of time with period $\mathcal{T} = 2\pi / \Omega$.

We are interested in the expression for the state in the long time-limit. For this purpose, we use the result
\begin{equation}
\lim_{t\rightarrow \infty} \frac{e^{i E t}}{i E} = \lim_{t\rightarrow \infty}\int_{-\infty}^t dt' e^{i E t'} =  2\pi \delta(E).
\end{equation}
By inserting the above expression into Eqs.~\eqref{eq:U6} and~\eqref{eq:V6}, one finds in the limit $\alpha \rightarrow 0$
\begin{align}
\Upsilon^{+}_{kk'} &= i 4\pi A_S \sum_{l}p^*_l(2q)\delta\left[E^{\sigma}_{N}(k)+E^{\bar{\sigma}}_{N}(k')-\left(l+2q\right)\Omega\right],\label{eq:U5}\\
\Upsilon^{-}_{kk'} &= -i4\pi A_S \sum_{l}p_l(2q)\delta\left[E^{\sigma}_{N}(k)+E^{\bar{\sigma}}_{N}(k')-\left(l+2q\right)\Omega\right].\label{eq:V5}
\end{align}

The energies appearing in the delta functions of Eq.~\eqref{eq:U5} (Eq.~\eqref{eq:V5}) are always positive (negative) for both $k$ and $k'$ because of the action of the fermion creation (annihilation) operators on the vacuum state of the normal regions. For a train of Lorentzian-shaped pulses for which $2q \in \mathbb{N}_+$, the photo-assisted coefficients obey the property $p_l = 0$ for $l<-2q$. As a result, both delta functions in Eqs.~\eqref{eq:U5} and \eqref{eq:V5} set the condition that the energies $E_N$ are necessarily positive for any $k$ and $k'$. The term with two annihilation operators must then vanish in Eq.~\eqref{eq:fs_spin}. We conclude that the final state generated by driving the system with a periodic train of Lorentzian-shaped pulses (with $2q \in \mathbb{N}_+$) is purely electronic and reads 
\begin{equation}\label{eq:fs5}
    \left\lvert\tilde{\mathcal{F}}^{(2)}(t)\right\rangle =\frac{\lambda^2}{2}\sum_{\substack{k,k'}}\Upsilon^+_{kk'}c_{k,\uparrow}^\dagger c_{k',\downarrow}^\dagger\left\lvert 0\right\rangle.
\end{equation}
We remark that if this state only contains electrons, it is due to the peculiar properties of Levitons. For any other type of drive the terms associated with the two holes cannot be neglected. Indeed, the state in Eq.~\eqref{eq:fs5} is a purely electronic energy-entangled state.

\section{Average backscattered charge}\label{sec:charge}
The time-dependent charge backscattered at the QPC for electrons with spin $\sigma $ is defined as 
\begin{equation}
Q_{\sigma}(t) = e\left\langle \tilde{\mathcal{F}}(t)\right\rvert \sum_{k}c^{\dagger}_{k,\sigma}c^{\phantom{\dagger}}_{k,\sigma}\left\lvert \tilde{\mathcal{F}}(t)\right\rangle.
\end{equation}
Despite the fact that this quantity is defined at the position of the QPC, for 2DTIs, according to the spin-momentum locking of the edge channels, the charge $Q_{\sigma}(t)$ corresponds to the one which is measured in the reservoir at the end of each channel. Moreover, we assumed that the total charge in the vacuum state is zero, such that
\begin{equation}
\left\langle 0 \left\lvert \sum_{k}c^{\dagger}_{k,\sigma}c^{\phantom{\dagger}}_{k,\sigma}\right\rvert 0 \right\rangle = 0\label{eq:Connected}
\end{equation}
We observe that the backscattered charge is induced solely by the presence of the driving voltage and that this quantity is related only to the transfer of Cooper pairs from the superconductor to the topological edge states. 

By using the expression for the two-particle state in Eq.~\eqref{eq:fs_spin} and the coefficients in Eqs.~\eqref{eq:yplus} and~\eqref{eq:yminus}, the average for the charge operator becomes
\begin{widetext}
\begin{align}
Q_{\sigma}(t) = \frac{e\lambda^4}{4}\sum_{\substack{k,k_1,k_1'\\k_2,k_2'}}\Bigg[ & \Upsilon^-_{k_1 k'_1}(t)\left[\Upsilon^-_{k_2 k'_2}(t)\right]^*\left\langle c^{\dagger}_{k'_1,\downarrow}c^{\dagger}_{k_1,\uparrow}c^{\dagger}_{k,\sigma}c_{k,\sigma}c_{k_2,\uparrow}c_{k'_2,\downarrow}\right\rangle \nonumber \\ 
& + \Upsilon^+_{k_2 k'_2}(t)\left[\Upsilon^+_{k_1 k'_1}(t)\right]^*\left\langle c_{k'_1,\downarrow}c_{k_1,\uparrow}c^{\dagger}_{k,\sigma}c_{k,\sigma}c^{\dagger}_{k_2,\uparrow}c^{\dagger}_{k'_2,\downarrow}\right\rangle\Bigg].
\end{align}
We use Wick's theorem to compute the averages. One has:
\begin{align}
 \left\langle c_{k'_1,\downarrow}c_{k_1,\uparrow}c^{\dagger}_{k,\sigma}c_{k,\sigma}c^{\dagger}_{k_2,\uparrow}c^{\dagger}_{k'_2,\downarrow}\right\rangle &=\left(\delta_{\sigma,\uparrow}\delta_{k,k_1}+\delta_{\sigma,\downarrow}\delta_{k,k'_1}\right) \delta_{k'_2,k'_1}\delta_{k_2,k_1}\Theta\left(E^{\uparrow}_N(k_1)\right) \Theta\left(E^{\downarrow}_N(k_1')\right) \\
 \left\langle c^{\dagger}_{k'_1,\downarrow}c^{\dagger}_{k_1,\uparrow}c^{\dagger}_{k,\sigma}c_{k,\sigma}c_{k_2,\uparrow}c_{k'_2,\downarrow}\right\rangle &=-\left(\delta_{\sigma,\uparrow}\delta_{k,k_1}+\delta_{\sigma,\downarrow}\delta_{k,k'_1}\right)\delta_{k'_2,k'_1}\delta_{k_2,k_1}\Theta\left(-E^{\uparrow}_N(k_1)\right) \Theta\left(-E^{\downarrow}_N(k_1')\right).
\end{align}
Here, we focused only on the connected contribution by removing all the average on operators with the same momentum by means of Eq.~\eqref{eq:Connected}.

The time-dependent charge becomes:
\begin{equation}
Q_{\sigma}(t) = \frac{e\lambda^4 }{4}\sum_{k,k'}\Bigg[\left|\Upsilon^+_{k k'}(t)\right|^2\Theta\left(E^{\uparrow}_N(k)\right)\Theta\left(E^{\downarrow}_N(k')\right) - \left|\Upsilon^-_{k k'}(t)\right|^2\Theta\left(-E^{\uparrow}_N(k)\right) \Theta\left(-E^{\downarrow}_N(k')\right)\Bigg].
\end{equation}
\end{widetext}
We replace the discrete sums with integrals over energies, by putting $E^{\uparrow}_N(k) = \omega$ and $E^{\downarrow}_N(k) = \omega'$. Here, we use the linear approximation for the energies of the normal part, which is exact in the case of helical edge states, such that:
\begin{equation}
E_N^{\sigma}(k) = v_{\sigma}\left(k-k_{\sigma}\right).
\end{equation}
The expression for {the backscattered} charge can be recast as
\begin{widetext}
\begin{equation}
Q_{\sigma}(t) = e\lambda^4 A_S^2{\nu_{\uparrow}\nu_{\downarrow}}\int \frac{d\omega}{2\pi} \int \frac{d\omega'}{2\pi}\sum_{l,m}\frac{p_l^*(2q)p_m(2q)e^{i\left(l-m\right)\Omega t} \left[\Theta(\omega)\Theta(\omega')-\Theta(-\omega)\Theta(-\omega')\right]}{\left[\omega + \omega' - \left(l+2q\right)\Omega-i\alpha\right]\left[\omega + \omega' - \left(m+2q\right)\Omega+i\alpha\right]},\label{eq:BackscatteredCharge}
\end{equation}
{where we introduce $\nu_{\sigma} = \sum_{k} \delta\left(\omega - E_N^{\sigma}(k)\right)$ as the density of states of the QSH edge states, which is constant for a system with linear energy dispersion.}
\end{widetext}
The details of the calculation of these integrals are given in Appendix~\ref{app:charge}.
Finally, the charge becomes:
\begin{align}
&Q_{\sigma}(t) =\lambda^4 A_S^2{\nu_{\uparrow}\nu_{\downarrow}}\frac{e^2}{2\pi}\int_{-\infty}^t dt'e^{\alpha\left(t'-t\right)} V(t').
\end{align}
We notice that $Q_{\uparrow}(t) = Q_{\downarrow}(t)$. Starting from the expression for the quantum states, we recover the intuitive result for the backscattered charge, i.e. that it is given by the integral of the drive until the time $t$. Nevertheless, no information about the entangled nature of the state can be extracted from this quantity. In the next section, we will derive an observable which can distinguish between product states and entangled states. We will apply this result to the case of the quantum state derived before and show how to detect the entanglement produced by our source in a modified setup.

\section{Cross-correlation as a witness of electron-electron entanglement}\label{sec:noise}
\begin{figure}[h]
	\centering
	\includegraphics[width=\linewidth]{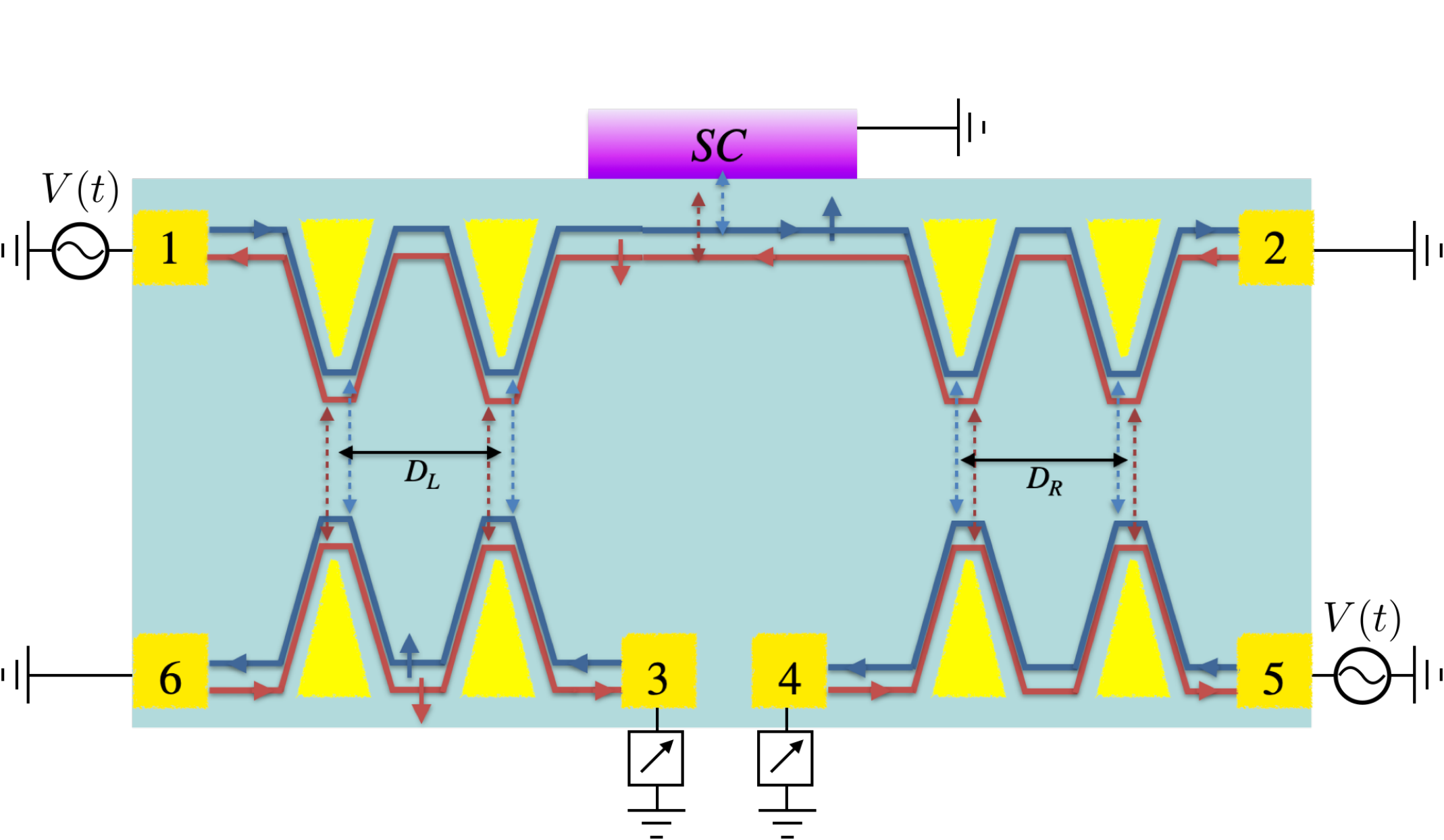}
 \caption{A quantum spin Hall bar in a six-terminal configuration with four QPCs. Terminals $1$ and $2$ are driven by Lorentzian-shaped periodic voltages, while terminals $3$ and $4$ are the detectors and the other terminals are grounded. A superconducting lead is coupled to the helical edge channels in the middle of the quantum spin Hall bar. One double-QPC barrier is placed on each on the two side of the superconductor. By measuring the cross-correlation noise between detectors, one is able to asses the entangled nature of the quantum state generated by the interplay with the BCS superconductivity.}
 	\label{fig:entanglement2}
\end{figure}

In this section we propose a way to detect the energy entanglement of the quantum state appearing in Eq.~\eqref{eq:fs_spin}. The choice of quantized Lorentzian-shaped pulses allows us, as shown in Sec.~\ref{sec:pert-calc}, to focus on the injection of a purely electronic entangled states, by setting $E_N(k)>0$ and $E_N(k')>0$. For the sake of completeness we report the corresponding electronic state here below
\begin{equation}\label{eq:fs5f}
	\left\lvert\tilde{\mathcal{F}}^{(2)}(t)\right\rangle =\frac{\lambda^2}{2}\sum_{k,k'}
	\Upsilon^+_{kk'}  c_{k,\uparrow}^\dagger c_{k',\downarrow}^\dagger\left\lvert 0\right\rangle.
\end{equation}
All the results can be straightforwardly extended to the case of a two-hole state obtained by setting $E_N(k)<0$ and $E_N(k')<0$.

In the following we present the setup and the experimental quantity that can be used to detect the entanglement of the state in Eq.~\eqref{eq:fs5f}. Let us present the principle of the probing device. The entanglement is characterized but the factor $\Upsilon_{k,k’}^{+}$ [see Eq.~\eqref{eq:U5}], which imposes that if the state contains an electron of spin up with energy $k$, the spin down electron has an energy $k’$ which can only take discrete values (these depend on $k$ and on the characteristics of the Leviton drive encoded in $q$, $\Omega$ and the $p_l$ coefficients).  One way to characterize the entanglement is then to take advantage of the helicity of the edge states: the two electrons emitted from the BCS lead travel naturally to the opposite sides of the device. One can then access separately the currents produced by the two electrons from an injected pair. Sending these currents into energy filtering devices will modify the correlations between the two currents in a way which is specific to the entanglement contained in $\Upsilon_{k,k’}^{+}$.   As a simple schematic example, suppose that the spin up current is filtered such that only electron with energy $k_0$ contributes to the current. Then the spin down current only contains electrons of energy $k’$ allowed by  $\Upsilon_{k_0,k’}^{+}$, and the current correlations will depend strongly on how the $k’$ energies are filtered.

The setup is presented in Fig.~\ref{fig:entanglement2}. On each side of the source of entangled electrons, we connect two QPCs in series, each of them is characterized by a constant reflectivity amplitude $r$. The two QPCs are placed at a distance $D_L$ and $D_R$ from each other, respectively in the left and right parts of the device. The presence of two QPCs produces the required energy filtering in the phase of electron states passing through these barriers~\cite{Vannucci2015,Ronetti2016,Ronetti2017}. A similar interferometric setup has already been proposed in Ref. ~\cite{Dasenbrook2015} as a Mach- Zender-like device to probe electron-hole entanglement.

It is instructive to remark that the proposed setup for witnessing energy entanglement described in this paper bears similarities (spin separation followed by energy filtering) with the protocol for detecting Bell inequalities violation in normal metal/BCS superconducting forks from energy entanglement proposed in Ref. \cite{Bayandin06} (which is itself inspired from the photonic case\cite{Rarity90}).

\begin{figure}[t]
\centering
    \subfloat{\includegraphics[width=.22\textwidth]{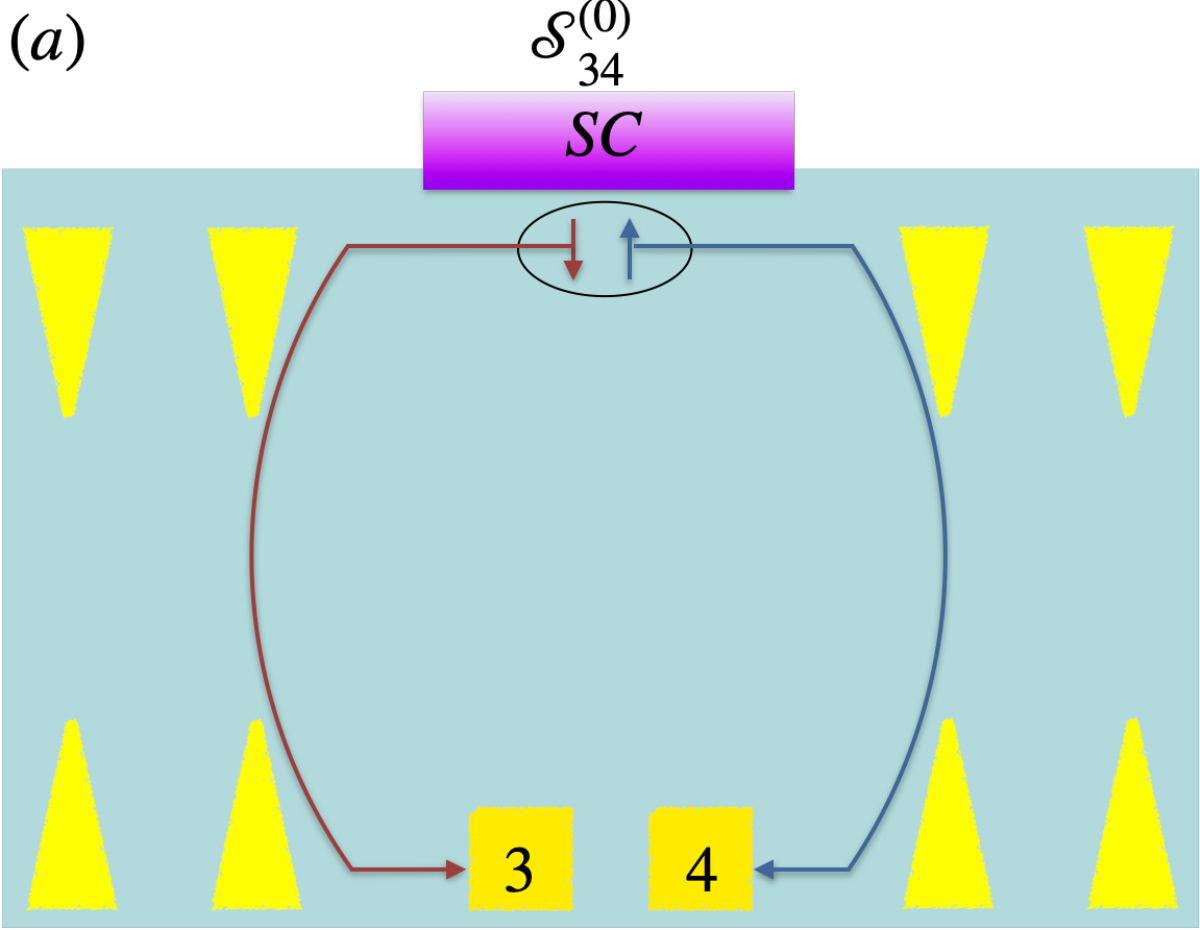}}
    \subfloat{\includegraphics[width=.22\textwidth]{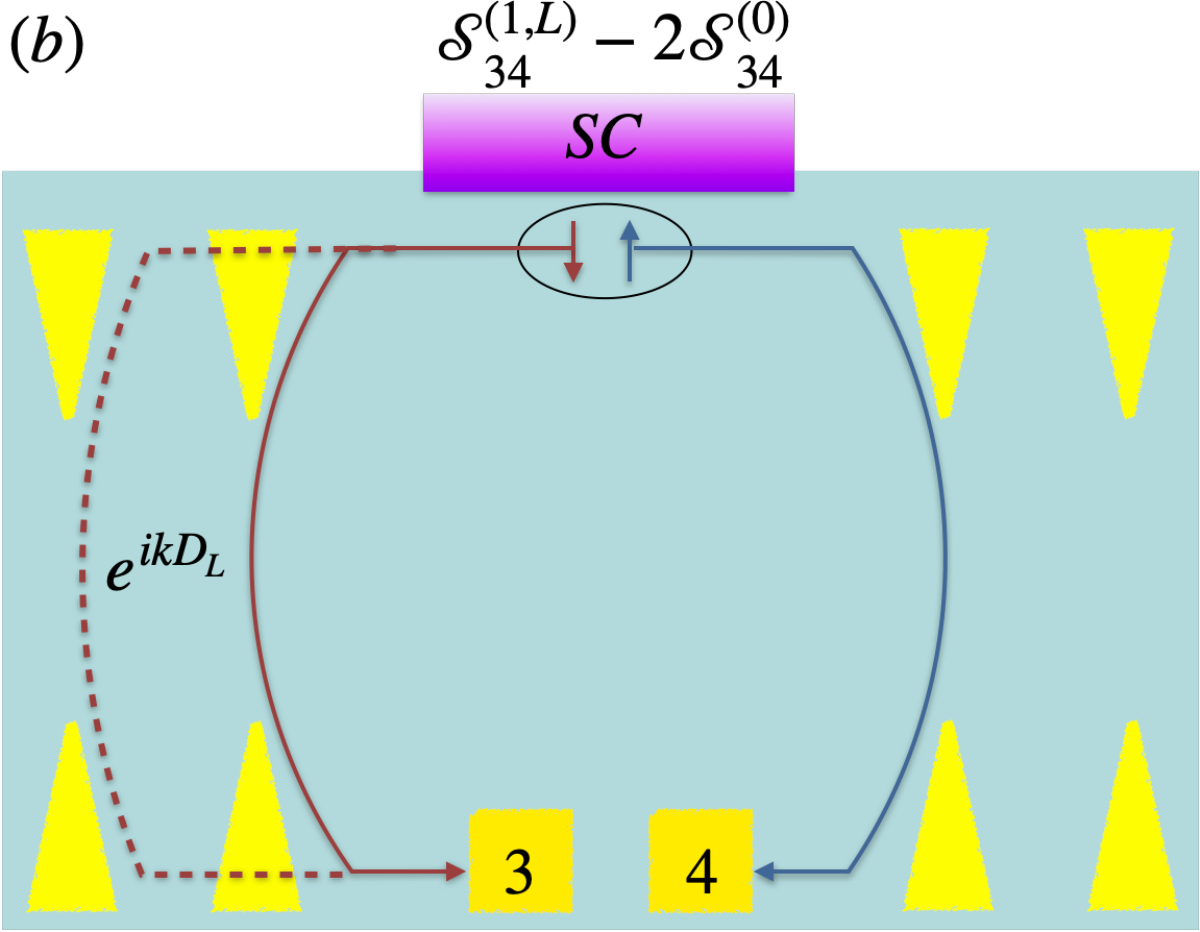}}
    
    \subfloat{\includegraphics[width=.22\textwidth]{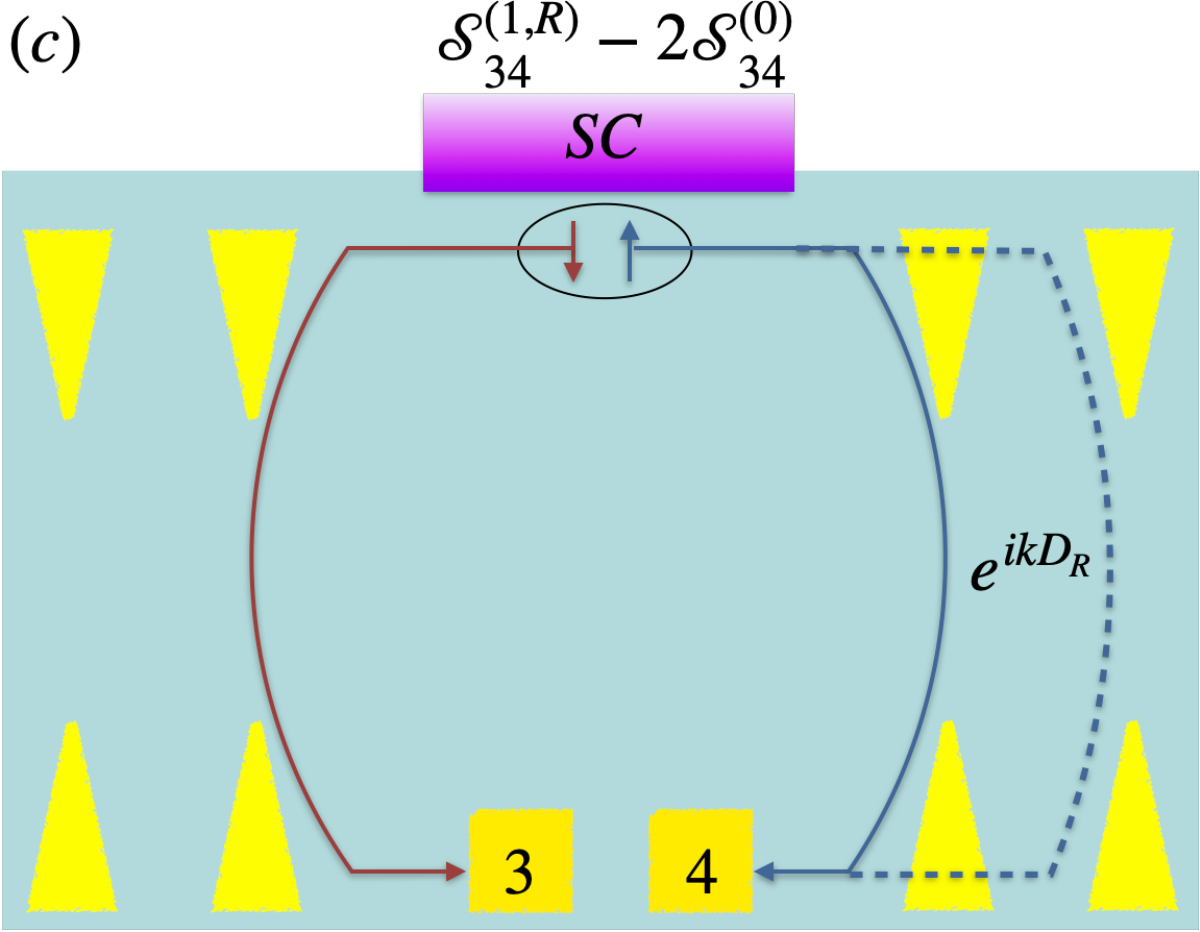}}
    \subfloat{\includegraphics[width=.22\textwidth]{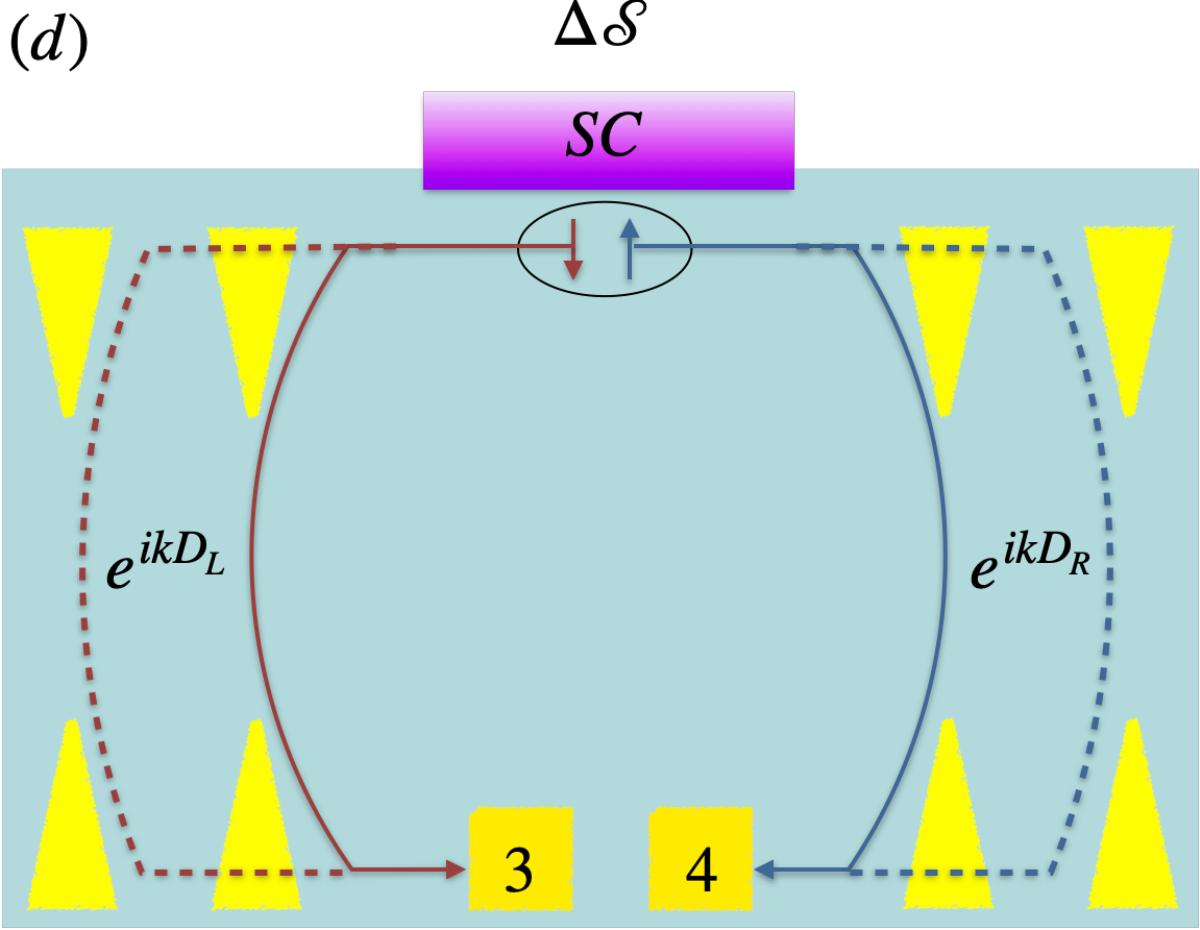}}
    \caption{Sketch of the different tunneling processes associated with different cross-correlators in the weak back-scattering regime. (a) The cross-correlator $\mathcal{S}^{(0)}_{34}$ contains information about tunneling processes where no interference occurs: the tunneling at each QPC is an independent event. (b) and (c) The cross-correlators $\mathcal{S}^{(1,L)}_{34} - \mathcal{S}^{(0)}_{34}$ and $\mathcal{S}^{(1,R)}_{34} - \mathcal{S}^{(0)}_{34}$ includes only the effect of the interference on one of the two double barrier, since the contributions coming from $\mathcal{S}^{(0)}_{34}$ have been subtracted. (d) The cross-correlator $\Delta S$ includes only the tunneling processes where both electrons take quantum interference paths, since all the previous processes have been subtracted according to Eq.~\eqref{eq:DefDS}. In this way each double barrier acts as an energy-filter thus allowing for the detection of energy entanglement between the two electrons.}
    \label{fig:interference}
\end{figure}

In the following, we will show that the entangled nature of the quantum state $\left|\tilde{\mathcal{F}}\right\rangle$ can be assessed by measuring the cross-correlation noise between terminals $3$ and $4$. The latter quantity is defined as
\begin{equation}
\mathcal{S}_{34} = \left\langle \tilde{\mathcal{F}}\right\lvert I_3I_4\left\lvert\tilde{\mathcal{F}}\right\rangle-\left\langle \tilde{\mathcal{F}}\right\lvert I_3\left\lvert\tilde{\mathcal{F}}\right\rangle\left\langle \tilde{\mathcal{F}}\right\lvert I_4\left\lvert\tilde{\mathcal{F}}\right\rangle,\label{eq:DefinitionNoise}
\end{equation}
where we introduced the current operators $I_3 = e v_{\downarrow}\sum_k d^{\dagger}_{k,3}d_{k,3}$ and $I_4 = e v_{\uparrow}\sum_k d^{\dagger}_{k,4}d_{k,4}$.  
In order to exploit the correlation induced by the double barriers to extract information about the state produced at the interface with the SC, we set each QPC to the weak-backscattering regime, i.e $|r|\ll 1$, such that their perturbation on the created entangled state is minimal. Moreover, in this regime, the tunneling paths through the system can be easily interpreted. In Fig.~\ref{fig:interference}, we associated the different tunneling paths at the lowest order in tunneling with a different contribution to the cross-correlation noise: this will be crucial to isolate the relevant contribution containing the necessary energy-filtering as discussed above.

In Fig.~\ref{fig:interference}(a), we present a tunneling path in the absence of interference between two QPCs. These paths give rise to the correlations which can be observed in a setup with no double barriers, but only a single QPC on each side: we term this contribution to the cross-correlation noise $\mathcal{S}^{(0)}_{34}$ and there are 4 such processes in total. Then, in Figs.~\ref{fig:interference}(b) and~\ref{fig:interference}(c), we depict the case where interference paths occur only on one side of the system, left or right, respectively. In this case, the associated cross-correlations are the ones which can be measured in a setup with only one double barrier on the left or the right side, respectively. We indicate these correlations with $\mathcal{S}^{(1,L/R)}_{34}$ and there are two such processes for each side of the system. Actually, the correlations $\mathcal{S}^{(1,L/R)}_{34}$ contains also the contributions coming from the tunneling paths with no interference, such that what is depicted in Figs.~\ref{fig:interference}(b) and~\ref{fig:interference}(c) are tunneling paths giving rise to some correlations quantified by $\mathcal{S}^{(1,L/R)}_{34} - 2\mathcal{S}^{(0)}_{34}$. Finally, in order to isolate the contributions containing only the interference paths on both double barriers, i.e. the ones where there is some energy filtering for both spin $\uparrow$ and $\downarrow$ electrons, we should subtract all the previous contribution from $\mathcal{S}^{34}$, thus defining the following quantity
\begin{equation}
\Delta \mathcal{S}\equiv \mathcal{S}_{34} - 2\left[\mathcal{S}^{(1,L)}_{34} + \mathcal{S}^{(1,R)}_{34} - 2\mathcal{S}^{(0)}_{34}\right],\label{eq:DefDS}
\end{equation}
whose associated tunneling paths are presented in Fig.~\ref{fig:interference}(d).

\subsection{Scattering-matrix description of the double barriers}
In order to compute the cross-correlation noise we employ a scattering matrix description for the double barriers. We will consider the case with a single double barrier or with no double barriers as particular cases.  
The reflectivity amplitudes are assumed to be equal for each QPC and corresponding to $r$. In this regard, the full scattering matrix of the double delta barriers reads~\cite{Lesovik11}
\begin{equation}
S^{(L/R)}_k = \left(\begin{matrix}
	r_{k,L/R} & t'_{k,L/R}\\ t_{k,L/R} & r'_{k,L/R}
\end{matrix}\right),
\end{equation}
where
\begin{align}
t_{k,L/R} = t'_{k, L/R} &= \frac{t^2 e^{i k D_{L/R}}}{1-r^2 e^{i 2k D_{L/R}}},\\
r_{k,L/R} =r'_{k,L/R} &= r\frac{1+(t^2 - r^2)e^{i 2k D_{L/R}}}{1-r^2 e^{i 2k D_{L/R}}},
\end{align}
where we have $R\equiv \left|r \right|^2$, $T\equiv \left|t\right|^2$ and $R + T = 1$. While we presented here the scattering matrix at all orders in tunnelling, all QPCs are assumed to be in the weak backscattering regime. This choice is motivated by the research of a quantity related to the quantum entanglement and it is not a necessary assumption  for the calculation, as the noise can be computed exactly in this configuration by means of the scattering matrix formalism. For the sake of clarity, we decided to name differently the widths of the left and right barriers, respectively as $D_L$ and $D_R$. Nevertheless, for the entanglement witness we will set $D_L=D_R\equiv D$.

In the weak backscattering regime one has $\left|r\right| \ll 1$ and $\left|t \right| \simeq 1$.  Therefore, the coefficients of the scattering matrix become
\begin{align}
\left| t_{k,L/R} \right| = \left| t'_{k,L/R} \right| &\simeq 1,\\
r_{k,L/R} = r'_{k,L/R} &\simeq r \left( 1 + e^{i 2k D_{L/R}}\right).\label{eq:ScatteringMatrixReflection}
\end{align}

The cross-correlation $\mathcal{S}_{34}^{(1,L/R)}$ can be obtained by the replacement 
\begin{equation}
r_{k,R/L} \rightarrow r,\label{eq:repl1}
\end{equation}
respectively. Therefore, these cross-correlators can be measured in the presented setup by fully opening one QPC in the right or left double barrier.  Similarly $\mathcal{S}_{34}^{(0)}$ corresponds to a scattering matrix with 
\begin{align}
r_{k,L}&\rightarrow r\\
r_{k,R}&\rightarrow r.\label{eq:repl2}
\end{align}
In terms of the corresponding experimental configuration, $\mathcal{S}_{34}^{(0)}$ can be measured by fully opening one QPC on both sides of the system.

We describe the tunneling at the double barrier in terms of a scattering problem of fermions exiting or entering all the terminals in the system. The fermions ``close'' to the SC region in between the two double barriers are $c_{k,\uparrow /\downarrow}$, as appearing in Eq.~\eqref{eq:fs5f}. The fermions exiting (entering) terminal $j$, with $j=1,\dots,6$, are termed $c_{k,j}$ ($d_{k,j}$). We relate the fermion operators appearing in the quantum state $\left\lvert\tilde{\mathcal{F}}\right\rangle$ to the ones exiting terminals $1,2,3,4$ as
\begin{align}
c_{k,\uparrow} &= t_{k,L} c_{k,1} + r_{k,L} c_{k,3}\\
c_{k,\downarrow} &= t_{k,R} c_{k,2} + r_{k,R} c_{k,4}.
\end{align}
In the weak backscattering regime, the product appearing in the entangled states reads
\begin{align}
c_{k,\uparrow}^\dagger c_{k',\downarrow}^\dagger &= c_{k,1}^\dagger c_{k',2}^\dagger  e^{-i (k D_L + k' D_R)} + \mathcal{O}\left(r_{1,2}\right).
\end{align}
We see already that, to lowest order in $r$, the combination of operators appearing in Eq.~\eqref{eq:fs5f} is related only to electrons exiting from terminals $1$ and $2$, those which are driven by Lorentzian pulses. In this sense, in the channels $\uparrow$ and $\downarrow$ the source is producing a quantum state which is equivalent to the one derived in the previous section, up to a phase. As a result, despite the presence of the additional QPCs, the presented configuration can be used to probe the properties of the state $\left|\tilde{\mathcal{F}}\right\rangle$. 

The operators appearing in the cross-correlator $S_{34}$ are
\begin{align}
\label{eq:ExitingFermionOperator3}
d_{k,4} &= r_{k,R} c_{k,\uparrow} + t_{k,R} c_{k,5} \sim r_{k,R} t_{k,L} c_{k,1} + t_{k,R} c_{k,5},\\
d_{k,3} &= r_{k,L} c_{k,\downarrow} + t_{k,L} c_{k,6} \sim r_{k,L} t_{k,R} c_{k,2} + t_{k,L} c_{k,6}.
\label{eq:ExitingFermionOperator4}
\end{align}
In the next part, we will use this scattering-matrix description to compute the cross-correlations $\mathcal{S}_{34} (D_L,D_R)$ ~\cite{Burset2023}.

\subsection{Calculation of the cross-correlations}
The cross-correlation noise in Eq.~\eqref{eq:DefinitionNoise} can be computed in the weak-backscattering regime and to lowest order in $\lambda$ as
\begin{widetext}
\begin{equation}
	\mathcal{S}_{34}\simeq \frac{\lambda^4}{4}{e^2v_{\uparrow}v_{\downarrow}}\sum_{k_3,k_4}\sum_{\substack{k,k'\\\overline{k},\overline{k}'}}\left[\left|r_{k_3,L}\right|^2 \left|r_{k_4,R}\right|^2 \Upsilon^+_{kk'}\left[\Upsilon^+_{\overline{k}\overline{k}'}\right]^*\left\langle c_{\overline{k},2} \left(c^{\dagger}_{k_3,2   }c_{k_3,2} - N_{k_3,2}\right) c^{\dagger}_{k,2}\right\rangle \left\langle c_{\overline{k}',1}\left(c^{\dagger}_{k_4,1}c_{k_4,1} - N_{k_4,1}\right)c^{\dagger}_{k',1}\right\rangle  \right]\, .\label{eq:WBNoise}
\end{equation}
\end{widetext}
where we defined $N_{k,j} = \left\langle 0\left\lvert c^{\dagger}_{k,j}c_{k,j}\right\lvert0\right\rangle$.  Moreover, since the operators $c_{k,5}$ and $c_{k,6}$ do not appear in $\left\lvert\tilde{\mathcal{F}}\right\rangle$ they will not contribute to the calculation of the cross-correlator, despite their presence in {Eqs.~\eqref{eq:ExitingFermionOperator3} and \eqref{eq:ExitingFermionOperator4}}. As a result, the cross-correlator is directly related to the electrons going out of terminals $1$ and $2$, \textit{upstream} with respect to the QPC, to lowest order in the tunneling amplitudes. This means that our observable is a probe of the source of entangled electrons in the absence of the double barriers, despite the presence of the QPCs. 

The cross-correlator in Eq.~\eqref{eq:WBNoise} can be simplified by using Wick's theorem, thus finding
\begin{align}
	\mathcal{S}_{34}=\frac{\lambda^4}{4}{e^2v_{\uparrow}v_{\downarrow}}\sum_{k,k'} & \left|r_{k,L}\right|^2 \left|r_{k',R}\right|^2 \nonumber \\
  & \times  \left|\Upsilon^+_{kk'}\right|^2   \Theta(E^{\uparrow}_N(k))\Theta(E^{\downarrow}_N(k')).
\end{align}
By using Eqs.~\eqref{eq:repl1} and~\eqref{eq:repl2}, one can find the cross-correlations in the other configurations as
\begin{align}
\mathcal{S}^{(1,L)}_{34}=&\frac{\lambda^4 \left|r\right|^2e^2v_{\uparrow}v_{\downarrow}}{4}\sum_{k,k'} \left|r_{k,L}\right|^2\left|\Upsilon^+_{kk'}\right|^2\nonumber \\
  & \times   \Theta(E^{\uparrow}_N(k))\Theta(E^{\downarrow}_N(k')),\\
\mathcal{S}^{(1,R)}_{34}=&\frac{\lambda^4\left|r\right|^2e^2v_{\uparrow}v_{\downarrow}}{4}\sum_{k,k'}  \left|r_{k',R}\right|^2 \left|\Upsilon^+_{kk'}\right|^2 \nonumber \\
  & \times  \Theta(E^{\uparrow}_N(k))\Theta(E^{\downarrow}_N(k')),\\
\mathcal{S}^{(0)}_{34}=&\frac{\lambda^4\left|r\right|^4e^2v_{\uparrow}v_{\downarrow}}{4}\sum_{k,k'}  \left|\Upsilon^+_{kk'}\right|^2   \Theta(E^{\uparrow}_N(k))\Theta(E^{\downarrow}_N(k')).
\end{align}
By inserting the above quantity in the definition of the entanglement witness in Eq.~\eqref{eq:DefDS} and the scattering matrix coefficients in the weak backscattering limit [see Eq.~\eqref{eq:ScatteringMatrixReflection}], we obtain
\begin{equation}
\begin{aligned}
	&\Delta \mathcal{S} = \lambda^4 R^2 {e^2v_{\uparrow}v_{\downarrow}}\\&\times\sum_{k,k'} \Theta(E^{\uparrow}_N(k))\Theta(E^{\downarrow}_N(k')) \cos \left(2 k D\right)\cos \left(2 k' D\right)\left|\Upsilon^+_{kk'}\right|^2
 \end{aligned}\label{eq:NoiseStart}
\end{equation}
In the rest of this section, we will characterize the latter quantity and show that it can be used as a witness of the electron-electron entanglement of the state in Eq.~\eqref{eq:fs5f} in systems with time-reversal symmetry (TRS), such as the QSH edge states.

\subsection{Witness of electron-electron entanglement}

For a separable state, the two-particle sector can always be decomposed into the product of single-particle states~\cite{Amico08}. As a result, the matrix appearing in Eq.~\eqref{eq:fs5f} would assume the form {$\Upsilon_{kk'} = f^{\uparrow}_k f^{\downarrow}_{k'}$}, such that
\begin{equation}
\left|\tilde{\mathcal{F}}^{(2)}(t)\right\rangle = \frac{\lambda^2}{2}\sum_k f^{\uparrow}_k c^{\dagger}_{k,\uparrow}\left|F_{\uparrow}\right\rangle \times
\sum_{k'} f^{\downarrow}_{k'} c^{\dagger}_{k',\downarrow}\left|F_{\downarrow}\right\rangle 
\end{equation}
\begin{figure}[t]
    \includegraphics[width=\linewidth]{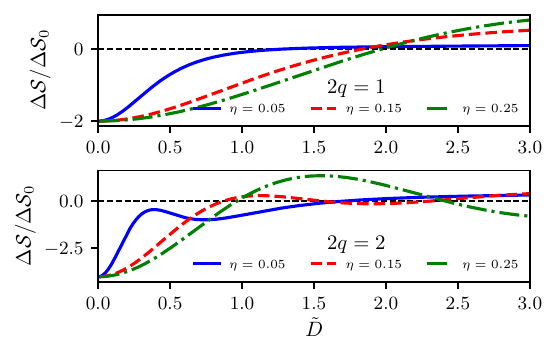}
    \caption{The entanglement witness $\Delta \mathcal{S}$ for $\alpha_F = 0$ in units of $\Delta \mathcal{S}_0$ as a function of $\tilde{D}$ for different values of the re-scaled Leviton half-height width $\eta = \gamma /\mathcal{T}$. The charge is set to $2q=1$ in the upper panel and to $2q=2$ in the lower panel.}
    \label{fig:Noise1}
\end{figure}For such separable state, the combination of cross-correlators defined in Eq.~\eqref{eq:DefDS} becomes
\begin{equation}
    \begin{aligned}
	&\Delta \mathcal{S} = \lambda^4 R^2 {e^2v_{\uparrow}v_{\downarrow}}\\&\times\sum_{k,k'}\Theta(E^{\uparrow}_N(k))\Theta(E^{\downarrow}_N(k')) \cos \left(2 k D\right)\cos \left(2 k' D\right)\left|f^{\uparrow}_k\right|^2\left|f^{\downarrow}_{k'}\right|^2.
\end{aligned}\label{eq:NoiseSeparableState}
\end{equation}
For systems with TRS, one has that $E_N^{\uparrow}(k) = E_N^{\downarrow}(\mp k)$, $v_{\uparrow} = -v_{\downarrow}$ and $f^{\uparrow}_k = f^{\downarrow}_{\mp k}$, respectively, such that the expression in Eq.~\eqref{eq:NoiseSeparableState} becomes
\begin{equation}
	\Delta \mathcal{S} = -\lambda^4 R^2 {e^2 v^2_{\uparrow}}\left|\sum_{k}\Theta(E^{\uparrow}_N(k)) \cos \left(2 k D\right)\left|f^{\uparrow}_k\right|^2\right|^2,
 \label{eq:NoiseSeparableStateTRS}
\end{equation}
which is manifestly a {negative or null} quantity for any choice of parameters. The condition $\Delta \mathcal{S} \le 0$ for separable state allows us to employ this quantity as a witness of electron-electron entanglement. For separable state, the quantum interference processes occuring on the left or right double barrier are completely independent. As a consequence the cross-correlations of two-particle are simply the square modulus of the same single-particle quantity, as it always happens for independent entities. 

\begin{figure}[t]
    \includegraphics[width=\linewidth]{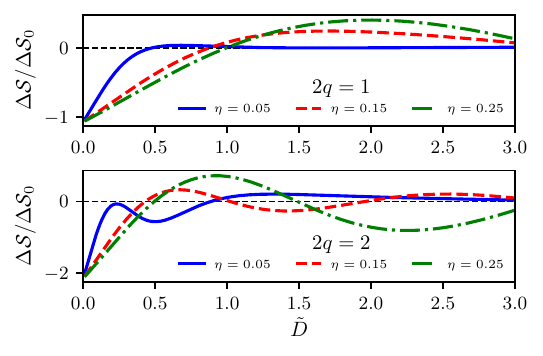}
    \caption{The entanglement witness $\Delta \mathcal{S}$ for $\alpha_F = 1.5$ in units of $\Delta \mathcal{S}_0$ as a function of $\tilde{D}$ for different values of the re-scaled Leviton half-height width $\eta = \gamma /\mathcal{T}$. The charge is set to $2q=1$ in the upper panel and to $2q=2$ in the lower panel.}
    \label{fig:Noise2}
\end{figure}

For the entangled-state of Levitons, we will show that the same quantity can change sign as a function of the parameters of the interferometers. We can compute this quantity explicitly by using the linear approximation for the energy dispersion
\begin{equation}
E_N^{\sigma}(k) = v_{\sigma}\left(k-k_{\sigma}\right)
\end{equation}
The main calculation is carried out in the Appendix~\ref{app:noise}. {For systems with TRS,} the final expression is
\begin{align}
\Delta \mathcal{S} & = -\Delta \mathcal{S}_0\sum_{l>-2q}\left|p_l\right|^2\left\{ (l+2 q) \cos \left[\alpha_F+\tilde{D} (l+2q)\right]\right. \nonumber \\
&+\left.\vphantom+ \tilde{D}^{-1}\sin \left[\tilde{D}(l+2
   q)\right]\right\}
\label{eq:NoiseFinal}
\end{align}
where we defined {$\Delta \mathcal{S}_0 = 2e^2 v^2_{\uparrow}\nu^2_{\uparrow}A_S^2\Omega \lambda^4R$}. Here we also introduced $\alpha_F = 4k_{\uparrow}D$ as the phase acquired by an electron travelling across one double barrier, and $\tilde{D} = D \Omega / v_{\uparrow}$ as the ratio between the length of the barrier and the average space separation between two consecutive Leviton pulses. The expression in Eq.~\eqref{eq:NoiseFinal} can be positive or negative as a function of the dimensionless system parameters. We conclude that observing a positive value of $\Delta S$ as a function of $\tilde{D}$ is an indicator of the entangled nature of the quantum state. Physically, this can be explained by the fact that, for entangled states, the tunneling processes occurring on the two barriers are not independent and give rise to additional quantum interference mechanisms which are absent for separable states. In other words, the intrinsic two-particle nature of these correlations becomes manifest in the interference oscillations that appear in Figs.~\ref{fig:Noise1} and~\ref{fig:Noise2}.

We remark that the expression in Eq.~\eqref{eq:NoiseFinal} is valid only in the case $2q \in \mathbb{Z}$. The values of the photo-assisted coefficients are presented in Eq.~\eqref{eq:PhotoAssistedCoefficients}. In particular, for $2q=1$, one can derive an analytical expression for $\Delta S$ (see Appendix~\ref{app:noise})
\begin{widetext}
\begin{equation}
\Delta \mathcal{S} =\Delta \mathcal{S}_0\left(1-e^{-4\pi\eta}\right)^2\frac{e^{8 \pi  \eta } \left[\tilde{D} \cos \left(\alpha_F +\tilde{D}\right)+\sin (\tilde{D})\right]-e^{4 \pi  \eta } \left[2 \tilde{D} \cos (\alpha_F
   )+\sin (2 \tilde{D})\right]+\tilde{D} \cos \left(\tilde{D}-\alpha_F \right)+\sin (D)}{4 D \left[\cos (D)-\cosh (4 \pi  \eta )\right]^2}
\end{equation}
\end{widetext}
The entanglement witness is presented for $2q=1$ and $2q=2$ as a function of $\tilde{D}$ in Figs.~\ref{fig:Noise1} and~\ref{fig:Noise2} for $\alpha_F = 0$ and $\alpha_F = 1.5$, respectively. In both cases, this quantity is changing sign as a function of the system parameters, thus proving that it can be used as a test for the entanglement generated by the proposed source.
\section{Conclusion}
Here, we considered a novel protocol to create entangled pairs of Levitons in the helical edge states of a two-dimensional quantum spin Hall systems. The entanglement naturally arises in our proposal by exploiting the proximity effect of a BCS superconductor, whose ground state is a condensate of Cooper pairs entangled in the energy domain. In particular, we proposed an on-demand periodic source of energy-entangled electron states. We focused on a two-dimensional topological insulator (2DTI) {whose edge states} are coupled via an adjustable quantum point contact (QPC) to the BCS superconductor. The choice of a two-dimensional topological insulator is illustrative and we speculate that our results can be extended to other systems with spin-polarized edge states, such as the chiral edge states of the quantum Hall effect at $\nu=2$. Two sources of Levitons are connected to the topological bar. We decide to focus on the case of voltage injection of Levitons, but our results are valid for any protocol of injection of Lorentzian-shaped pulses {(see Ref.~\cite{Aluffi2023}).}

We employed perturbation theory up to second order in the tunneling amplitude to compute the quantum state emitted in this configuration. Our focus is on the regime where the superconducting gap is the largest energy scale, i.e. the Andreev regime. In this limit, the BCS ground state is unperturbed and BCS excitations are excluded {in the final outcome}, thus also validating the mean field approach here considered.

Moreover, we computed analytically the charge locally backscattered at the QPC by a quantum average over the emitted state and we found the intuitive result that the charge is proportional to the integral of the voltage source, thus validating our perturbative approach. Secondly, we showed that the entangled nature of the quantum state can be tested in a multiple-QPC setup by computing a quantity related to the cross-correlations in this setup. We proved that it is always monotonous and negative for separable states, while it can change sign as a function of the system parameters for entangled states. In future extensions of this paper, we plan to explore ways to quantify the generated entanglement using measures such as concurrence.~\cite{Dolcetto2016}.

The generated energy entangled states can be exploited in a variety of quantum protocols based on nanoscale devices, such as quantum teleportation~\cite{Bennett1993}, quantum key distribution~\cite{Long2002}, secure cryptography~\cite{Ekert1991,Bennett1992,Pan2020}, or quantum dense coding~\cite{Bennett1992b}.
\acknowledgments{This work received support from the French government under the France 2030 investment plan, as part of the Initiative d’Excellence d’Aix-Marseille Université A*MIDEX. We acknowledge support from the institutes IPhU (AMX-19-IET008) and AMUtech (AMX-19-IET-01X). This work has benefited from State aid managed by the Agence Nationale de la Recherche under the France 2030 programme, reference ``ANR-22-PETQ-0012".}

\appendix

\begin{widetext}
\section{Electron-hole creation term}
\label{app:eh}
In this Appendix, we show that the electron-hole term $\left\lvert\mathcal{F}_{\rm 1p}(t)\right\rangle$ appearing in Eq.~\eqref{eq:FormalQuantumState} does not depend on the driving voltage and that its only contribution is to renormalize the coefficient in front of the zeroth-order term.

As a first step, we write the electron-hole term by making explicit the time-dependence of the fermion operator
\begin{equation}
  \begin{aligned}
    \left\lvert\mathcal{F}_{\rm 1p}(t)\right\rangle = -\frac{\lambda^2}{4}\sum_{\substack{k_N,k_N'\\k_S,\sigma}}& \int_{-\infty}^{t}\mathrm{d}t_1\, \int_{-\infty}^{t_1}\mathrm{d}t_2\, \\
    & \times e^{\alpha t_2}\left[e^{-i\phi(t_1)}e^{i\phi(t_2)}e^{i\left[E^{\sigma}_N(k_N)-E_S(k_S)\right]t_1}e^{i\left[-E^{\sigma}_N(k_N')+E_S(k_S)\right]t_2}u_{k_S}^2 c_{k_N,\sigma}^\dagger c^{\phantom{\dagger}}_{k_N',\sigma} \right. \\
    & \left. + e^{i\phi(t_1)}e^{-i\phi(t_2)}e^{i\left[-E^{\sigma}_N(k_N)-E_S(-k_S)\right]t_1}e^{i\left[E^{\sigma}_N(k_N')+E_S(-k_S)\right]t_2}v_{k_S}^2 c^{\phantom{\dagger}}_{k_N,\sigma}c_{k_N',\sigma}^\dagger\right]\left\lvert 0\right\rangle\,.
  \end{aligned}
\end{equation}
where $\alpha$ is a positive regularisation parameter that will be taken to zero, ensuring that the contribution at $-\infty$ vanishes in the integral over $t_2$. Next, we express the voltage phase in terms of the photo-assisted Fourier series in Eq.~\eqref{eq:pl}, thus finding
\begin{equation}
  \begin{aligned}
    \left\lvert\mathcal{F}_{\rm 1p}(t)\right\rangle = -\frac{\lambda^2}{4}\sum_{\substack{k_N,k_N'\\k_S,\sigma}}\sum_{l,m} & \int_{-\infty}^{t}\mathrm{d}t_1\, \int_{-\infty}^{t_1}\mathrm{d}t_2\, \\
    & \times e^{\alpha t_2}\left[ p^*_lp_me^{i\left[E^{\sigma}_N(k_N)-E_S(k_S)-(l+q)\Omega\right]t_1}e^{i\left[-E^{\sigma}_N(k_N')+E_S(k_S)+(m+q)\Omega\right]t_2}u_{k_S}^2 c_{k_N,\sigma}^\dagger c^{\phantom{\dagger}}_{k_N',\sigma} \right.\\
    & \left. + p_lp^*_me^{i\left[-E^{\sigma}_N(k_N)-E_S(-k_S)+(l+q)\Omega\right]t_1}e^{i\left[E^{\sigma}_N(k_N')+E_S(-k_S)-(m+q)\Omega\right]t_2}v_{k_S}^2 c^{\phantom{\dagger}}_{k_N,\sigma}c_{k_N',\sigma}^\dagger\right]\left\lvert 0\right\rangle\,.
  \end{aligned}
\end{equation}
Using the integral in Eq.~\eqref{eq:regu}, we perform the integral over $t_2$ and take the limit $\alpha\to 0$
\begin{equation}
  \begin{aligned}
    \left\lvert\mathcal{F}_{\rm 1p}(t)\right\rangle &= -\frac{\lambda^2}{4}\sum_{\substack{k_N,k_N'\\k_S\sigma}}\sum_{l,m}\int_{-\infty}^{t}\mathrm{d}t_1\, \Bigg[\frac{p^*_lp_me^{i\left[E^{\sigma}_N(k_N)-E^{\sigma}_N(k_N')-(l-m)\Omega\right]t_1}u_{k_S}^2}{E^{\sigma}_N(k_N')-E_S(k_S)-(m+q)\Omega}c_{k_N,\sigma}^\dagger c^{\phantom{\dagger}}_{k_N',\sigma} \\
    +&\frac{p_lp^*_me^{i\left[-E^{\sigma}_N(k_N)+E^{\sigma}_N(k'_N)+(l-m)\Omega\right]t_1}v_{k_S}^2 }{E^{\sigma}_N(k_N')+E_S(-k_S)-(m+q)\Omega}c^{\phantom{\dagger}}_{k_N,\sigma}c_{k_N',\sigma}^\dagger\Bigg]\left\lvert 0\right\rangle\,.
  \end{aligned}
\end{equation}
In the limit of large superconducting gap, we have in particular that $E_S \gg \gamma^{-1}$, where $\gamma$ is the temporal width of the Lorentzian pulses, defined in Eq.~\eqref{eq:Levitons}. This condition imposes an upper bound to the values of $l$ and $m$, given the exponentially-decaying form of the photo-assisted coefficient for a train of Lorentzian-shaped pulses of width $\gamma$. This fact together with the other condition $E_S \gg \Omega$ entails that the term $(m+q)\Omega$ appearing in the denominator of the above expression can be dropped in the Andreev limit. Therefore, one has
\begin{equation}
  \begin{aligned}
    \left\lvert\mathcal{F}_{\rm 1p}(t)\right\rangle&=-\frac{\lambda^2}{4}\sum_{\substack{k_N,k_N'\\k_S\sigma}}\sum_{l,m}\int_{-\infty}^{t}\mathrm{d}t_1\, \Bigg[\frac{p^*_lp_me^{i\left[E^{\sigma}_N(k_N)-E^{\sigma}_N(k_N')-(l-m)\Omega\right]t_1}u_{k_S}^2}{E^{\sigma}_N(k_N')-E_S(k_S)}c_{k_N,\sigma}^\dagger c^{\phantom{\dagger}}_{k_N',\sigma}\\
    +&\frac{p_lp^*_me^{i\left[-E^{\sigma}_N(k_N)+E^{\sigma}_N(k'_N)+(l-m)\Omega\right]t_1}v_{k_S}^2 }{E^{\sigma}_N(k_N')+E_S(-k_S)}c^{\phantom{\dagger}}_{k_N,\sigma}c_{k_N',\sigma}^\dagger\Bigg]\left\lvert 0\right\rangle\,.
  \end{aligned}
\end{equation}
Finally, one can use the following property
\begin{equation}
\sum_{l,m}p^*_lp_m e^{-i(l-m)\Omega t_1} = e^{-i\phi (t_1)}e^{i\phi (t_1)}=1
\end{equation}
to show that
\begin{equation}
  \begin{aligned}
    \left\lvert\mathcal{F}_{\rm 1p}(t)\right\rangle&=-\frac{\lambda^2}{4}\sum_{\substack{k_N,k_N'\\k_S\sigma}}\int_{-\infty}^{t}\mathrm{d}t_1\, \Bigg[\frac{e^{i\left[E^{\sigma}_N(k_N)-E^{\sigma}_N(k_N')\right]t_1}u_{k_S}^2}{E^{\sigma}_N(k_N')-E_S(k_S)}c_{k_N,\sigma}^\dagger c^{\phantom{\dagger}}_{k_N',\sigma}+\frac{e^{i\left[-E^{\sigma}_N(k_N)+E^{\sigma}_N(k'_N)\right]t_1}v_{k_S}^2 }{E^{\sigma}_N(k_N')+E_S(-k_S)}c^{\phantom{\dagger}}_{k_N,\sigma}c_{k_N',\sigma}^\dagger\Bigg]\left\lvert 0\right\rangle\,.
  \end{aligned}
\end{equation}
The latter expression is clearly independent of the driving voltage. As a result, this is an equilibrium term which cannot affect the transport properties of the system at zero temperature. Since this term is not affected by the train of Lorentzian pulses, we just use the identification in Eq.~\eqref{eq:Identification} between the perturbative quantum state $\left\lvert\tilde{\mathcal{F}}^{(2)}(t)\right\rangle$ and the two-particle term $\left\lvert\mathcal{F}_{2p}(t)\right\rangle$.

\section{Backscattered charge calculations}\label{app:charge}
In this Appendix we provide the details for the calculations of the integrals appearing in the backscattered charge in Eq.~\eqref{eq:BackscatteredCharge}. For the sake of completeness, we report here the expression appearing in the main text
\begin{equation}
Q_{\sigma}(t) = {e\lambda^4 A_S^2  \nu_{\uparrow} \nu_{\downarrow}}\int \frac{d\omega}{2\pi} \int \frac{d\omega'}{2\pi}\sum_{l,m}\frac{p_l^*(2q)p_m(2q)e^{i\left(l-m\right)\Omega t}e^{\alpha t} \left[\Theta(\omega)\Theta(\omega')-\Theta(-\omega)\Theta(-\omega')\right]}{\left[\omega + \omega' - \left(l+2q\right)\Omega-i\alpha\right]\left[\omega + \omega' - \left(m+2q\right)\Omega+i\alpha\right]}\label{eq:BackscatteredChargeApp}
\end{equation}
In order to simplify the expression inside the above integral we use the following general identity
\begin{equation}
\frac{1}{x-i\alpha}\frac{1}{y+i\alpha} = \frac{1}{y-x+2i\alpha}\left(\frac{1}{x-i\alpha}-\frac{1}{y+i\alpha}\right),
\end{equation}
which by using $1/\left(x-i\alpha\right) = i\pi \delta\left(x\right) + x/\left(x^2+\alpha^2\right)$, valid in the limit $\alpha \rightarrow 0^+$, becomes
\begin{equation}
\frac{1}{x-i\alpha}\frac{1}{y+i\alpha} = \frac{1}{y-x+2i\alpha}\left(i\pi \delta\left(x\right)+i\pi \delta\left(y\right)+\frac{x}{x^2+\alpha^2}-\frac{y}{y^2+\alpha^2}\right).\label{eq:IdentityAppCharge}
\end{equation}

By using the general identity in Eq.~\eqref{eq:IdentityAppCharge}, the integral appearing in Eq.~\eqref{eq:BackscatteredChargeApp} can be recast as
\begin{align}
Q_{\sigma}(t) = {e\lambda^4 A_S^2  \nu_{\uparrow} \nu_{\downarrow}} \sum_{l,m}\, & \int \frac{d\omega}{2\pi}\int \frac{d\omega'}{2\pi}~\frac{p^*_{l}(2q)p_{m}(2q)e^{\alpha t}e^{i(l-m)\Omega t}}{{(l-m)\Omega}+2i\alpha}\left[\Theta(\omega)\Theta(\omega')-\Theta(-\omega)\Theta(-\omega')\right] \nonumber \\
& \times\left\{ i\pi \delta\left[\omega+\omega' -(l+2q)\Omega\right]+i\pi \delta\left[\omega+\omega' -(m+2q)\Omega\right] \vphantom{+ \frac{\left[\omega+\omega' -(l+2q)\Omega\right]}{\left[\omega+\omega' -(l+2q)\Omega\right]^2+\alpha^2}-\frac{\left[\omega+\omega' -(m+2q)\Omega\right]}{\left[\omega+\omega' -(m+2q)\Omega\right]^2+\alpha^2}} \right. \nonumber \\ 
& \left. + \frac{\left[\omega+\omega' -(l+2q)\Omega\right]}{\left[\omega+\omega' -(l+2q)\Omega\right]^2+\alpha^2}-\frac{\left[\omega+\omega' -(m+2q)\Omega\right]}{\left[\omega+\omega' -(m+2q)\Omega\right]^2+\alpha^2}\right\} 
\label{eq:BackscatteredChargeApp2}
\end{align}
We focus on the second contribution to the integral 
\begin{align}
\frac{1}{(l-m)\Omega+2i\alpha}\int d\omega\int d\omega'\left[\Theta(\omega)\Theta(\omega')-\Theta(-\omega)\Theta(-\omega')\right]\left\{\frac{\left[\omega+\omega' -(l+2q)\Omega\right]}{\left[\omega+\omega' -(l+2q)\Omega\right]^2+\alpha^2}-\frac{\left[\omega+\omega' -(m+2q)\Omega\right]}{\left[\omega+\omega' -(m+2q)\Omega\right]^2+\alpha^2}\right\}
\end{align}
and we will show that it vanishes. We compute the integral over $\omega$, thus obtaining
\begin{align}
&\frac{1}{2}\frac{1}{(l-m)\Omega+2i\alpha}\int d\omega'\left[\Theta(\omega')+\Theta(-\omega')\right]\left\{\log\left\{\left[\omega' -(l+2q)\Omega\right]^2+\alpha^2\right\}-\log\left\{\left[\omega' -(m+2q)\Omega\right]^2+\alpha^2\right\}\right\}\nonumber\\
& = \frac{1}{2}\frac{1}{(l-m)\Omega+2i\alpha}\int d\omega'\left\{\log\left\{\left[\omega' -(l+2q)\Omega\right]^2+\alpha^2\right\}-\log\left\{\left[\omega' -(m+2q)\Omega\right]^2+ \alpha^2\right\}\right\}.
\end{align}
For the last integral, one finds
\begin{align}
&\frac{1}{2}\frac{1}{(l-m)\Omega+2i\alpha}\int d\omega'\left\{\log\left\{\left[\omega' -(l+2q)\Omega\right]^2+\alpha^2\right\}-\log\left\{\left[\omega' -(m+2q)\Omega\right]^2+\alpha^2\right\}\right\}\nonumber\\&=\frac{1}{2}\frac{1}{(l-m)\Omega+2i\alpha}\int d\omega'\left\{\log\left[\omega'^2+\alpha^2\right]-\log\left[\omega'^2+\alpha^2\right]\right\}=0.
\end{align}
Let us now focus on the term with the delta functions in Eq.~\eqref{eq:BackscatteredChargeApp2}. Since for Levitons one has $l+2q>0$ and $m+2q>0$, for the integrals with the Heaviside function $\Theta(-\omega)\Theta(-\omega')$, the argument of the delta functions cannot be zero and, therefore, they vanish everywhere on the integration region $\omega<0$ and $\omega'<0$. By putting all these results together, the charge becomes
\begin{equation}
Q_{\sigma}(t) = \frac{ie\lambda^4 A_S^2  \nu_{\uparrow} \nu_{\downarrow}}{4\pi} \sum_{l,m}\, \frac{p^*_{l}(2q)p_{m}(2q)e^{\alpha t}e^{i(l-m)\Omega t}}{(l-m)\Omega+2i\alpha}\int_0^{\infty} d\omega\int_0^{\infty} d\omega'~\left\{\delta\left[\omega+\omega' -(l+2q)\Omega\right]+\delta\left[\omega+\omega' -(m+2q)\Omega\right]\right\}.\label{eq:BackscatteredChargeApp3}
\end{equation}
We compute the integral over $\omega'$
\begin{equation}
Q_{\sigma}(t) = \frac{ie\lambda^4 A_S^2  \nu_{\uparrow} \nu_{\downarrow}}{4\pi} \sum_{l,m}\, \frac{p^*_{l}(2q)p_{m}(2q)e^{\alpha t}e^{i(l-m)\Omega t}}{(l-m)\Omega + 2 i \alpha}~\left[\int_0^{(l+2q)\Omega}d\omega+\int_0^{(m+2q)\Omega}d\omega\right].\label{eq:BackscatteredChargeApp4}
\end{equation}
The integrals can be simplified to
\begin{equation}
Q_{\sigma}(t) = \frac{ie\lambda^4 A_S^2  \nu_{\uparrow} \nu_{\downarrow}}{4\pi} \sum_{l,m}\, \frac{p^*_{l}(2q)p_{m}(2q)e^{\alpha t}e^{i(l-m)\Omega t}}{(l-m)\Omega + 2 i \alpha}~\left[(l+2q)\Omega+(m+2q)\Omega\right]
\end{equation}
Finally, the charge becomes
\begin{align}
&Q_{\sigma}(t) = \frac{e\lambda^4 A_S^2  \nu_{\uparrow} \nu_{\downarrow}}{4\pi}\sum_{l,m}\, \frac{p^*_{l}(2q)p_{m}(2q)e^{-\alpha t}e^{2\alpha t}e^{i(l-m)\Omega t}}{(l-m)\Omega + 2i\alpha}\Bigg\{i\left(l+2q\right)\Omega+i\left(m+2q\right)\Omega\Bigg\}\\
&=-\frac{ie\lambda^4 A_S^2  \nu_{\uparrow} \nu_{\downarrow}}{4\pi}\int_{-\infty}^t dt'~\sum_{l,m}\, p^*_{l}(2q)p_{m}(2q)e^{\alpha\left(t'-t\right)} \left\{\left[-\frac{d}{dt'}e^{-i(l+2q)\Omega t'}\right]e^{i(m+2q)\Omega t'}+\left[\frac{d}{dt'}e^{i(m+2q)\Omega t'}\right]e^{-i(l+2q)\Omega t'}\right\}\\
&=\frac{\lambda^4 A_S^2  \nu_{\uparrow} \nu_{\downarrow}}{2\pi}\int_{-\infty}^t dt' e^{\alpha\left(t'-t\right)}e^2V(t')
\end{align}

\section{Witness of the entanglement}\label{app:noise}
In this Appendix, we give the intermediate steps between Eq.~\eqref{eq:NoiseStart} and Eq.~\eqref{eq:NoiseFinal} for the quantum state of Levitons in Eq.~\eqref{eq:fs5f}. We will consider a dispersion relation linear in $k$
\begin{equation}
    E_N^{\sigma}(k) = v_{\sigma}\left(k-k_{\sigma}\right),
\end{equation}
which is exact in the case of QSH edge states and an approximation for other spin systems.

The expression reads
\begin{equation}
\begin{aligned}
	\Delta \mathcal{S} &= {e^2 v_{\uparrow}v_{\downarrow}}\lambda^4 R^2 \sum_{k,k'}\Theta\left[E_N^{\uparrow}(k)\right]\Theta\left[E_N^{\downarrow}(k')\right]\cos \left(2 k D\right)\cos \left(2 k' D\right)\left|\Upsilon_{kk'}\right|^2 =\\& ={4e^2 v_{\uparrow}v_{\downarrow}}{\nu_{\uparrow}\nu_{\downarrow}A_S^2\lambda^4 R^2} \int_0^{\infty} d\epsilon_{\uparrow}\int_0^{\infty} d\epsilon_{\downarrow}\cos \left[2 \left(\frac{\epsilon_{\uparrow}}{v_{\uparrow}} + k_{\uparrow}\right) D\right]\cos \left[2 \left(\frac{\epsilon_{\downarrow}}{v_{\downarrow}} + k_{\downarrow}\right) D\right] \\&\times\sum_{l,m}p^*_lp_m\delta \left[\epsilon_{\uparrow}+\epsilon_{\downarrow}-(l+2q)\Omega\right]\delta\left[\epsilon_{\uparrow}+\epsilon_{\downarrow}-(m+2q)\Omega\right]
 \end{aligned}
 \end{equation}
 We compute the integral over $\epsilon_{\downarrow}$ by using one of the delta function
\begin{align}
\Delta \mathcal{S} &={4e^2 v_{\uparrow}v_{\downarrow}}{\nu_{\uparrow}\nu_{\downarrow}A_S^2\lambda^4 R^2} \sum_{l,m} \int_0^{\infty} d\epsilon_{\uparrow}\cos \left[2 \left(\frac{\epsilon_{\uparrow}}{v_{\uparrow}} + k_{\uparrow}\right) D\right]\cos \left\{2 \left[\frac{\epsilon_{\uparrow}-\left(l+2q\right)\Omega}{v_{\downarrow}} - k_{\downarrow}\right] D\right\}\\
&\times p^*_lp_m\delta_{l,m}\Theta\left[-\epsilon_{\uparrow}+(m+2q)\Omega\right]\\
& ={4e^2 v_{\uparrow}v_{\downarrow}}{\nu_{\uparrow}\nu_{\downarrow} A_S^2\lambda^4 R^2}
\sum_{l}\left|p_l\right|^2 \int_0^{\left(l+2q\right)\Omega} d\epsilon_{\uparrow}\cos \left[2 \left(\frac{\epsilon_{\uparrow}}{v_{\uparrow}} + k_{\uparrow}\right) D\right]\cos \left\{2 \left[\frac{\epsilon_{\uparrow}-\left(l+2q\right)\Omega}{v_{\downarrow}} - k_{\downarrow}\right] D\right\}.
\end{align}
From now on, we focus on system with time-reversal symmetry ($v_{\uparrow} = -v_{\downarrow}$, $\nu_{\uparrow} = \nu_{\downarrow}$ and $k_{\uparrow} = -k_{\downarrow}$), thus obtaining
\begin{equation}
\Delta \mathcal{S} =-{4e^2v^2_{\uparrow}}{\nu^2_{\uparrow}A_S^2\lambda^4 R^2}\sum_{l}\left|p_l\right|^2 \int_0^{\left(l+2q\right)\Omega} d\epsilon_{\uparrow}\cos \left[2 \left(\frac{\epsilon_{\uparrow}}{v_{\uparrow}} + k_{\uparrow}\right) D\right]\cos \left\{2 \left[\frac{\epsilon_{\uparrow}-\left(l+2q\right)\Omega}{v_{\uparrow}} - k_{\uparrow}\right] D\right\}.
\end{equation}
Then, we evaluate the integral over $\epsilon_{\uparrow}$
\begin{equation}
\Delta \mathcal{S} =-{e^2v^2_{\uparrow}}{\frac{\nu^2_{\uparrow}A_S^2\lambda^4 R^2}{D}}\sum_{l}\left|p_l\right|^2\left\{2 D \Omega  (l+2 q) \cos \left[4 k_{\uparrow} D+\frac{2 D \Omega  (l+2
   q)}{v_{\uparrow}}\right]+ v_{\uparrow} \sin \left[\frac{2 D \Omega  (l+2
   q)}{v_{\uparrow}}\right]\right\}.
\end{equation}
This expression can be recast as
\begin{align}
\Delta \mathcal{S} &=-\Delta \mathcal{S}_0\sum_{l>-2q}\left|p_l\right|^2\left\{ (l+2 q) \cos \left[\alpha_F+\tilde{D} (l+2q) \right]
+ \tilde{D}^{-1}\sin \left[\tilde{D}(l+2 q)\right]\right\}
\end{align}
by defining
\begin{align}
\Delta \mathcal{S}_0 & ={2e^2v^2_{\uparrow}}{\nu^2_{\uparrow}A_S^2 \Omega \lambda^4 R^2},\\
\tilde{D}& = \frac{2D\Omega}{v_{\uparrow}},\\
\alpha_F &= 4 k_F D.
\end{align}
Moreover, we notice that 
\begin{equation}
(l+2 q) \cos \left[\alpha_F+\tilde{D} (l+2
   q)\right]=\frac{d}{d\tilde{D}}\sin \left[\alpha_F+\tilde{D} (l+2
   q)\right]
\end{equation}
such that
\begin{align}
\Delta \mathcal{S} &=-\Delta \mathcal{S}_0\sum_{l>-2q}\left|p_l\right|^2\left\{\frac{d}{d\tilde{D}}\sin \left[\alpha_F+\tilde{D} (l+2
   q)\right]+ \tilde{D}^{-1}\sin \left[\tilde{D}(l+2
   q)\right]\right\}
\label{eq:AppNoiseFinal}
\end{align}
We remark that this expression is valid for Levitons with $2q \in \mathbb{Z}$. For $2q=1$, the photo-assisted coefficients can be computed analytically and for $l\ge 0$ they are given by
\begin{equation}
    \left|p_l\right|^2=e^{-4\pi \eta l}\left(1-e^{-4\pi \eta}\right)^2
\end{equation}
By using the above expression, the sum over $l$ can be computed explicitly. In particular, one has
\begin{align}
\sum_{l\ge0}e^{-4\pi \eta l}\sin \left[\alpha_F+\tilde{D} (l+1)\right] = \frac{\sin\left(\alpha_F\right)-e^{4\pi \eta}\sin\left(\tilde{D} + \alpha_F\right)}{2\cos\left(\tilde{D}\right)-2 \cosh\left(4\pi \eta\right)}
\end{align}
which can be plugged into Eq.~\eqref{eq:AppNoiseFinal}, thus obtaining
\begin{equation}
\Delta \mathcal{S} =\Delta \mathcal{S}_0\left(1-e^{-4\pi\eta}\right)^2\frac{e^{8 \pi  \eta } \left[\tilde{D} \cos \left(\alpha_F +\tilde{D}\right)+\sin (\tilde{D})\right]-e^{4 \pi  \eta } \left[2 \tilde{D} \cos (\alpha_F
   )+\sin (2 \tilde{D})\right]+\tilde{D} \cos \left(\tilde{D}-\alpha_F \right)+\sin (D)}{4 D \left[\cos (D)-\cosh (4 \pi  \eta )\right]^2}
\end{equation}

\end{widetext}


\begin{thebibliography}{94}%
	\makeatletter
	\providecommand \@ifxundefined [1]{%
		\@ifx{#1\undefined}
	}%
	\providecommand \@ifnum [1]{%
		\ifnum #1\expandafter \@firstoftwo
		\else \expandafter \@secondoftwo
		\fi
	}%
	\providecommand \@ifx [1]{%
		\ifx #1\expandafter \@firstoftwo
		\else \expandafter \@secondoftwo
		\fi
	}%
	\providecommand \natexlab [1]{#1}%
	\providecommand \enquote  [1]{``#1''}%
	\providecommand \bibnamefont  [1]{#1}%
	\providecommand \bibfnamefont [1]{#1}%
	\providecommand \citenamefont [1]{#1}%
	\providecommand \href@noop [0]{\@secondoftwo}%
	\providecommand \href [0]{\begingroup \@sanitize@url \@href}%
	\providecommand \@href[1]{\@@startlink{#1}\@@href}%
	\providecommand \@@href[1]{\endgroup#1\@@endlink}%
	\providecommand \@sanitize@url [0]{\catcode `\\12\catcode `\$12\catcode
		`\&12\catcode `\#12\catcode `\^12\catcode `\_12\catcode `\%12\relax}%
	\providecommand \@@startlink[1]{}%
	\providecommand \@@endlink[0]{}%
	\providecommand \url  [0]{\begingroup\@sanitize@url \@url }%
	\providecommand \@url [1]{\endgroup\@href {#1}{\urlprefix }}%
	\providecommand \urlprefix  [0]{URL }%
	\providecommand \Eprint [0]{\href }%
	\providecommand \doibase [0]{https://doi.org/}%
	\providecommand \selectlanguage [0]{\@gobble}%
	\providecommand \bibinfo  [0]{\@secondoftwo}%
	\providecommand \bibfield  [0]{\@secondoftwo}%
	\providecommand \translation [1]{[#1]}%
	\providecommand \BibitemOpen [0]{}%
	\providecommand \bibitemStop [0]{}%
	\providecommand \bibitemNoStop [0]{.\EOS\space}%
	\providecommand \EOS [0]{\spacefactor3000\relax}%
	\providecommand \BibitemShut  [1]{\csname bibitem#1\endcsname}%
	\let\auto@bib@innerbib\@empty
	\bibitem [{\citenamefont {Josephson}(1962)}]{Josephson1962}%
	\BibitemOpen
	\bibfield  {author} {\bibinfo {author} {\bibfnamefont {B.}~\bibnamefont
			{Josephson}},\ }\bibfield  {title} {\bibinfo {title} {Possible new effects in
			superconductive tunnelling},\ }\href
	{https://doi.org/https://doi.org/10.1016/0031-9163(62)91369-0} {\bibfield
		{journal} {\bibinfo  {journal} {Physics Letters}\ }\textbf {\bibinfo {volume}
			{1}},\ \bibinfo {pages} {251} (\bibinfo {year} {1962})}\BibitemShut {NoStop}%
	\bibitem [{\citenamefont {Liu}\ \emph {et~al.}(1998)\citenamefont {Liu},
		\citenamefont {Odom}, \citenamefont {Yamamoto},\ and\ \citenamefont
		{Tarucha}}]{Liu1998}%
	\BibitemOpen
	\bibfield  {author} {\bibinfo {author} {\bibfnamefont {R.~C.}\ \bibnamefont
			{Liu}}, \bibinfo {author} {\bibfnamefont {B.}~\bibnamefont {Odom}}, \bibinfo
		{author} {\bibfnamefont {Y.}~\bibnamefont {Yamamoto}},\ and\ \bibinfo
		{author} {\bibfnamefont {S.}~\bibnamefont {Tarucha}},\ }\bibfield  {title}
	{\bibinfo {title} {Quantum interference in electron collision},\ }\href
	{https://doi.org/10.1038/34611} {\bibfield  {journal} {\bibinfo  {journal}
			{Nature}\ }\textbf {\bibinfo {volume} {391}},\ \bibinfo {pages} {263}
		(\bibinfo {year} {1998})}\BibitemShut {NoStop}%
	\bibitem [{\citenamefont {Laflorencie}(2016)}]{Laflorencie2016}%
	\BibitemOpen
	\bibfield  {author} {\bibinfo {author} {\bibfnamefont {N.}~\bibnamefont
			{Laflorencie}},\ }\bibfield  {title} {\bibinfo {title} {Quantum entanglement
			in condensed matter systems},\ }\href
	{https://doi.org/https://doi.org/10.1016/j.physrep.2016.06.008} {\bibfield
		{journal} {\bibinfo  {journal} {Physics Reports}\ }\textbf {\bibinfo {volume}
			{646}},\ \bibinfo {pages} {1} (\bibinfo {year} {2016})}\BibitemShut {NoStop}%
	\bibitem [{Nat(2016)}]{Nature2016}%
	\BibitemOpen
	\bibfield  {title} {\bibinfo {title} {The rise of quantum materials},\ }\href
	{https://doi.org/10.1038/nphys3668} {\bibfield  {journal} {\bibinfo
			{journal} {Nature Physics}\ }\textbf {\bibinfo {volume} {12}},\ \bibinfo
		{pages} {105} (\bibinfo {year} {2016})}\BibitemShut {NoStop}%
	\bibitem [{\citenamefont {Fève}\ \emph {et~al.}(2007)\citenamefont {Fève},
		\citenamefont {Mahé}, \citenamefont {Berroir}, \citenamefont {Kontos},
		\citenamefont {Plaçais}, \citenamefont {Glattli}, \citenamefont {Cavanna},
		\citenamefont {Etienne},\ and\ \citenamefont {Jin}}]{Feve07}%
	\BibitemOpen
	\bibfield  {author} {\bibinfo {author} {\bibfnamefont {G.}~\bibnamefont
			{Fève}}, \bibinfo {author} {\bibfnamefont {A.}~\bibnamefont {Mahé}},
		\bibinfo {author} {\bibfnamefont {J.-M.}\ \bibnamefont {Berroir}}, \bibinfo
		{author} {\bibfnamefont {T.}~\bibnamefont {Kontos}}, \bibinfo {author}
		{\bibfnamefont {B.}~\bibnamefont {Plaçais}}, \bibinfo {author}
		{\bibfnamefont {D.~C.}\ \bibnamefont {Glattli}}, \bibinfo {author}
		{\bibfnamefont {A.}~\bibnamefont {Cavanna}}, \bibinfo {author} {\bibfnamefont
			{B.}~\bibnamefont {Etienne}},\ and\ \bibinfo {author} {\bibfnamefont
			{Y.}~\bibnamefont {Jin}},\ }\bibfield  {title} {\bibinfo {title} {An
			on-demand coherent single-electron source},\ }\href
	{https://doi.org/10.1126/science.1141243} {\bibfield  {journal} {\bibinfo
			{journal} {Science}\ }\textbf {\bibinfo {volume} {316}},\ \bibinfo {pages}
		{1169} (\bibinfo {year} {2007})}\BibitemShut {NoStop}%
	\bibitem [{\citenamefont {Bocquillon}\ \emph {et~al.}(2013)\citenamefont
		{Bocquillon}, \citenamefont {Freulon}, \citenamefont {Berroir}, \citenamefont
		{Degiovanni}, \citenamefont {Plaçais}, \citenamefont {Cavanna},
		\citenamefont {Jin},\ and\ \citenamefont {Fève}}]{Bocquillon2013}%
	\BibitemOpen
	\bibfield  {author} {\bibinfo {author} {\bibfnamefont {E.}~\bibnamefont
			{Bocquillon}}, \bibinfo {author} {\bibfnamefont {V.}~\bibnamefont {Freulon}},
		\bibinfo {author} {\bibfnamefont {J.-M.}\ \bibnamefont {Berroir}}, \bibinfo
		{author} {\bibfnamefont {P.}~\bibnamefont {Degiovanni}}, \bibinfo {author}
		{\bibfnamefont {B.}~\bibnamefont {Plaçais}}, \bibinfo {author}
		{\bibfnamefont {A.}~\bibnamefont {Cavanna}}, \bibinfo {author} {\bibfnamefont
			{Y.}~\bibnamefont {Jin}},\ and\ \bibinfo {author} {\bibfnamefont
			{G.}~\bibnamefont {Fève}},\ }\bibfield  {title} {\bibinfo {title} {Coherence
			and indistinguishability of single electrons emitted by independent
			sources},\ }\href {https://doi.org/10.1126/science.1232572} {\bibfield
		{journal} {\bibinfo  {journal} {Science}\ }\textbf {\bibinfo {volume}
			{339}},\ \bibinfo {pages} {1054} (\bibinfo {year} {2013})}\BibitemShut
	{NoStop}%
	\bibitem [{\citenamefont {Pekola}\ \emph {et~al.}(2013)\citenamefont {Pekola},
		\citenamefont {Saira}, \citenamefont {Maisi}, \citenamefont {Kemppinen},
		\citenamefont {M\"ott\"onen}, \citenamefont {Pashkin},\ and\ \citenamefont
		{Averin}}]{Pekola2013}%
	\BibitemOpen
	\bibfield  {author} {\bibinfo {author} {\bibfnamefont {J.~P.}\ \bibnamefont
			{Pekola}}, \bibinfo {author} {\bibfnamefont {O.-P.}\ \bibnamefont {Saira}},
		\bibinfo {author} {\bibfnamefont {V.~F.}\ \bibnamefont {Maisi}}, \bibinfo
		{author} {\bibfnamefont {A.}~\bibnamefont {Kemppinen}}, \bibinfo {author}
		{\bibfnamefont {M.}~\bibnamefont {M\"ott\"onen}}, \bibinfo {author}
		{\bibfnamefont {Y.~A.}\ \bibnamefont {Pashkin}},\ and\ \bibinfo {author}
		{\bibfnamefont {D.~V.}\ \bibnamefont {Averin}},\ }\bibfield  {title}
	{\bibinfo {title} {Single-electron current sources: Toward a refined
			definition of the ampere},\ }\href
	{https://doi.org/10.1103/RevModPhys.85.1421} {\bibfield  {journal} {\bibinfo
			{journal} {Rev. Mod. Phys.}\ }\textbf {\bibinfo {volume} {85}},\ \bibinfo
		{pages} {1421} (\bibinfo {year} {2013})}\BibitemShut {NoStop}%
	\bibitem [{\citenamefont {Bäuerle}\ \emph {et~al.}(2018)\citenamefont
		{Bäuerle}, \citenamefont {Glattli}, \citenamefont {Meunier}, \citenamefont
		{Portier}, \citenamefont {Roche}, \citenamefont {Roulleau}, \citenamefont
		{Takada},\ and\ \citenamefont {Waintal}}]{Bauerle2018}%
	\BibitemOpen
	\bibfield  {author} {\bibinfo {author} {\bibfnamefont {C.}~\bibnamefont
			{Bäuerle}}, \bibinfo {author} {\bibfnamefont {D.~C.}\ \bibnamefont
			{Glattli}}, \bibinfo {author} {\bibfnamefont {T.}~\bibnamefont {Meunier}},
		\bibinfo {author} {\bibfnamefont {F.}~\bibnamefont {Portier}}, \bibinfo
		{author} {\bibfnamefont {P.}~\bibnamefont {Roche}}, \bibinfo {author}
		{\bibfnamefont {P.}~\bibnamefont {Roulleau}}, \bibinfo {author}
		{\bibfnamefont {S.}~\bibnamefont {Takada}},\ and\ \bibinfo {author}
		{\bibfnamefont {X.}~\bibnamefont {Waintal}},\ }\bibfield  {title} {\bibinfo
		{title} {Coherent control of single electrons: a review of current
			progress},\ }\href {https://doi.org/10.1088/1361-6633/aaa98a} {\bibfield
		{journal} {\bibinfo  {journal} {Reports on Progress in Physics}\ }\textbf
		{\bibinfo {volume} {81}},\ \bibinfo {pages} {056503} (\bibinfo {year}
		{2018})}\BibitemShut {NoStop}%
	\bibitem [{\citenamefont {Edlbauer}\ \emph {et~al.}(2022)\citenamefont
		{Edlbauer}, \citenamefont {Wang}, \citenamefont {Crozes}, \citenamefont
		{Perrier}, \citenamefont {Ouacel}, \citenamefont {Geffroy}, \citenamefont
		{Georgiou}, \citenamefont {Chatzikyriakou}, \citenamefont {Lacerda-Santos},
		\citenamefont {Waintal}, \citenamefont {Glattli}, \citenamefont {Roulleau},
		\citenamefont {Nath}, \citenamefont {Kataoka}, \citenamefont
		{Splettstoesser}, \citenamefont {Acciai}, \citenamefont {da~Silva~Figueira},
		\citenamefont {{\"O}ztas}, \citenamefont {Trellakis}, \citenamefont {Grange},
		\citenamefont {Yevtushenko}, \citenamefont {Birner},\ and\ \citenamefont
		{B{\"a}uerle}}]{Edlbauer2022}%
	\BibitemOpen
	\bibfield  {author} {\bibinfo {author} {\bibfnamefont {H.}~\bibnamefont
			{Edlbauer}}, \bibinfo {author} {\bibfnamefont {J.}~\bibnamefont {Wang}},
		\bibinfo {author} {\bibfnamefont {T.}~\bibnamefont {Crozes}}, \bibinfo
		{author} {\bibfnamefont {P.}~\bibnamefont {Perrier}}, \bibinfo {author}
		{\bibfnamefont {S.}~\bibnamefont {Ouacel}}, \bibinfo {author} {\bibfnamefont
			{C.}~\bibnamefont {Geffroy}}, \bibinfo {author} {\bibfnamefont
			{G.}~\bibnamefont {Georgiou}}, \bibinfo {author} {\bibfnamefont
			{E.}~\bibnamefont {Chatzikyriakou}}, \bibinfo {author} {\bibfnamefont
			{A.}~\bibnamefont {Lacerda-Santos}}, \bibinfo {author} {\bibfnamefont
			{X.}~\bibnamefont {Waintal}}, \bibinfo {author} {\bibfnamefont {D.~C.}\
			\bibnamefont {Glattli}}, \bibinfo {author} {\bibfnamefont {P.}~\bibnamefont
			{Roulleau}}, \bibinfo {author} {\bibfnamefont {J.}~\bibnamefont {Nath}},
		\bibinfo {author} {\bibfnamefont {M.}~\bibnamefont {Kataoka}}, \bibinfo
		{author} {\bibfnamefont {J.}~\bibnamefont {Splettstoesser}}, \bibinfo
		{author} {\bibfnamefont {M.}~\bibnamefont {Acciai}}, \bibinfo {author}
		{\bibfnamefont {M.~C.}\ \bibnamefont {da~Silva~Figueira}}, \bibinfo {author}
		{\bibfnamefont {K.}~\bibnamefont {{\"O}ztas}}, \bibinfo {author}
		{\bibfnamefont {A.}~\bibnamefont {Trellakis}}, \bibinfo {author}
		{\bibfnamefont {T.}~\bibnamefont {Grange}}, \bibinfo {author} {\bibfnamefont
			{O.~M.}\ \bibnamefont {Yevtushenko}}, \bibinfo {author} {\bibfnamefont
			{S.}~\bibnamefont {Birner}},\ and\ \bibinfo {author} {\bibfnamefont
			{C.}~\bibnamefont {B{\"a}uerle}},\ }\bibfield  {title} {\bibinfo {title}
		{Semiconductor-based electron flying qubits: review on recent progress
			accelerated by numerical modelling},\ }\href
	{https://doi.org/10.1140/epjqt/s40507-022-00139-w} {\bibfield  {journal}
		{\bibinfo  {journal} {EPJ Quantum Technology}\ }\textbf {\bibinfo {volume}
			{9}},\ \bibinfo {pages} {21} (\bibinfo {year} {2022})}\BibitemShut {NoStop}%
	\bibitem [{\citenamefont {Hermelin}\ \emph {et~al.}(2011)\citenamefont
		{Hermelin}, \citenamefont {Takada}, \citenamefont {Yamamoto}, \citenamefont
		{Tarucha}, \citenamefont {Wieck}, \citenamefont {Saminadayar}, \citenamefont
		{B{\"a}uerle},\ and\ \citenamefont {Meunier}}]{Hermelin2011}%
	\BibitemOpen
	\bibfield  {author} {\bibinfo {author} {\bibfnamefont {S.}~\bibnamefont
			{Hermelin}}, \bibinfo {author} {\bibfnamefont {S.}~\bibnamefont {Takada}},
		\bibinfo {author} {\bibfnamefont {M.}~\bibnamefont {Yamamoto}}, \bibinfo
		{author} {\bibfnamefont {S.}~\bibnamefont {Tarucha}}, \bibinfo {author}
		{\bibfnamefont {A.~D.}\ \bibnamefont {Wieck}}, \bibinfo {author}
		{\bibfnamefont {L.}~\bibnamefont {Saminadayar}}, \bibinfo {author}
		{\bibfnamefont {C.}~\bibnamefont {B{\"a}uerle}},\ and\ \bibinfo {author}
		{\bibfnamefont {T.}~\bibnamefont {Meunier}},\ }\bibfield  {title} {\bibinfo
		{title} {Electrons surfing on a sound wave as a platform for quantum optics
			with flying electrons},\ }\href {https://doi.org/10.1038/nature10416}
	{\bibfield  {journal} {\bibinfo  {journal} {Nature}\ }\textbf {\bibinfo
			{volume} {477}},\ \bibinfo {pages} {435} (\bibinfo {year}
		{2011})}\BibitemShut {NoStop}%
	\bibitem [{\citenamefont {Takada}\ \emph {et~al.}(2019)\citenamefont {Takada},
		\citenamefont {Edlbauer}, \citenamefont {Lepage}, \citenamefont {Wang},
		\citenamefont {Mortemousque}, \citenamefont {Georgiou}, \citenamefont
		{Barnes}, \citenamefont {Ford}, \citenamefont {Yuan}, \citenamefont {Santos},
		\citenamefont {Waintal}, \citenamefont {Ludwig}, \citenamefont {Wieck},
		\citenamefont {Urdampilleta}, \citenamefont {Meunier},\ and\ \citenamefont
		{B{\"a}uerle}}]{Takada2019}%
	\BibitemOpen
	\bibfield  {author} {\bibinfo {author} {\bibfnamefont {S.}~\bibnamefont
			{Takada}}, \bibinfo {author} {\bibfnamefont {H.}~\bibnamefont {Edlbauer}},
		\bibinfo {author} {\bibfnamefont {H.~V.}\ \bibnamefont {Lepage}}, \bibinfo
		{author} {\bibfnamefont {J.}~\bibnamefont {Wang}}, \bibinfo {author}
		{\bibfnamefont {P.-A.}\ \bibnamefont {Mortemousque}}, \bibinfo {author}
		{\bibfnamefont {G.}~\bibnamefont {Georgiou}}, \bibinfo {author}
		{\bibfnamefont {C.~H.~W.}\ \bibnamefont {Barnes}}, \bibinfo {author}
		{\bibfnamefont {C.~J.~B.}\ \bibnamefont {Ford}}, \bibinfo {author}
		{\bibfnamefont {M.}~\bibnamefont {Yuan}}, \bibinfo {author} {\bibfnamefont
			{P.~V.}\ \bibnamefont {Santos}}, \bibinfo {author} {\bibfnamefont
			{X.}~\bibnamefont {Waintal}}, \bibinfo {author} {\bibfnamefont
			{A.}~\bibnamefont {Ludwig}}, \bibinfo {author} {\bibfnamefont {A.~D.}\
			\bibnamefont {Wieck}}, \bibinfo {author} {\bibfnamefont {M.}~\bibnamefont
			{Urdampilleta}}, \bibinfo {author} {\bibfnamefont {T.}~\bibnamefont
			{Meunier}},\ and\ \bibinfo {author} {\bibfnamefont {C.}~\bibnamefont
			{B{\"a}uerle}},\ }\bibfield  {title} {\bibinfo {title} {Sound-driven
			single-electron transfer in a circuit of coupled quantum rails},\ }\href
	{https://doi.org/10.1038/s41467-019-12514-w} {\bibfield  {journal} {\bibinfo
			{journal} {Nature Communications}\ }\textbf {\bibinfo {volume} {10}},\
		\bibinfo {pages} {4557} (\bibinfo {year} {2019})}\BibitemShut {NoStop}%
	\bibitem [{\citenamefont {Mah\'e}\ \emph {et~al.}(2010)\citenamefont {Mah\'e},
		\citenamefont {Parmentier}, \citenamefont {Bocquillon}, \citenamefont
		{Berroir}, \citenamefont {Glattli}, \citenamefont {Kontos}, \citenamefont
		{Pla\ifmmode~\mbox{\c{c}}\else \c{c}\fi{}ais}, \citenamefont {F\`eve},
		\citenamefont {Cavanna},\ and\ \citenamefont {Jin}}]{Mahe2010}%
	\BibitemOpen
	\bibfield  {author} {\bibinfo {author} {\bibfnamefont {A.}~\bibnamefont
			{Mah\'e}}, \bibinfo {author} {\bibfnamefont {F.~D.}\ \bibnamefont
			{Parmentier}}, \bibinfo {author} {\bibfnamefont {E.}~\bibnamefont
			{Bocquillon}}, \bibinfo {author} {\bibfnamefont {J.-M.}\ \bibnamefont
			{Berroir}}, \bibinfo {author} {\bibfnamefont {D.~C.}\ \bibnamefont
			{Glattli}}, \bibinfo {author} {\bibfnamefont {T.}~\bibnamefont {Kontos}},
		\bibinfo {author} {\bibfnamefont {B.}~\bibnamefont
			{Pla\ifmmode~\mbox{\c{c}}\else \c{c}\fi{}ais}}, \bibinfo {author}
		{\bibfnamefont {G.}~\bibnamefont {F\`eve}}, \bibinfo {author} {\bibfnamefont
			{A.}~\bibnamefont {Cavanna}},\ and\ \bibinfo {author} {\bibfnamefont
			{Y.}~\bibnamefont {Jin}},\ }\bibfield  {title} {\bibinfo {title} {Current
			correlations of an on-demand single-electron emitter},\ }\href
	{https://doi.org/10.1103/PhysRevB.82.201309} {\bibfield  {journal} {\bibinfo
			{journal} {Phys. Rev. B}\ }\textbf {\bibinfo {volume} {82}},\ \bibinfo
		{pages} {201309} (\bibinfo {year} {2010})}\BibitemShut {NoStop}%
	\bibitem [{\citenamefont {Grenier}\ \emph
		{et~al.}(2011{\natexlab{a}})\citenamefont {Grenier}, \citenamefont {Hervé},
		\citenamefont {Bocquillon}, \citenamefont {Parmentier}, \citenamefont
		{Plaçais}, \citenamefont {Berroir}, \citenamefont {Fève},\ and\
		\citenamefont {Degiovanni}}]{Grenier2011}%
	\BibitemOpen
	\bibfield  {author} {\bibinfo {author} {\bibfnamefont {C.}~\bibnamefont
			{Grenier}}, \bibinfo {author} {\bibfnamefont {R.}~\bibnamefont {Hervé}},
		\bibinfo {author} {\bibfnamefont {E.}~\bibnamefont {Bocquillon}}, \bibinfo
		{author} {\bibfnamefont {F.~D.}\ \bibnamefont {Parmentier}}, \bibinfo
		{author} {\bibfnamefont {B.}~\bibnamefont {Plaçais}}, \bibinfo {author}
		{\bibfnamefont {J.~M.}\ \bibnamefont {Berroir}}, \bibinfo {author}
		{\bibfnamefont {G.}~\bibnamefont {Fève}},\ and\ \bibinfo {author}
		{\bibfnamefont {P.}~\bibnamefont {Degiovanni}},\ }\bibfield  {title}
	{\bibinfo {title} {Single-electron quantum tomography in quantum hall edge
			channels},\ }\href {https://doi.org/10.1088/1367-2630/13/9/093007} {\bibfield
		{journal} {\bibinfo  {journal} {New Journal of Physics}\ }\textbf {\bibinfo
			{volume} {13}},\ \bibinfo {pages} {093007} (\bibinfo {year}
		{2011}{\natexlab{a}})}\BibitemShut {NoStop}%
	\bibitem [{\citenamefont {Parmentier}\ \emph {et~al.}(2012)\citenamefont
		{Parmentier}, \citenamefont {Bocquillon}, \citenamefont {Berroir},
		\citenamefont {Glattli}, \citenamefont {Pla\ifmmode~\mbox{\c{c}}\else
			\c{c}\fi{}ais}, \citenamefont {F\`eve}, \citenamefont {Albert}, \citenamefont
		{Flindt},\ and\ \citenamefont {B\"uttiker}}]{Parmentier2012}%
	\BibitemOpen
	\bibfield  {author} {\bibinfo {author} {\bibfnamefont {F.~D.}\ \bibnamefont
			{Parmentier}}, \bibinfo {author} {\bibfnamefont {E.}~\bibnamefont
			{Bocquillon}}, \bibinfo {author} {\bibfnamefont {J.-M.}\ \bibnamefont
			{Berroir}}, \bibinfo {author} {\bibfnamefont {D.~C.}\ \bibnamefont
			{Glattli}}, \bibinfo {author} {\bibfnamefont {B.}~\bibnamefont
			{Pla\ifmmode~\mbox{\c{c}}\else \c{c}\fi{}ais}}, \bibinfo {author}
		{\bibfnamefont {G.}~\bibnamefont {F\`eve}}, \bibinfo {author} {\bibfnamefont
			{M.}~\bibnamefont {Albert}}, \bibinfo {author} {\bibfnamefont
			{C.}~\bibnamefont {Flindt}},\ and\ \bibinfo {author} {\bibfnamefont
			{M.}~\bibnamefont {B\"uttiker}},\ }\bibfield  {title} {\bibinfo {title}
		{Current noise spectrum of a single-particle emitter: Theory and
			experiment},\ }\href {https://doi.org/10.1103/PhysRevB.85.165438} {\bibfield
		{journal} {\bibinfo  {journal} {Phys. Rev. B}\ }\textbf {\bibinfo {volume}
			{85}},\ \bibinfo {pages} {165438} (\bibinfo {year} {2012})}\BibitemShut
	{NoStop}%
	\bibitem [{\citenamefont {Glattli}\ and\ \citenamefont
		{Roulleau}(2017{\natexlab{a}})}]{Glattli2016}%
	\BibitemOpen
	\bibfield  {author} {\bibinfo {author} {\bibfnamefont {D.~C.}\ \bibnamefont
			{Glattli}}\ and\ \bibinfo {author} {\bibfnamefont {P.~S.}\ \bibnamefont
			{Roulleau}},\ }\bibfield  {title} {\bibinfo {title} {Levitons for electron
			quantum optics},\ }\href
	{https://doi.org/https://doi.org/10.1002/pssb.201600650} {\bibfield
		{journal} {\bibinfo  {journal} {physica status solidi (b)}\ }\textbf
		{\bibinfo {volume} {254}},\ \bibinfo {pages} {1600650} (\bibinfo {year}
		{2017}{\natexlab{a}})}\BibitemShut {NoStop}%
	\bibitem [{\citenamefont {Levitov}\ \emph {et~al.}(1996)\citenamefont
		{Levitov}, \citenamefont {Lee},\ and\ \citenamefont {Lesovik}}]{Levitov1996}%
	\BibitemOpen
	\bibfield  {author} {\bibinfo {author} {\bibfnamefont {L.~S.}\ \bibnamefont
			{Levitov}}, \bibinfo {author} {\bibfnamefont {H.}~\bibnamefont {Lee}},\ and\
		\bibinfo {author} {\bibfnamefont {G.~B.}\ \bibnamefont {Lesovik}},\
	}\bibfield  {title} {\bibinfo {title} {{Electron counting statistics and
				coherent states of electric current}},\ }\href
	{https://doi.org/10.1063/1.531672} {\bibfield  {journal} {\bibinfo  {journal}
			{Journal of Mathematical Physics}\ }\textbf {\bibinfo {volume} {37}},\
		\bibinfo {pages} {4845} (\bibinfo {year} {1996})}\BibitemShut {NoStop}%
	\bibitem [{\citenamefont {Ivanov}\ \emph {et~al.}(1997)\citenamefont {Ivanov},
		\citenamefont {Lee},\ and\ \citenamefont {Levitov}}]{Ivanov1997}%
	\BibitemOpen
	\bibfield  {author} {\bibinfo {author} {\bibfnamefont {D.~A.}\ \bibnamefont
			{Ivanov}}, \bibinfo {author} {\bibfnamefont {H.~W.}\ \bibnamefont {Lee}},\
		and\ \bibinfo {author} {\bibfnamefont {L.~S.}\ \bibnamefont {Levitov}},\
	}\bibfield  {title} {\bibinfo {title} {Coherent states of alternating
			current},\ }\href {https://doi.org/10.1103/PhysRevB.56.6839} {\bibfield
		{journal} {\bibinfo  {journal} {Phys. Rev. B}\ }\textbf {\bibinfo {volume}
			{56}},\ \bibinfo {pages} {6839} (\bibinfo {year} {1997})}\BibitemShut
	{NoStop}%
	\bibitem [{\citenamefont {Keeling}\ \emph {et~al.}(2006)\citenamefont
		{Keeling}, \citenamefont {Klich},\ and\ \citenamefont
		{Levitov}}]{Keeling2006}%
	\BibitemOpen
	\bibfield  {author} {\bibinfo {author} {\bibfnamefont {J.}~\bibnamefont
			{Keeling}}, \bibinfo {author} {\bibfnamefont {I.}~\bibnamefont {Klich}},\
		and\ \bibinfo {author} {\bibfnamefont {L.~S.}\ \bibnamefont {Levitov}},\
	}\bibfield  {title} {\bibinfo {title} {Minimal excitation states of electrons
			in one-dimensional wires},\ }\href
	{https://doi.org/10.1103/PhysRevLett.97.116403} {\bibfield  {journal}
		{\bibinfo  {journal} {Phys. Rev. Lett.}\ }\textbf {\bibinfo {volume} {97}},\
		\bibinfo {pages} {116403} (\bibinfo {year} {2006})}\BibitemShut {NoStop}%
	\bibitem [{\citenamefont {Flindt}(2013)}]{Flindt2013}%
	\BibitemOpen
	\bibfield  {author} {\bibinfo {author} {\bibfnamefont {C.}~\bibnamefont
			{Flindt}},\ }\bibfield  {title} {\bibinfo {title} {Single electrons pop out
			of the fermi sea},\ }\href {https://doi.org/10.1038/nature12699} {\bibfield
		{journal} {\bibinfo  {journal} {Nature}\ }\textbf {\bibinfo {volume} {502}},\
		\bibinfo {pages} {630} (\bibinfo {year} {2013})}\BibitemShut {NoStop}%
	\bibitem [{\citenamefont {Glattli}\ and\ \citenamefont
		{Roulleau}(2017{\natexlab{b}})}]{Glattli2017}%
	\BibitemOpen
	\bibfield  {author} {\bibinfo {author} {\bibfnamefont {D.~C.}\ \bibnamefont
			{Glattli}}\ and\ \bibinfo {author} {\bibfnamefont {P.~S.}\ \bibnamefont
			{Roulleau}},\ }\bibfield  {title} {\bibinfo {title} {Levitons for electron
			quantum optics},\ }\href
	{https://doi.org/https://doi.org/10.1002/pssb.201600650} {\bibfield
		{journal} {\bibinfo  {journal} {physica status solidi (b)}\ }\textbf
		{\bibinfo {volume} {254}},\ \bibinfo {pages} {1600650} (\bibinfo {year}
		{2017}{\natexlab{b}})}\BibitemShut {NoStop}%
	\bibitem [{\citenamefont {Moskalets}(2018)}]{Moskalets2018}%
	\BibitemOpen
	\bibfield  {author} {\bibinfo {author} {\bibfnamefont {M.}~\bibnamefont
			{Moskalets}},\ }\bibfield  {title} {\bibinfo {title} {High-temperature fusion
			of a multielectron leviton},\ }\href
	{https://doi.org/10.1103/PhysRevB.97.155411} {\bibfield  {journal} {\bibinfo
			{journal} {Phys. Rev. B}\ }\textbf {\bibinfo {volume} {97}},\ \bibinfo
		{pages} {155411} (\bibinfo {year} {2018})}\BibitemShut {NoStop}%
	\bibitem [{\citenamefont {Ferraro}\ \emph {et~al.}(2014)\citenamefont
		{Ferraro}, \citenamefont {Roussel}, \citenamefont {Cabart}, \citenamefont
		{Thibierge}, \citenamefont {F\`eve}, \citenamefont {Grenier},\ and\
		\citenamefont {Degiovanni}}]{Ferraro2014}%
	\BibitemOpen
	\bibfield  {author} {\bibinfo {author} {\bibfnamefont {D.}~\bibnamefont
			{Ferraro}}, \bibinfo {author} {\bibfnamefont {B.}~\bibnamefont {Roussel}},
		\bibinfo {author} {\bibfnamefont {C.}~\bibnamefont {Cabart}}, \bibinfo
		{author} {\bibfnamefont {E.}~\bibnamefont {Thibierge}}, \bibinfo {author}
		{\bibfnamefont {G.}~\bibnamefont {F\`eve}}, \bibinfo {author} {\bibfnamefont
			{C.}~\bibnamefont {Grenier}},\ and\ \bibinfo {author} {\bibfnamefont
			{P.}~\bibnamefont {Degiovanni}},\ }\bibfield  {title} {\bibinfo {title}
		{Real-time decoherence of landau and levitov quasiparticles in quantum hall
			edge channels},\ }\href {https://doi.org/10.1103/PhysRevLett.113.166403}
	{\bibfield  {journal} {\bibinfo  {journal} {Phys. Rev. Lett.}\ }\textbf
		{\bibinfo {volume} {113}},\ \bibinfo {pages} {166403} (\bibinfo {year}
		{2014})}\BibitemShut {NoStop}%
	\bibitem [{\citenamefont {Dubois}\ \emph {et~al.}(2013)\citenamefont {Dubois},
		\citenamefont {Jullien}, \citenamefont {Portier}, \citenamefont {Roche},
		\citenamefont {Cavanna}, \citenamefont {Jin}, \citenamefont {Wegscheider},
		\citenamefont {Roulleau},\ and\ \citenamefont {Glattli}}]{Dubois2013}%
	\BibitemOpen
	\bibfield  {author} {\bibinfo {author} {\bibfnamefont {J.}~\bibnamefont
			{Dubois}}, \bibinfo {author} {\bibfnamefont {T.}~\bibnamefont {Jullien}},
		\bibinfo {author} {\bibfnamefont {F.}~\bibnamefont {Portier}}, \bibinfo
		{author} {\bibfnamefont {P.}~\bibnamefont {Roche}}, \bibinfo {author}
		{\bibfnamefont {A.}~\bibnamefont {Cavanna}}, \bibinfo {author} {\bibfnamefont
			{Y.}~\bibnamefont {Jin}}, \bibinfo {author} {\bibfnamefont {W.}~\bibnamefont
			{Wegscheider}}, \bibinfo {author} {\bibfnamefont {P.}~\bibnamefont
			{Roulleau}},\ and\ \bibinfo {author} {\bibfnamefont {D.~C.}\ \bibnamefont
			{Glattli}},\ }\bibfield  {title} {\bibinfo {title} {Minimal-excitation states
			for electron quantum optics using levitons},\ }\href
	{https://doi.org/10.1038/nature12713} {\bibfield  {journal} {\bibinfo
			{journal} {Nature}\ }\textbf {\bibinfo {volume} {502}},\ \bibinfo {pages}
		{659} (\bibinfo {year} {2013})}\BibitemShut {NoStop}%
	\bibitem [{\citenamefont {Assouline}\ \emph {et~al.}(2023)\citenamefont
		{Assouline}, \citenamefont {Pugliese}, \citenamefont {Chakraborti},
		\citenamefont {Lee}, \citenamefont {Bernabeu}, \citenamefont {Jo},
		\citenamefont {Watanabe}, \citenamefont {Taniguchi}, \citenamefont {Glattli},
		\citenamefont {Kumada}, \citenamefont {Sim}, \citenamefont {Parmentier},\
		and\ \citenamefont {Roulleau}}]{Assouline2023}%
	\BibitemOpen
	\bibfield  {author} {\bibinfo {author} {\bibfnamefont {A.}~\bibnamefont
			{Assouline}}, \bibinfo {author} {\bibfnamefont {L.}~\bibnamefont {Pugliese}},
		\bibinfo {author} {\bibfnamefont {H.}~\bibnamefont {Chakraborti}}, \bibinfo
		{author} {\bibfnamefont {S.}~\bibnamefont {Lee}}, \bibinfo {author}
		{\bibfnamefont {L.}~\bibnamefont {Bernabeu}}, \bibinfo {author}
		{\bibfnamefont {M.}~\bibnamefont {Jo}}, \bibinfo {author} {\bibfnamefont
			{K.}~\bibnamefont {Watanabe}}, \bibinfo {author} {\bibfnamefont
			{T.}~\bibnamefont {Taniguchi}}, \bibinfo {author} {\bibfnamefont {D.~C.}\
			\bibnamefont {Glattli}}, \bibinfo {author} {\bibfnamefont {N.}~\bibnamefont
			{Kumada}}, \bibinfo {author} {\bibfnamefont {H.-S.}\ \bibnamefont {Sim}},
		\bibinfo {author} {\bibfnamefont {F.~D.}\ \bibnamefont {Parmentier}},\ and\
		\bibinfo {author} {\bibfnamefont {P.}~\bibnamefont {Roulleau}},\ }\bibfield
	{title} {\bibinfo {title} {Emission and coherent control of levitons in
			graphene},\ }\href {https://doi.org/10.1126/science.adf9887} {\bibfield
		{journal} {\bibinfo  {journal} {Science}\ }\textbf {\bibinfo {volume}
			{382}},\ \bibinfo {pages} {1260} (\bibinfo {year} {2023})}\BibitemShut
	{NoStop}%
	\bibitem [{\citenamefont {Jullien}\ \emph {et~al.}(2014)\citenamefont
		{Jullien}, \citenamefont {Roulleau}, \citenamefont {Roche}, \citenamefont
		{Cavanna}, \citenamefont {Jin},\ and\ \citenamefont {Glattli}}]{Jullien2014}%
	\BibitemOpen
	\bibfield  {author} {\bibinfo {author} {\bibfnamefont {T.}~\bibnamefont
			{Jullien}}, \bibinfo {author} {\bibfnamefont {P.}~\bibnamefont {Roulleau}},
		\bibinfo {author} {\bibfnamefont {B.}~\bibnamefont {Roche}}, \bibinfo
		{author} {\bibfnamefont {A.}~\bibnamefont {Cavanna}}, \bibinfo {author}
		{\bibfnamefont {Y.}~\bibnamefont {Jin}},\ and\ \bibinfo {author}
		{\bibfnamefont {D.~C.}\ \bibnamefont {Glattli}},\ }\bibfield  {title}
	{\bibinfo {title} {Quantum tomography of an electron},\ }\href
	{https://doi.org/10.1038/nature13821} {\bibfield  {journal} {\bibinfo
			{journal} {Nature}\ }\textbf {\bibinfo {volume} {514}},\ \bibinfo {pages}
		{603} (\bibinfo {year} {2014})}\BibitemShut {NoStop}%
	\bibitem [{\citenamefont {Bisognin}\ \emph {et~al.}(2019)\citenamefont
		{Bisognin}, \citenamefont {Marguerite}, \citenamefont {Roussel},
		\citenamefont {Kumar}, \citenamefont {Cabart}, \citenamefont {Chapdelaine},
		\citenamefont {Mohammad-Djafari}, \citenamefont {Berroir}, \citenamefont
		{Bocquillon}, \citenamefont {Pla{\c c}ais}, \citenamefont {Cavanna},
		\citenamefont {Gennser}, \citenamefont {Jin}, \citenamefont {Degiovanni},\
		and\ \citenamefont {F{\`e}ve}}]{Bisognin2019}%
	\BibitemOpen
	\bibfield  {author} {\bibinfo {author} {\bibfnamefont {R.}~\bibnamefont
			{Bisognin}}, \bibinfo {author} {\bibfnamefont {A.}~\bibnamefont
			{Marguerite}}, \bibinfo {author} {\bibfnamefont {B.}~\bibnamefont {Roussel}},
		\bibinfo {author} {\bibfnamefont {M.}~\bibnamefont {Kumar}}, \bibinfo
		{author} {\bibfnamefont {C.}~\bibnamefont {Cabart}}, \bibinfo {author}
		{\bibfnamefont {C.}~\bibnamefont {Chapdelaine}}, \bibinfo {author}
		{\bibfnamefont {A.}~\bibnamefont {Mohammad-Djafari}}, \bibinfo {author}
		{\bibfnamefont {J.~M.}\ \bibnamefont {Berroir}}, \bibinfo {author}
		{\bibfnamefont {E.}~\bibnamefont {Bocquillon}}, \bibinfo {author}
		{\bibfnamefont {B.}~\bibnamefont {Pla{\c c}ais}}, \bibinfo {author}
		{\bibfnamefont {A.}~\bibnamefont {Cavanna}}, \bibinfo {author} {\bibfnamefont
			{U.}~\bibnamefont {Gennser}}, \bibinfo {author} {\bibfnamefont
			{Y.}~\bibnamefont {Jin}}, \bibinfo {author} {\bibfnamefont {P.}~\bibnamefont
			{Degiovanni}},\ and\ \bibinfo {author} {\bibfnamefont {G.}~\bibnamefont
			{F{\`e}ve}},\ }\bibfield  {title} {\bibinfo {title} {Quantum tomography of
			electrical currents},\ }\href {https://doi.org/10.1038/s41467-019-11369-5}
	{\bibfield  {journal} {\bibinfo  {journal} {Nature Communications}\ }\textbf
		{\bibinfo {volume} {10}},\ \bibinfo {pages} {3379} (\bibinfo {year}
		{2019})}\BibitemShut {NoStop}%
	\bibitem [{\citenamefont {Roussel}\ \emph {et~al.}(2021)\citenamefont
		{Roussel}, \citenamefont {Cabart}, \citenamefont {F\`eve},\ and\
		\citenamefont {Degiovanni}}]{Roussel2021}%
	\BibitemOpen
	\bibfield  {author} {\bibinfo {author} {\bibfnamefont {B.}~\bibnamefont
			{Roussel}}, \bibinfo {author} {\bibfnamefont {C.}~\bibnamefont {Cabart}},
		\bibinfo {author} {\bibfnamefont {G.}~\bibnamefont {F\`eve}},\ and\ \bibinfo
		{author} {\bibfnamefont {P.}~\bibnamefont {Degiovanni}},\ }\bibfield  {title}
	{\bibinfo {title} {Processing quantum signals carried by electrical
			currents},\ }\href {https://doi.org/10.1103/PRXQuantum.2.020314} {\bibfield
		{journal} {\bibinfo  {journal} {PRX Quantum}\ }\textbf {\bibinfo {volume}
			{2}},\ \bibinfo {pages} {020314} (\bibinfo {year} {2021})}\BibitemShut
	{NoStop}%
	\bibitem [{\citenamefont {Moskalets}(2016)}]{Moskalets2017}%
	\BibitemOpen
	\bibfield  {author} {\bibinfo {author} {\bibfnamefont {M.}~\bibnamefont
			{Moskalets}},\ }\bibfield  {title} {\bibinfo {title} {Fractionally charged
			zero-energy single-particle excitations in a driven fermi sea},\ }\href
	{https://doi.org/10.1103/PhysRevLett.117.046801} {\bibfield  {journal}
		{\bibinfo  {journal} {Phys. Rev. Lett.}\ }\textbf {\bibinfo {volume} {117}},\
		\bibinfo {pages} {046801} (\bibinfo {year} {2016})}\BibitemShut {NoStop}%
	\bibitem [{\citenamefont {Dasenbrook}\ and\ \citenamefont
		{Flindt}(2015)}]{Dasenbrook2015}%
	\BibitemOpen
	\bibfield  {author} {\bibinfo {author} {\bibfnamefont {D.}~\bibnamefont
			{Dasenbrook}}\ and\ \bibinfo {author} {\bibfnamefont {C.}~\bibnamefont
			{Flindt}},\ }\bibfield  {title} {\bibinfo {title} {Dynamical generation and
			detection of entanglement in neutral leviton pairs},\ }\href
	{https://doi.org/10.1103/PhysRevB.92.161412} {\bibfield  {journal} {\bibinfo
			{journal} {Phys. Rev. B}\ }\textbf {\bibinfo {volume} {92}},\ \bibinfo
		{pages} {161412} (\bibinfo {year} {2015})}\BibitemShut {NoStop}%
	\bibitem [{\citenamefont {Hofer}\ \emph {et~al.}(2017)\citenamefont {Hofer},
		\citenamefont {Dasenbrook},\ and\ \citenamefont {Flindt}}]{Hofer2016}%
	\BibitemOpen
	\bibfield  {author} {\bibinfo {author} {\bibfnamefont {P.~P.}\ \bibnamefont
			{Hofer}}, \bibinfo {author} {\bibfnamefont {D.}~\bibnamefont {Dasenbrook}},\
		and\ \bibinfo {author} {\bibfnamefont {C.}~\bibnamefont {Flindt}},\
	}\bibfield  {title} {\bibinfo {title} {On-demand entanglement generation
			using dynamic single-electron sources},\ }\href
	{https://doi.org/https://doi.org/10.1002/pssb.201600582} {\bibfield
		{journal} {\bibinfo  {journal} {physica status solidi (b)}\ }\textbf
		{\bibinfo {volume} {254}},\ \bibinfo {pages} {1600582} (\bibinfo {year}
		{2017})}\BibitemShut {NoStop}%
	\bibitem [{\citenamefont {Ronetti}\ \emph {et~al.}(2024)\citenamefont
		{Ronetti}, \citenamefont {Bertin-Johannet}, \citenamefont {Popoff},
		\citenamefont {Rech}, \citenamefont {Jonckheere}, \citenamefont {Grémaud},
		\citenamefont {Raymond},\ and\ \citenamefont {Martin}}]{Ronetti2024}%
	\BibitemOpen
	\bibfield  {author} {\bibinfo {author} {\bibfnamefont {F.}~\bibnamefont
			{Ronetti}}, \bibinfo {author} {\bibfnamefont {B.}~\bibnamefont
			{Bertin-Johannet}}, \bibinfo {author} {\bibfnamefont {A.}~\bibnamefont
			{Popoff}}, \bibinfo {author} {\bibfnamefont {J.}~\bibnamefont {Rech}},
		\bibinfo {author} {\bibfnamefont {T.}~\bibnamefont {Jonckheere}}, \bibinfo
		{author} {\bibfnamefont {B.}~\bibnamefont {Grémaud}}, \bibinfo {author}
		{\bibfnamefont {L.}~\bibnamefont {Raymond}},\ and\ \bibinfo {author}
		{\bibfnamefont {T.}~\bibnamefont {Martin}},\ }\bibfield  {title} {\bibinfo
		{title} {{Levitons in correlated nano-scale systems}},\ }\href
	{https://doi.org/10.1063/5.0199567} {\bibfield  {journal} {\bibinfo
			{journal} {Chaos: An Interdisciplinary Journal of Nonlinear Science}\
		}\textbf {\bibinfo {volume} {34}},\ \bibinfo {pages} {042103} (\bibinfo
		{year} {2024})}\BibitemShut {NoStop}%
	\bibitem [{\citenamefont {Stormer}\ \emph {et~al.}(1999)\citenamefont
		{Stormer}, \citenamefont {Tsui},\ and\ \citenamefont
		{Gossard}}]{Stormer1999}%
	\BibitemOpen
	\bibfield  {author} {\bibinfo {author} {\bibfnamefont {H.~L.}\ \bibnamefont
			{Stormer}}, \bibinfo {author} {\bibfnamefont {D.~C.}\ \bibnamefont {Tsui}},\
		and\ \bibinfo {author} {\bibfnamefont {A.~C.}\ \bibnamefont {Gossard}},\
	}\bibfield  {title} {\bibinfo {title} {The fractional quantum hall effect},\
	}\href {https://doi.org/10.1103/RevModPhys.71.S298} {\bibfield  {journal}
		{\bibinfo  {journal} {Rev. Mod. Phys.}\ }\textbf {\bibinfo {volume} {71}},\
		\bibinfo {pages} {S298} (\bibinfo {year} {1999})}\BibitemShut {NoStop}%
	\bibitem [{\citenamefont {Rech}\ \emph {et~al.}(2017)\citenamefont {Rech},
		\citenamefont {Ferraro}, \citenamefont {Jonckheere}, \citenamefont
		{Vannucci}, \citenamefont {Sassetti},\ and\ \citenamefont
		{Martin}}]{Rech2017}%
	\BibitemOpen
	\bibfield  {author} {\bibinfo {author} {\bibfnamefont {J.}~\bibnamefont
			{Rech}}, \bibinfo {author} {\bibfnamefont {D.}~\bibnamefont {Ferraro}},
		\bibinfo {author} {\bibfnamefont {T.}~\bibnamefont {Jonckheere}}, \bibinfo
		{author} {\bibfnamefont {L.}~\bibnamefont {Vannucci}}, \bibinfo {author}
		{\bibfnamefont {M.}~\bibnamefont {Sassetti}},\ and\ \bibinfo {author}
		{\bibfnamefont {T.}~\bibnamefont {Martin}},\ }\bibfield  {title} {\bibinfo
		{title} {Minimal excitations in the fractional quantum hall regime},\ }\href
	{https://doi.org/10.1103/PhysRevLett.118.076801} {\bibfield  {journal}
		{\bibinfo  {journal} {Phys. Rev. Lett.}\ }\textbf {\bibinfo {volume} {118}},\
		\bibinfo {pages} {076801} (\bibinfo {year} {2017})}\BibitemShut {NoStop}%
	\bibitem [{\citenamefont {Vannucci}\ \emph {et~al.}(2017)\citenamefont
		{Vannucci}, \citenamefont {Ronetti}, \citenamefont {Rech}, \citenamefont
		{Ferraro}, \citenamefont {Jonckheere}, \citenamefont {Martin},\ and\
		\citenamefont {Sassetti}}]{Vannucci2017}%
	\BibitemOpen
	\bibfield  {author} {\bibinfo {author} {\bibfnamefont {L.}~\bibnamefont
			{Vannucci}}, \bibinfo {author} {\bibfnamefont {F.}~\bibnamefont {Ronetti}},
		\bibinfo {author} {\bibfnamefont {J.}~\bibnamefont {Rech}}, \bibinfo {author}
		{\bibfnamefont {D.}~\bibnamefont {Ferraro}}, \bibinfo {author} {\bibfnamefont
			{T.}~\bibnamefont {Jonckheere}}, \bibinfo {author} {\bibfnamefont
			{T.}~\bibnamefont {Martin}},\ and\ \bibinfo {author} {\bibfnamefont
			{M.}~\bibnamefont {Sassetti}},\ }\bibfield  {title} {\bibinfo {title}
		{Minimal excitation states for heat transport in driven quantum hall
			systems},\ }\href {https://doi.org/10.1103/PhysRevB.95.245415} {\bibfield
		{journal} {\bibinfo  {journal} {Phys. Rev. B}\ }\textbf {\bibinfo {volume}
			{95}},\ \bibinfo {pages} {245415} (\bibinfo {year} {2017})}\BibitemShut
	{NoStop}%
	\bibitem [{\citenamefont {Ronetti}\ \emph {et~al.}(2019)\citenamefont
		{Ronetti}, \citenamefont {Vannucci}, \citenamefont {Ferraro}, \citenamefont
		{Jonckheere}, \citenamefont {Rech}, \citenamefont {Martin},\ and\
		\citenamefont {Sassetti}}]{Ronetti2019}%
	\BibitemOpen
	\bibfield  {author} {\bibinfo {author} {\bibfnamefont {F.}~\bibnamefont
			{Ronetti}}, \bibinfo {author} {\bibfnamefont {L.}~\bibnamefont {Vannucci}},
		\bibinfo {author} {\bibfnamefont {D.}~\bibnamefont {Ferraro}}, \bibinfo
		{author} {\bibfnamefont {T.}~\bibnamefont {Jonckheere}}, \bibinfo {author}
		{\bibfnamefont {J.}~\bibnamefont {Rech}}, \bibinfo {author} {\bibfnamefont
			{T.}~\bibnamefont {Martin}},\ and\ \bibinfo {author} {\bibfnamefont
			{M.}~\bibnamefont {Sassetti}},\ }\bibfield  {title} {\bibinfo {title}
		{Hong-ou-mandel heat noise in the quantum hall regime},\ }\href
	{https://doi.org/10.1103/PhysRevB.99.205406} {\bibfield  {journal} {\bibinfo
			{journal} {Phys. Rev. B}\ }\textbf {\bibinfo {volume} {99}},\ \bibinfo
		{pages} {205406} (\bibinfo {year} {2019})}\BibitemShut {NoStop}%
	\bibitem [{\citenamefont {Bertin-Johannet}\ \emph
		{et~al.}(2024{\natexlab{a}})\citenamefont {Bertin-Johannet}, \citenamefont
		{Popoff}, \citenamefont {Ronetti}, \citenamefont {Rech}, \citenamefont
		{Jonckheere}, \citenamefont {Raymond}, \citenamefont {Gr\'emaud},\ and\
		\citenamefont {Martin}}]{Bertin2024}%
	\BibitemOpen
	\bibfield  {author} {\bibinfo {author} {\bibfnamefont {B.}~\bibnamefont
			{Bertin-Johannet}}, \bibinfo {author} {\bibfnamefont {A.}~\bibnamefont
			{Popoff}}, \bibinfo {author} {\bibfnamefont {F.}~\bibnamefont {Ronetti}},
		\bibinfo {author} {\bibfnamefont {J.}~\bibnamefont {Rech}}, \bibinfo {author}
		{\bibfnamefont {T.}~\bibnamefont {Jonckheere}}, \bibinfo {author}
		{\bibfnamefont {L.}~\bibnamefont {Raymond}}, \bibinfo {author} {\bibfnamefont
			{B.}~\bibnamefont {Gr\'emaud}},\ and\ \bibinfo {author} {\bibfnamefont
			{T.}~\bibnamefont {Martin}},\ }\bibfield  {title} {\bibinfo {title}
		{Correlated two-leviton states in the fractional quantum hall regime},\
	}\href {https://doi.org/10.1103/PhysRevB.109.035436} {\bibfield  {journal}
		{\bibinfo  {journal} {Phys. Rev. B}\ }\textbf {\bibinfo {volume} {109}},\
		\bibinfo {pages} {035436} (\bibinfo {year} {2024}{\natexlab{a}})}\BibitemShut
	{NoStop}%
	\bibitem [{\citenamefont {Ronetti}\ \emph {et~al.}(2018)\citenamefont
		{Ronetti}, \citenamefont {Vannucci}, \citenamefont {Ferraro}, \citenamefont
		{Jonckheere}, \citenamefont {Rech}, \citenamefont {Martin},\ and\
		\citenamefont {Sassetti}}]{Ronetti2018}%
	\BibitemOpen
	\bibfield  {author} {\bibinfo {author} {\bibfnamefont {F.}~\bibnamefont
			{Ronetti}}, \bibinfo {author} {\bibfnamefont {L.}~\bibnamefont {Vannucci}},
		\bibinfo {author} {\bibfnamefont {D.}~\bibnamefont {Ferraro}}, \bibinfo
		{author} {\bibfnamefont {T.}~\bibnamefont {Jonckheere}}, \bibinfo {author}
		{\bibfnamefont {J.}~\bibnamefont {Rech}}, \bibinfo {author} {\bibfnamefont
			{T.}~\bibnamefont {Martin}},\ and\ \bibinfo {author} {\bibfnamefont
			{M.}~\bibnamefont {Sassetti}},\ }\bibfield  {title} {\bibinfo {title}
		{Crystallization of levitons in the fractional quantum hall regime},\ }\href
	{https://doi.org/10.1103/PhysRevB.98.075401} {\bibfield  {journal} {\bibinfo
			{journal} {Phys. Rev. B}\ }\textbf {\bibinfo {volume} {98}},\ \bibinfo
		{pages} {075401} (\bibinfo {year} {2018})}\BibitemShut {NoStop}%
	\bibitem [{\citenamefont {Ferraro}\ \emph {et~al.}(2018)\citenamefont
		{Ferraro}, \citenamefont {Ronetti}, \citenamefont {Vannucci}, \citenamefont
		{Acciai}, \citenamefont {Rech}, \citenamefont {Jockheere}, \citenamefont
		{Martin},\ and\ \citenamefont {Sassetti}}]{Ferraro2018}%
	\BibitemOpen
	\bibfield  {author} {\bibinfo {author} {\bibfnamefont {D.}~\bibnamefont
			{Ferraro}}, \bibinfo {author} {\bibfnamefont {F.}~\bibnamefont {Ronetti}},
		\bibinfo {author} {\bibfnamefont {L.}~\bibnamefont {Vannucci}}, \bibinfo
		{author} {\bibfnamefont {M.}~\bibnamefont {Acciai}}, \bibinfo {author}
		{\bibfnamefont {J.}~\bibnamefont {Rech}}, \bibinfo {author} {\bibfnamefont
			{T.}~\bibnamefont {Jockheere}}, \bibinfo {author} {\bibfnamefont
			{T.}~\bibnamefont {Martin}},\ and\ \bibinfo {author} {\bibfnamefont
			{M.}~\bibnamefont {Sassetti}},\ }\bibfield  {title} {\bibinfo {title}
		{Hong-ou-mandel characterization of multiply charged levitons},\ }\href
	{https://doi.org/10.1140/epjst/e2018-800074-1} {\bibfield  {journal}
		{\bibinfo  {journal} {The European Physical Journal Special Topics}\ }\textbf
		{\bibinfo {volume} {227}},\ \bibinfo {pages} {1345} (\bibinfo {year}
		{2018})}\BibitemShut {NoStop}%
	\bibitem [{\citenamefont {Vannucci}\ \emph {et~al.}(2018)\citenamefont
		{Vannucci}, \citenamefont {Ronetti}, \citenamefont {Ferraro}, \citenamefont
		{Rech}, \citenamefont {Jonckheere}, \citenamefont {Martin},\ and\
		\citenamefont {Sassetti}}]{Vannucci2018}%
	\BibitemOpen
	\bibfield  {author} {\bibinfo {author} {\bibfnamefont {L.}~\bibnamefont
			{Vannucci}}, \bibinfo {author} {\bibfnamefont {F.}~\bibnamefont {Ronetti}},
		\bibinfo {author} {\bibfnamefont {D.}~\bibnamefont {Ferraro}}, \bibinfo
		{author} {\bibfnamefont {J.}~\bibnamefont {Rech}}, \bibinfo {author}
		{\bibfnamefont {T.}~\bibnamefont {Jonckheere}}, \bibinfo {author}
		{\bibfnamefont {T.}~\bibnamefont {Martin}},\ and\ \bibinfo {author}
		{\bibfnamefont {M.}~\bibnamefont {Sassetti}},\ }\bibfield  {title} {\bibinfo
		{title} {Photoassisted shot noise spectroscopy at fractional filling
			factor},\ }\href {https://doi.org/10.1088/1742-6596/969/1/012143} {\bibfield
		{journal} {\bibinfo  {journal} {Journal of Physics: Conference Series}\
		}\textbf {\bibinfo {volume} {969}},\ \bibinfo {pages} {012143} (\bibinfo
		{year} {2018})}\BibitemShut {NoStop}%
	\bibitem [{\citenamefont {Acciai}\ \emph {et~al.}(2019)\citenamefont {Acciai},
		\citenamefont {Ronetti}, \citenamefont {Ferraro}, \citenamefont {Rech},
		\citenamefont {Jonckheere}, \citenamefont {Sassetti},\ and\ \citenamefont
		{Martin}}]{Acciai2019}%
	\BibitemOpen
	\bibfield  {author} {\bibinfo {author} {\bibfnamefont {M.}~\bibnamefont
			{Acciai}}, \bibinfo {author} {\bibfnamefont {F.}~\bibnamefont {Ronetti}},
		\bibinfo {author} {\bibfnamefont {D.}~\bibnamefont {Ferraro}}, \bibinfo
		{author} {\bibfnamefont {J.}~\bibnamefont {Rech}}, \bibinfo {author}
		{\bibfnamefont {T.}~\bibnamefont {Jonckheere}}, \bibinfo {author}
		{\bibfnamefont {M.}~\bibnamefont {Sassetti}},\ and\ \bibinfo {author}
		{\bibfnamefont {T.}~\bibnamefont {Martin}},\ }\bibfield  {title} {\bibinfo
		{title} {Levitons in superconducting point contacts},\ }\href
	{https://doi.org/10.1103/PhysRevB.100.085418} {\bibfield  {journal} {\bibinfo
			{journal} {Phys. Rev. B}\ }\textbf {\bibinfo {volume} {100}},\ \bibinfo
		{pages} {085418} (\bibinfo {year} {2019})}\BibitemShut {NoStop}%
	\bibitem [{\citenamefont {Ronetti}\ \emph {et~al.}(2020)\citenamefont
		{Ronetti}, \citenamefont {Carrega},\ and\ \citenamefont
		{Sassetti}}]{Ronetti2020}%
	\BibitemOpen
	\bibfield  {author} {\bibinfo {author} {\bibfnamefont {F.}~\bibnamefont
			{Ronetti}}, \bibinfo {author} {\bibfnamefont {M.}~\bibnamefont {Carrega}},\
		and\ \bibinfo {author} {\bibfnamefont {M.}~\bibnamefont {Sassetti}},\
	}\bibfield  {title} {\bibinfo {title} {Levitons in helical liquids with
			rashba spin-orbit coupling probed by a superconducting contact},\ }\href
	{https://doi.org/10.1103/PhysRevResearch.2.013203} {\bibfield  {journal}
		{\bibinfo  {journal} {Phys. Rev. Res.}\ }\textbf {\bibinfo {volume} {2}},\
		\bibinfo {pages} {013203} (\bibinfo {year} {2020})}\BibitemShut {NoStop}%
	\bibitem [{\citenamefont {Bertin-Johannet}\ \emph {et~al.}(2022)\citenamefont
		{Bertin-Johannet}, \citenamefont {Rech}, \citenamefont {Jonckheere},
		\citenamefont {Gr\'emaud}, \citenamefont {Raymond},\ and\ \citenamefont
		{Martin}}]{Bertin2022}%
	\BibitemOpen
	\bibfield  {author} {\bibinfo {author} {\bibfnamefont {B.}~\bibnamefont
			{Bertin-Johannet}}, \bibinfo {author} {\bibfnamefont {J.}~\bibnamefont
			{Rech}}, \bibinfo {author} {\bibfnamefont {T.}~\bibnamefont {Jonckheere}},
		\bibinfo {author} {\bibfnamefont {B.}~\bibnamefont {Gr\'emaud}}, \bibinfo
		{author} {\bibfnamefont {L.}~\bibnamefont {Raymond}},\ and\ \bibinfo {author}
		{\bibfnamefont {T.}~\bibnamefont {Martin}},\ }\bibfield  {title} {\bibinfo
		{title} {Microscopic theory of photoassisted electronic transport in
			normal-metal/bcs-superconductor junctions},\ }\href
	{https://doi.org/10.1103/PhysRevB.105.115112} {\bibfield  {journal} {\bibinfo
			{journal} {Phys. Rev. B}\ }\textbf {\bibinfo {volume} {105}},\ \bibinfo
		{pages} {115112} (\bibinfo {year} {2022})}\BibitemShut {NoStop}%
	\bibitem [{\citenamefont {Bertin-Johannet}\ \emph {et~al.}(2023)\citenamefont
		{Bertin-Johannet}, \citenamefont {Raymond}, \citenamefont {Ronetti},
		\citenamefont {Rech}, \citenamefont {Jonckheere}, \citenamefont {Grémaud},\
		and\ \citenamefont {Martin}}]{Bertin2023}%
	\BibitemOpen
	\bibfield  {author} {\bibinfo {author} {\bibfnamefont {B.}~\bibnamefont
			{Bertin-Johannet}}, \bibinfo {author} {\bibfnamefont {L.}~\bibnamefont
			{Raymond}}, \bibinfo {author} {\bibfnamefont {F.}~\bibnamefont {Ronetti}},
		\bibinfo {author} {\bibfnamefont {J.}~\bibnamefont {Rech}}, \bibinfo {author}
		{\bibfnamefont {T.}~\bibnamefont {Jonckheere}}, \bibinfo {author}
		{\bibfnamefont {B.}~\bibnamefont {Grémaud}},\ and\ \bibinfo {author}
		{\bibfnamefont {T.}~\bibnamefont {Martin}},\ }\bibfield  {title} {\bibinfo
		{title} {{An on-demand source of energy-entangled electrons using
				levitons}},\ }\href {https://doi.org/10.1063/5.0148041} {\bibfield  {journal}
		{\bibinfo  {journal} {Applied Physics Letters}\ }\textbf {\bibinfo {volume}
			{122}},\ \bibinfo {pages} {202601} (\bibinfo {year} {2023})}\BibitemShut
	{NoStop}%
	\bibitem [{\citenamefont {Bertin-Johannet}\ \emph
		{et~al.}(2024{\natexlab{b}})\citenamefont {Bertin-Johannet}, \citenamefont
		{Gr\'emaud}, \citenamefont {Ronetti}, \citenamefont {Raymond}, \citenamefont
		{Rech}, \citenamefont {Jonckheere},\ and\ \citenamefont
		{Martin}}]{Bertin2024b}%
	\BibitemOpen
	\bibfield  {author} {\bibinfo {author} {\bibfnamefont {B.}~\bibnamefont
			{Bertin-Johannet}}, \bibinfo {author} {\bibfnamefont {B.}~\bibnamefont
			{Gr\'emaud}}, \bibinfo {author} {\bibfnamefont {F.}~\bibnamefont {Ronetti}},
		\bibinfo {author} {\bibfnamefont {L.}~\bibnamefont {Raymond}}, \bibinfo
		{author} {\bibfnamefont {J.}~\bibnamefont {Rech}}, \bibinfo {author}
		{\bibfnamefont {T.}~\bibnamefont {Jonckheere}},\ and\ \bibinfo {author}
		{\bibfnamefont {T.}~\bibnamefont {Martin}},\ }\href
	{https://doi.org/10.1103/PhysRevB.109.174514} {\bibfield  {journal} {\bibinfo
			{journal} {Phys. Rev. B}\ }\textbf {\bibinfo {volume} {109}},\ \bibinfo
		{pages} {174514} (\bibinfo {year} {2024}{\natexlab{b}})}\BibitemShut
	{NoStop}%
	\bibitem [{\citenamefont {Grenier}\ \emph
		{et~al.}(2011{\natexlab{b}})\citenamefont {Grenier}, \citenamefont
		{Herv\'{e}}, \citenamefont {F\`{e}ve},\ and\ \citenamefont
		{Degiovanni}}]{Grenier2011b}%
	\BibitemOpen
	\bibfield  {author} {\bibinfo {author} {\bibfnamefont {C.}~\bibnamefont
			{Grenier}}, \bibinfo {author} {\bibfnamefont {R.}~\bibnamefont {Herv\'{e}}},
		\bibinfo {author} {\bibfnamefont {G.}~\bibnamefont {F\`{e}ve}},\ and\
		\bibinfo {author} {\bibfnamefont {P.}~\bibnamefont {Degiovanni}},\ }\bibfield
	{title} {\bibinfo {title} {Electron quantum optics in quantum hall edge
			channels},\ }\href {https://doi.org/10.1142/S0217984911026772} {\bibfield
		{journal} {\bibinfo  {journal} {Modern Physics Letters B}\ }\textbf {\bibinfo
			{volume} {25}},\ \bibinfo {pages} {1053} (\bibinfo {year}
		{2011}{\natexlab{b}})}\BibitemShut {NoStop}%
	\bibitem [{\citenamefont {Bocquillon}\ \emph {et~al.}(2012)\citenamefont
		{Bocquillon}, \citenamefont {Parmentier}, \citenamefont {Grenier},
		\citenamefont {Berroir}, \citenamefont {Degiovanni}, \citenamefont {Glattli},
		\citenamefont {Pla\ifmmode~\mbox{\c{c}}\else \c{c}\fi{}ais}, \citenamefont
		{Cavanna}, \citenamefont {Jin},\ and\ \citenamefont
		{F\`eve}}]{Bocquillon2012}%
	\BibitemOpen
	\bibfield  {author} {\bibinfo {author} {\bibfnamefont {E.}~\bibnamefont
			{Bocquillon}}, \bibinfo {author} {\bibfnamefont {F.~D.}\ \bibnamefont
			{Parmentier}}, \bibinfo {author} {\bibfnamefont {C.}~\bibnamefont {Grenier}},
		\bibinfo {author} {\bibfnamefont {J.-M.}\ \bibnamefont {Berroir}}, \bibinfo
		{author} {\bibfnamefont {P.}~\bibnamefont {Degiovanni}}, \bibinfo {author}
		{\bibfnamefont {D.~C.}\ \bibnamefont {Glattli}}, \bibinfo {author}
		{\bibfnamefont {B.}~\bibnamefont {Pla\ifmmode~\mbox{\c{c}}\else
				\c{c}\fi{}ais}}, \bibinfo {author} {\bibfnamefont {A.}~\bibnamefont
			{Cavanna}}, \bibinfo {author} {\bibfnamefont {Y.}~\bibnamefont {Jin}},\ and\
		\bibinfo {author} {\bibfnamefont {G.}~\bibnamefont {F\`eve}},\ }\bibfield
	{title} {\bibinfo {title} {Electron quantum optics: Partitioning electrons
			one by one},\ }\href {https://doi.org/10.1103/PhysRevLett.108.196803}
	{\bibfield  {journal} {\bibinfo  {journal} {Phys. Rev. Lett.}\ }\textbf
		{\bibinfo {volume} {108}},\ \bibinfo {pages} {196803} (\bibinfo {year}
		{2012})}\BibitemShut {NoStop}%
	\bibitem [{\citenamefont {Bocquillon}\ \emph {et~al.}(2014)\citenamefont
		{Bocquillon}, \citenamefont {Freulon}, \citenamefont {Parmentier},
		\citenamefont {Berroir}, \citenamefont {Plaçais}, \citenamefont {Wahl},
		\citenamefont {Rech}, \citenamefont {Jonckheere}, \citenamefont {Martin},
		\citenamefont {Grenier}, \citenamefont {Ferraro}, \citenamefont
		{Degiovanni},\ and\ \citenamefont {Fève}}]{Bocquillon2014}%
	\BibitemOpen
	\bibfield  {author} {\bibinfo {author} {\bibfnamefont {E.}~\bibnamefont
			{Bocquillon}}, \bibinfo {author} {\bibfnamefont {V.}~\bibnamefont {Freulon}},
		\bibinfo {author} {\bibfnamefont {F.~D.}\ \bibnamefont {Parmentier}},
		\bibinfo {author} {\bibfnamefont {J.-M.}\ \bibnamefont {Berroir}}, \bibinfo
		{author} {\bibfnamefont {B.}~\bibnamefont {Plaçais}}, \bibinfo {author}
		{\bibfnamefont {C.}~\bibnamefont {Wahl}}, \bibinfo {author} {\bibfnamefont
			{J.}~\bibnamefont {Rech}}, \bibinfo {author} {\bibfnamefont {T.}~\bibnamefont
			{Jonckheere}}, \bibinfo {author} {\bibfnamefont {T.}~\bibnamefont {Martin}},
		\bibinfo {author} {\bibfnamefont {C.}~\bibnamefont {Grenier}}, \bibinfo
		{author} {\bibfnamefont {D.}~\bibnamefont {Ferraro}}, \bibinfo {author}
		{\bibfnamefont {P.}~\bibnamefont {Degiovanni}},\ and\ \bibinfo {author}
		{\bibfnamefont {G.}~\bibnamefont {Fève}},\ }\bibfield  {title} {\bibinfo
		{title} {Electron quantum optics in ballistic chiral conductors},\ }\href
	{https://doi.org/https://doi.org/10.1002/andp.201300181} {\bibfield
		{journal} {\bibinfo  {journal} {Annalen der Physik}\ }\textbf {\bibinfo
			{volume} {526}},\ \bibinfo {pages} {1} (\bibinfo {year} {2014})}\BibitemShut
	{NoStop}%
	\bibitem [{\citenamefont {Roussel}\ \emph {et~al.}(2017)\citenamefont
		{Roussel}, \citenamefont {Cabart}, \citenamefont {Fève}, \citenamefont
		{Thibierge},\ and\ \citenamefont {Degiovanni}}]{Roussel2017}%
	\BibitemOpen
	\bibfield  {author} {\bibinfo {author} {\bibfnamefont {B.}~\bibnamefont
			{Roussel}}, \bibinfo {author} {\bibfnamefont {C.}~\bibnamefont {Cabart}},
		\bibinfo {author} {\bibfnamefont {G.}~\bibnamefont {Fève}}, \bibinfo
		{author} {\bibfnamefont {E.}~\bibnamefont {Thibierge}},\ and\ \bibinfo
		{author} {\bibfnamefont {P.}~\bibnamefont {Degiovanni}},\ }\bibfield  {title}
	{\bibinfo {title} {Electron quantum optics as quantum signal processing},\
	}\href {https://doi.org/https://doi.org/10.1002/pssb.201600621} {\bibfield
		{journal} {\bibinfo  {journal} {physica status solidi (b)}\ }\textbf
		{\bibinfo {volume} {254}},\ \bibinfo {pages} {1600621} (\bibinfo {year}
		{2017})}\BibitemShut {NoStop}%
	\bibitem [{\citenamefont {Marguerite}\ \emph {et~al.}(2016)\citenamefont
		{Marguerite}, \citenamefont {Cabart}, \citenamefont {Wahl}, \citenamefont
		{Roussel}, \citenamefont {Freulon}, \citenamefont {Ferraro}, \citenamefont
		{Grenier}, \citenamefont {Berroir}, \citenamefont
		{Pla\ifmmode~\mbox{\c{c}}\else \c{c}\fi{}ais}, \citenamefont {Jonckheere},
		\citenamefont {Rech}, \citenamefont {Martin}, \citenamefont {Degiovanni},
		\citenamefont {Cavanna}, \citenamefont {Jin},\ and\ \citenamefont
		{F\`eve}}]{Marguerite2016}%
	\BibitemOpen
	\bibfield  {author} {\bibinfo {author} {\bibfnamefont {A.}~\bibnamefont
			{Marguerite}}, \bibinfo {author} {\bibfnamefont {C.}~\bibnamefont {Cabart}},
		\bibinfo {author} {\bibfnamefont {C.}~\bibnamefont {Wahl}}, \bibinfo {author}
		{\bibfnamefont {B.}~\bibnamefont {Roussel}}, \bibinfo {author} {\bibfnamefont
			{V.}~\bibnamefont {Freulon}}, \bibinfo {author} {\bibfnamefont
			{D.}~\bibnamefont {Ferraro}}, \bibinfo {author} {\bibfnamefont
			{C.}~\bibnamefont {Grenier}}, \bibinfo {author} {\bibfnamefont {J.-M.}\
			\bibnamefont {Berroir}}, \bibinfo {author} {\bibfnamefont {B.}~\bibnamefont
			{Pla\ifmmode~\mbox{\c{c}}\else \c{c}\fi{}ais}}, \bibinfo {author}
		{\bibfnamefont {T.}~\bibnamefont {Jonckheere}}, \bibinfo {author}
		{\bibfnamefont {J.}~\bibnamefont {Rech}}, \bibinfo {author} {\bibfnamefont
			{T.}~\bibnamefont {Martin}}, \bibinfo {author} {\bibfnamefont
			{P.}~\bibnamefont {Degiovanni}}, \bibinfo {author} {\bibfnamefont
			{A.}~\bibnamefont {Cavanna}}, \bibinfo {author} {\bibfnamefont
			{Y.}~\bibnamefont {Jin}},\ and\ \bibinfo {author} {\bibfnamefont
			{G.}~\bibnamefont {F\`eve}},\ }\bibfield  {title} {\bibinfo {title}
		{Decoherence and relaxation of a single electron in a one-dimensional
			conductor},\ }\href {https://doi.org/10.1103/PhysRevB.94.115311} {\bibfield
		{journal} {\bibinfo  {journal} {Phys. Rev. B}\ }\textbf {\bibinfo {volume}
			{94}},\ \bibinfo {pages} {115311} (\bibinfo {year} {2016})}\BibitemShut
	{NoStop}%
	\bibitem [{\citenamefont {Wang}\ \emph {et~al.}(2023)\citenamefont {Wang},
		\citenamefont {Edlbauer}, \citenamefont {Richard}, \citenamefont {Ota},
		\citenamefont {Park}, \citenamefont {Shim}, \citenamefont {Ludwig},
		\citenamefont {Wieck}, \citenamefont {Sim}, \citenamefont {Urdampilleta},
		\citenamefont {Meunier}, \citenamefont {Kodera}, \citenamefont {Kaneko},
		\citenamefont {Sellier}, \citenamefont {Waintal}, \citenamefont {Takada},\
		and\ \citenamefont {B{\"a}uerle}}]{Wang2023}%
	\BibitemOpen
	\bibfield  {author} {\bibinfo {author} {\bibfnamefont {J.}~\bibnamefont
			{Wang}}, \bibinfo {author} {\bibfnamefont {H.}~\bibnamefont {Edlbauer}},
		\bibinfo {author} {\bibfnamefont {A.}~\bibnamefont {Richard}}, \bibinfo
		{author} {\bibfnamefont {S.}~\bibnamefont {Ota}}, \bibinfo {author}
		{\bibfnamefont {W.}~\bibnamefont {Park}}, \bibinfo {author} {\bibfnamefont
			{J.}~\bibnamefont {Shim}}, \bibinfo {author} {\bibfnamefont {A.}~\bibnamefont
			{Ludwig}}, \bibinfo {author} {\bibfnamefont {A.~D.}\ \bibnamefont {Wieck}},
		\bibinfo {author} {\bibfnamefont {H.-S.}\ \bibnamefont {Sim}}, \bibinfo
		{author} {\bibfnamefont {M.}~\bibnamefont {Urdampilleta}}, \bibinfo {author}
		{\bibfnamefont {T.}~\bibnamefont {Meunier}}, \bibinfo {author} {\bibfnamefont
			{T.}~\bibnamefont {Kodera}}, \bibinfo {author} {\bibfnamefont {N.-H.}\
			\bibnamefont {Kaneko}}, \bibinfo {author} {\bibfnamefont {H.}~\bibnamefont
			{Sellier}}, \bibinfo {author} {\bibfnamefont {X.}~\bibnamefont {Waintal}},
		\bibinfo {author} {\bibfnamefont {S.}~\bibnamefont {Takada}},\ and\ \bibinfo
		{author} {\bibfnamefont {C.}~\bibnamefont {B{\"a}uerle}},\ }\bibfield
	{title} {\bibinfo {title} {Coulomb-mediated antibunching of an electron pair
			surfing on sound},\ }\href {https://doi.org/10.1038/s41565-023-01368-5}
	{\bibfield  {journal} {\bibinfo  {journal} {Nature Nanotechnology}\ }\textbf
		{\bibinfo {volume} {18}},\ \bibinfo {pages} {721} (\bibinfo {year}
		{2023})}\BibitemShut {NoStop}%
	\bibitem [{\citenamefont {Ubbelohde}\ \emph {et~al.}(2023)\citenamefont
		{Ubbelohde}, \citenamefont {Freise}, \citenamefont {Pavlovska}, \citenamefont
		{Silvestrov}, \citenamefont {Recher}, \citenamefont {Kokainis}, \citenamefont
		{Barinovs}, \citenamefont {Hohls}, \citenamefont {Weimann}, \citenamefont
		{Pierz},\ and\ \citenamefont {Kashcheyevs}}]{Ubbelohde2023}%
	\BibitemOpen
	\bibfield  {author} {\bibinfo {author} {\bibfnamefont {N.}~\bibnamefont
			{Ubbelohde}}, \bibinfo {author} {\bibfnamefont {L.}~\bibnamefont {Freise}},
		\bibinfo {author} {\bibfnamefont {E.}~\bibnamefont {Pavlovska}}, \bibinfo
		{author} {\bibfnamefont {P.~G.}\ \bibnamefont {Silvestrov}}, \bibinfo
		{author} {\bibfnamefont {P.}~\bibnamefont {Recher}}, \bibinfo {author}
		{\bibfnamefont {M.}~\bibnamefont {Kokainis}}, \bibinfo {author}
		{\bibfnamefont {G.}~\bibnamefont {Barinovs}}, \bibinfo {author}
		{\bibfnamefont {F.}~\bibnamefont {Hohls}}, \bibinfo {author} {\bibfnamefont
			{T.}~\bibnamefont {Weimann}}, \bibinfo {author} {\bibfnamefont
			{K.}~\bibnamefont {Pierz}},\ and\ \bibinfo {author} {\bibfnamefont
			{V.}~\bibnamefont {Kashcheyevs}},\ }\bibfield  {title} {\bibinfo {title} {Two
			electrons interacting at a mesoscopic beam splitter},\ }\href
	{https://doi.org/10.1038/s41565-023-01370-x} {\bibfield  {journal} {\bibinfo
			{journal} {Nature Nanotechnology}\ }\textbf {\bibinfo {volume} {18}},\
		\bibinfo {pages} {733} (\bibinfo {year} {2023})}\BibitemShut {NoStop}%
	\bibitem [{\citenamefont {Fletcher}\ \emph {et~al.}(2023)\citenamefont
		{Fletcher}, \citenamefont {Park}, \citenamefont {Ryu}, \citenamefont {See},
		\citenamefont {Griffiths}, \citenamefont {Jones}, \citenamefont {Farrer},
		\citenamefont {Ritchie}, \citenamefont {Sim},\ and\ \citenamefont
		{Kataoka}}]{Fletcher2023}%
	\BibitemOpen
	\bibfield  {author} {\bibinfo {author} {\bibfnamefont {J.~D.}\ \bibnamefont
			{Fletcher}}, \bibinfo {author} {\bibfnamefont {W.}~\bibnamefont {Park}},
		\bibinfo {author} {\bibfnamefont {S.}~\bibnamefont {Ryu}}, \bibinfo {author}
		{\bibfnamefont {P.}~\bibnamefont {See}}, \bibinfo {author} {\bibfnamefont
			{J.~P.}\ \bibnamefont {Griffiths}}, \bibinfo {author} {\bibfnamefont
			{G.~A.~C.}\ \bibnamefont {Jones}}, \bibinfo {author} {\bibfnamefont
			{I.}~\bibnamefont {Farrer}}, \bibinfo {author} {\bibfnamefont {D.~A.}\
			\bibnamefont {Ritchie}}, \bibinfo {author} {\bibfnamefont {H.~S.}\
			\bibnamefont {Sim}},\ and\ \bibinfo {author} {\bibfnamefont {M.}~\bibnamefont
			{Kataoka}},\ }\bibfield  {title} {\bibinfo {title} {Time-resolved coulomb
			collision of single electrons},\ }\href@noop {} {\bibfield  {journal}
		{\bibinfo  {journal} {Nature Nanotechnology}\ }\textbf {\bibinfo {volume}
			{18}},\ \bibinfo {pages} {727} (\bibinfo {year} {2023})}\BibitemShut
	{NoStop}%
	\bibitem [{\citenamefont {Roussely}\ \emph {et~al.}(2018)\citenamefont
		{Roussely}, \citenamefont {Arrighi}, \citenamefont {Georgiou}, \citenamefont
		{Takada}, \citenamefont {Schalk}, \citenamefont {Urdampilleta}, \citenamefont
		{Ludwig}, \citenamefont {Wieck}, \citenamefont {Armagnat}, \citenamefont
		{Kloss}, \citenamefont {Waintal}, \citenamefont {Meunier},\ and\
		\citenamefont {B{\"a}uerle}}]{Roussely2018}%
	\BibitemOpen
	\bibfield  {author} {\bibinfo {author} {\bibfnamefont {G.}~\bibnamefont
			{Roussely}}, \bibinfo {author} {\bibfnamefont {E.}~\bibnamefont {Arrighi}},
		\bibinfo {author} {\bibfnamefont {G.}~\bibnamefont {Georgiou}}, \bibinfo
		{author} {\bibfnamefont {S.}~\bibnamefont {Takada}}, \bibinfo {author}
		{\bibfnamefont {M.}~\bibnamefont {Schalk}}, \bibinfo {author} {\bibfnamefont
			{M.}~\bibnamefont {Urdampilleta}}, \bibinfo {author} {\bibfnamefont
			{A.}~\bibnamefont {Ludwig}}, \bibinfo {author} {\bibfnamefont {A.~D.}\
			\bibnamefont {Wieck}}, \bibinfo {author} {\bibfnamefont {P.}~\bibnamefont
			{Armagnat}}, \bibinfo {author} {\bibfnamefont {T.}~\bibnamefont {Kloss}},
		\bibinfo {author} {\bibfnamefont {X.}~\bibnamefont {Waintal}}, \bibinfo
		{author} {\bibfnamefont {T.}~\bibnamefont {Meunier}},\ and\ \bibinfo {author}
		{\bibfnamefont {C.}~\bibnamefont {B{\"a}uerle}},\ }\bibfield  {title}
	{\bibinfo {title} {Unveiling the bosonic nature of an ultrashort few-electron
			pulse},\ }\href@noop {} {\bibfield  {journal} {\bibinfo  {journal} {Nature
				Communications}\ }\textbf {\bibinfo {volume} {9}},\ \bibinfo {pages} {2811}
		(\bibinfo {year} {2018})}\BibitemShut {NoStop}%
	\bibitem [{\citenamefont {Aluffi}\ \emph {et~al.}(2023)\citenamefont {Aluffi},
		\citenamefont {Vasselon}, \citenamefont {Ouacel}, \citenamefont {Edlbauer},
		\citenamefont {Geffroy}, \citenamefont {Roulleau}, \citenamefont {Glattli},
		\citenamefont {Georgiou},\ and\ \citenamefont {B\"auerle}}]{Aluffi2023}%
	\BibitemOpen
	\bibfield  {author} {\bibinfo {author} {\bibfnamefont {M.}~\bibnamefont
			{Aluffi}}, \bibinfo {author} {\bibfnamefont {T.}~\bibnamefont {Vasselon}},
		\bibinfo {author} {\bibfnamefont {S.}~\bibnamefont {Ouacel}}, \bibinfo
		{author} {\bibfnamefont {H.}~\bibnamefont {Edlbauer}}, \bibinfo {author}
		{\bibfnamefont {C.}~\bibnamefont {Geffroy}}, \bibinfo {author} {\bibfnamefont
			{P.}~\bibnamefont {Roulleau}}, \bibinfo {author} {\bibfnamefont {D.~C.}\
			\bibnamefont {Glattli}}, \bibinfo {author} {\bibfnamefont {G.}~\bibnamefont
			{Georgiou}},\ and\ \bibinfo {author} {\bibfnamefont {C.}~\bibnamefont
			{B\"auerle}},\ }\bibfield  {title} {\bibinfo {title} {Ultrashort electron
			wave packets via frequency-comb synthesis},\ }\href
	{https://doi.org/10.1103/PhysRevApplied.20.034005} {\bibfield  {journal}
		{\bibinfo  {journal} {Phys. Rev. Appl.}\ }\textbf {\bibinfo {volume} {20}},\
		\bibinfo {pages} {034005} (\bibinfo {year} {2023})}\BibitemShut {NoStop}%
	\bibitem [{\citenamefont {Wang}\ \emph {et~al.}(2024)\citenamefont {Wang},
		\citenamefont {Edlbauer}, \citenamefont {Jadot}, \citenamefont
		{Mortemousque}, \citenamefont {Takada}, \citenamefont {Bäuerle},\ and\
		\citenamefont {Sellier}}]{Wang2024}%
	\BibitemOpen
	\bibfield  {author} {\bibinfo {author} {\bibfnamefont {J.}~\bibnamefont
			{Wang}}, \bibinfo {author} {\bibfnamefont {H.}~\bibnamefont {Edlbauer}},
		\bibinfo {author} {\bibfnamefont {B.}~\bibnamefont {Jadot}}, \bibinfo
		{author} {\bibfnamefont {P.-A.}\ \bibnamefont {Mortemousque}}, \bibinfo
		{author} {\bibfnamefont {S.}~\bibnamefont {Takada}}, \bibinfo {author}
		{\bibfnamefont {C.}~\bibnamefont {Bäuerle}},\ and\ \bibinfo {author}
		{\bibfnamefont {H.}~\bibnamefont {Sellier}},\ }\href@noop {} {\bibinfo
		{title} {Electron qubits surfing on acoustic waves: review of recent
			progress}} (\bibinfo {year} {2024}),\ \Eprint
	{https://arxiv.org/abs/2402.04748} {arXiv:2402.04748 [cond-mat.mes-hall]}
	\BibitemShut {NoStop}%
	\bibitem [{\citenamefont {Takeda}\ and\ \citenamefont
		{Furusawa}(2017)}]{Takeda2017}%
	\BibitemOpen
	\bibfield  {author} {\bibinfo {author} {\bibfnamefont {S.}~\bibnamefont
			{Takeda}}\ and\ \bibinfo {author} {\bibfnamefont {A.}~\bibnamefont
			{Furusawa}},\ }\bibfield  {title} {\bibinfo {title} {Universal quantum
			computing with measurement-induced continuous-variable gate sequence in a
			loop-based architecture},\ }\href
	{https://doi.org/10.1103/PhysRevLett.119.120504} {\bibfield  {journal}
		{\bibinfo  {journal} {Phys. Rev. Lett.}\ }\textbf {\bibinfo {volume} {119}},\
		\bibinfo {pages} {120504} (\bibinfo {year} {2017})}\BibitemShut {NoStop}%
	\bibitem [{\citenamefont {Wang}\ \emph {et~al.}(2022)\citenamefont {Wang},
		\citenamefont {Ota}, \citenamefont {Edlbauer}, \citenamefont {Jadot},
		\citenamefont {Mortemousque}, \citenamefont {Richard}, \citenamefont
		{Okazaki}, \citenamefont {Nakamura}, \citenamefont {Ludwig}, \citenamefont
		{Wieck}, \citenamefont {Urdampilleta}, \citenamefont {Meunier}, \citenamefont
		{Kodera}, \citenamefont {Kaneko}, \citenamefont {Takada},\ and\ \citenamefont
		{B\"auerle}}]{Wang2022Rev}%
	\BibitemOpen
	\bibfield  {author} {\bibinfo {author} {\bibfnamefont {J.}~\bibnamefont
			{Wang}}, \bibinfo {author} {\bibfnamefont {S.}~\bibnamefont {Ota}}, \bibinfo
		{author} {\bibfnamefont {H.}~\bibnamefont {Edlbauer}}, \bibinfo {author}
		{\bibfnamefont {B.}~\bibnamefont {Jadot}}, \bibinfo {author} {\bibfnamefont
			{P.-A.}\ \bibnamefont {Mortemousque}}, \bibinfo {author} {\bibfnamefont
			{A.}~\bibnamefont {Richard}}, \bibinfo {author} {\bibfnamefont
			{Y.}~\bibnamefont {Okazaki}}, \bibinfo {author} {\bibfnamefont
			{S.}~\bibnamefont {Nakamura}}, \bibinfo {author} {\bibfnamefont
			{A.}~\bibnamefont {Ludwig}}, \bibinfo {author} {\bibfnamefont {A.~D.}\
			\bibnamefont {Wieck}}, \bibinfo {author} {\bibfnamefont {M.}~\bibnamefont
			{Urdampilleta}}, \bibinfo {author} {\bibfnamefont {T.}~\bibnamefont
			{Meunier}}, \bibinfo {author} {\bibfnamefont {T.}~\bibnamefont {Kodera}},
		\bibinfo {author} {\bibfnamefont {N.-H.}\ \bibnamefont {Kaneko}}, \bibinfo
		{author} {\bibfnamefont {S.}~\bibnamefont {Takada}},\ and\ \bibinfo {author}
		{\bibfnamefont {C.}~\bibnamefont {B\"auerle}},\ }\bibfield  {title} {\bibinfo
		{title} {Generation of a single-cycle acoustic pulse: A scalable solution for
			transport in single-electron circuits},\ }\href
	{https://doi.org/10.1103/PhysRevX.12.031035} {\bibfield  {journal} {\bibinfo
			{journal} {Phys. Rev. X}\ }\textbf {\bibinfo {volume} {12}},\ \bibinfo
		{pages} {031035} (\bibinfo {year} {2022})}\BibitemShut {NoStop}%
	\bibitem [{\citenamefont {Zhang}\ \emph {et~al.}(2024)\citenamefont {Zhang},
		\citenamefont {Hong}, \citenamefont {Alkalay}, \citenamefont {Umansky},
		\citenamefont {Heiblum}, \citenamefont {Gornyi},\ and\ \citenamefont
		{Gefen}}]{Zhang2024}%
	\BibitemOpen
	\bibfield  {author} {\bibinfo {author} {\bibfnamefont {G.}~\bibnamefont
			{Zhang}}, \bibinfo {author} {\bibfnamefont {C.}~\bibnamefont {Hong}},
		\bibinfo {author} {\bibfnamefont {T.}~\bibnamefont {Alkalay}}, \bibinfo
		{author} {\bibfnamefont {V.}~\bibnamefont {Umansky}}, \bibinfo {author}
		{\bibfnamefont {M.}~\bibnamefont {Heiblum}}, \bibinfo {author} {\bibfnamefont
			{I.}~\bibnamefont {Gornyi}},\ and\ \bibinfo {author} {\bibfnamefont
			{Y.}~\bibnamefont {Gefen}},\ }\bibfield  {title} {\bibinfo {title} {Measuring
			statistics-induced entanglement entropy with a hong--ou--mandel
			interferometer},\ }\href {https://doi.org/10.1038/s41467-024-47335-z}
	{\bibfield  {journal} {\bibinfo  {journal} {Nature Communications}\ }\textbf
		{\bibinfo {volume} {15}},\ \bibinfo {pages} {3428} (\bibinfo {year}
		{2024})}\BibitemShut {NoStop}%
	\bibitem [{\citenamefont {Vyshnevyy}\ \emph {et~al.}(2013)\citenamefont
		{Vyshnevyy}, \citenamefont {Lesovik}, \citenamefont {Jonckheere},\ and\
		\citenamefont {Martin}}]{Vyshnevyy2013}%
	\BibitemOpen
	\bibfield  {author} {\bibinfo {author} {\bibfnamefont {A.~A.}\ \bibnamefont
			{Vyshnevyy}}, \bibinfo {author} {\bibfnamefont {G.~B.}\ \bibnamefont
			{Lesovik}}, \bibinfo {author} {\bibfnamefont {T.}~\bibnamefont
			{Jonckheere}},\ and\ \bibinfo {author} {\bibfnamefont {T.}~\bibnamefont
			{Martin}},\ }\bibfield  {title} {\bibinfo {title} {Setup of three
			mach-zehnder interferometers for production and observation of
			greenberger-horne-zeilinger entanglement of electrons},\ }\href
	{https://doi.org/10.1103/PhysRevB.87.165417} {\bibfield  {journal} {\bibinfo
			{journal} {Phys. Rev. B}\ }\textbf {\bibinfo {volume} {87}},\ \bibinfo
		{pages} {165417} (\bibinfo {year} {2013})}\BibitemShut {NoStop}%
	\bibitem [{\citenamefont {Chtchelkatchev}\ \emph {et~al.}(2002)\citenamefont
		{Chtchelkatchev}, \citenamefont {Blatter}, \citenamefont {Lesovik},\ and\
		\citenamefont {Martin}}]{Chtchelkatchev2002}%
	\BibitemOpen
	\bibfield  {author} {\bibinfo {author} {\bibfnamefont {N.~M.}\ \bibnamefont
			{Chtchelkatchev}}, \bibinfo {author} {\bibfnamefont {G.}~\bibnamefont
			{Blatter}}, \bibinfo {author} {\bibfnamefont {G.~B.}\ \bibnamefont
			{Lesovik}},\ and\ \bibinfo {author} {\bibfnamefont {T.}~\bibnamefont
			{Martin}},\ }\bibfield  {title} {\bibinfo {title} {Bell inequalities and
			entanglement in solid-state devices},\ }\href
	{https://doi.org/10.1103/PhysRevB.66.161320} {\bibfield  {journal} {\bibinfo
			{journal} {Phys. Rev. B}\ }\textbf {\bibinfo {volume} {66}},\ \bibinfo
		{pages} {161320} (\bibinfo {year} {2002})}\BibitemShut {NoStop}%
	\bibitem [{\citenamefont {Recher}\ \emph {et~al.}(2001)\citenamefont {Recher},
		\citenamefont {Sukhorukov},\ and\ \citenamefont {Loss}}]{Recher2001}%
	\BibitemOpen
	\bibfield  {author} {\bibinfo {author} {\bibfnamefont {P.}~\bibnamefont
			{Recher}}, \bibinfo {author} {\bibfnamefont {E.~V.}\ \bibnamefont
			{Sukhorukov}},\ and\ \bibinfo {author} {\bibfnamefont {D.}~\bibnamefont
			{Loss}},\ }\bibfield  {title} {\bibinfo {title} {Andreev tunneling, coulomb
			blockade, and resonant transport of nonlocal spin-entangled electrons},\
	}\href {https://doi.org/10.1103/PhysRevB.63.165314} {\bibfield  {journal}
		{\bibinfo  {journal} {Phys. Rev. B}\ }\textbf {\bibinfo {volume} {63}},\
		\bibinfo {pages} {165314} (\bibinfo {year} {2001})}\BibitemShut {NoStop}%
	\bibitem [{\citenamefont {Samuelsson}\ \emph {et~al.}(2003)\citenamefont
		{Samuelsson}, \citenamefont {Sukhorukov},\ and\ \citenamefont
		{B\"uttiker}}]{Samuelsson2003}%
	\BibitemOpen
	\bibfield  {author} {\bibinfo {author} {\bibfnamefont {P.}~\bibnamefont
			{Samuelsson}}, \bibinfo {author} {\bibfnamefont {E.~V.}\ \bibnamefont
			{Sukhorukov}},\ and\ \bibinfo {author} {\bibfnamefont {M.}~\bibnamefont
			{B\"uttiker}},\ }\bibfield  {title} {\bibinfo {title} {Orbital entanglement
			and violation of bell inequalities in mesoscopic conductors},\ }\href
	{https://doi.org/10.1103/PhysRevLett.91.157002} {\bibfield  {journal}
		{\bibinfo  {journal} {Phys. Rev. Lett.}\ }\textbf {\bibinfo {volume} {91}},\
		\bibinfo {pages} {157002} (\bibinfo {year} {2003})}\BibitemShut {NoStop}%
	\bibitem [{\citenamefont {Sauret}\ \emph {et~al.}(2005)\citenamefont {Sauret},
		\citenamefont {Martin},\ and\ \citenamefont {Feinberg}}]{Sauret2005}%
	\BibitemOpen
	\bibfield  {author} {\bibinfo {author} {\bibfnamefont {O.}~\bibnamefont
			{Sauret}}, \bibinfo {author} {\bibfnamefont {T.}~\bibnamefont {Martin}},\
		and\ \bibinfo {author} {\bibfnamefont {D.}~\bibnamefont {Feinberg}},\
	}\bibfield  {title} {\bibinfo {title} {Spin-current noise and bell
			inequalities in a realistic superconductor-quantum dot entangler},\ }\href
	{https://doi.org/10.1103/PhysRevB.72.024544} {\bibfield  {journal} {\bibinfo
			{journal} {Phys. Rev. B}\ }\textbf {\bibinfo {volume} {72}},\ \bibinfo
		{pages} {024544} (\bibinfo {year} {2005})}\BibitemShut {NoStop}%
	\bibitem [{\citenamefont {Martin}(1996)}]{Martin1996}%
	\BibitemOpen
	\bibfield  {author} {\bibinfo {author} {\bibfnamefont {T.}~\bibnamefont
			{Martin}},\ }\bibfield  {title} {\bibinfo {title} {Wave packet approach to
			noise in n-s junctions},\ }\href
	{https://doi.org/https://doi.org/10.1016/0375-9601(96)00484-7} {\bibfield
		{journal} {\bibinfo  {journal} {Physics Letters A}\ }\textbf {\bibinfo
			{volume} {220}},\ \bibinfo {pages} {137} (\bibinfo {year}
		{1996})}\BibitemShut {NoStop}%
	\bibitem [{\citenamefont {Anantram}\ and\ \citenamefont
		{Datta}(1996)}]{Anantram1996}%
	\BibitemOpen
	\bibfield  {author} {\bibinfo {author} {\bibfnamefont {M.~P.}\ \bibnamefont
			{Anantram}}\ and\ \bibinfo {author} {\bibfnamefont {S.}~\bibnamefont
			{Datta}},\ }\bibfield  {title} {\bibinfo {title} {Current fluctuations in
			mesoscopic systems with andreev scattering},\ }\href
	{https://doi.org/10.1103/PhysRevB.53.16390} {\bibfield  {journal} {\bibinfo
			{journal} {Phys. Rev. B}\ }\textbf {\bibinfo {volume} {53}},\ \bibinfo
		{pages} {16390} (\bibinfo {year} {1996})}\BibitemShut {NoStop}%
	\bibitem [{\citenamefont {Torr\`es}\ \emph {et~al.}(2001)\citenamefont
		{Torr\`es}, \citenamefont {Martin},\ and\ \citenamefont
		{Lesovik}}]{Torres2001}%
	\BibitemOpen
	\bibfield  {author} {\bibinfo {author} {\bibfnamefont {J.}~\bibnamefont
			{Torr\`es}}, \bibinfo {author} {\bibfnamefont {T.}~\bibnamefont {Martin}},\
		and\ \bibinfo {author} {\bibfnamefont {G.~B.}\ \bibnamefont {Lesovik}},\
	}\bibfield  {title} {\bibinfo {title} {Effective charges and statistical
			signatures in the noise of normal metal--superconductor junctions at
			arbitrary bias},\ }\href {https://doi.org/10.1103/PhysRevB.63.134517}
	{\bibfield  {journal} {\bibinfo  {journal} {Phys. Rev. B}\ }\textbf {\bibinfo
			{volume} {63}},\ \bibinfo {pages} {134517} (\bibinfo {year}
		{2001})}\BibitemShut {NoStop}%
	\bibitem [{\citenamefont {Sauret}\ and\ \citenamefont
		{Feinberg}(2004)}]{Sauret2004}%
	\BibitemOpen
	\bibfield  {author} {\bibinfo {author} {\bibfnamefont {O.}~\bibnamefont
			{Sauret}}\ and\ \bibinfo {author} {\bibfnamefont {D.}~\bibnamefont
			{Feinberg}},\ }\bibfield  {title} {\bibinfo {title} {Spin-current shot noise
			as a probe of interactions in mesoscopic systems},\ }\href
	{https://doi.org/10.1103/PhysRevLett.92.106601} {\bibfield  {journal}
		{\bibinfo  {journal} {Phys. Rev. Lett.}\ }\textbf {\bibinfo {volume} {92}},\
		\bibinfo {pages} {106601} (\bibinfo {year} {2004})}\BibitemShut {NoStop}%
	\bibitem [{\citenamefont {Taddei}\ \emph {et~al.}(2005)\citenamefont {Taddei},
		\citenamefont {Giazotto},\ and\ \citenamefont {Fazio}}]{Taddei2005}%
	\BibitemOpen
	\bibfield  {author} {\bibinfo {author} {\bibfnamefont {F.}~\bibnamefont
			{Taddei}}, \bibinfo {author} {\bibfnamefont {F.}~\bibnamefont {Giazotto}},\
		and\ \bibinfo {author} {\bibfnamefont {R.}~\bibnamefont {Fazio}},\ }\bibfield
	{title} {\bibinfo {title} {{Properties of Mesoscopic Hybrid Superconducting
				Systems}},\ }\href {https://doi.org/10.1166/jctn.2005.201} {\bibfield
		{journal} {\bibinfo  {journal} {Journal of Computational and Theoretical
				Nanoscience}\ }\textbf {\bibinfo {volume} {2}},\ \bibinfo {pages} {329}
		(\bibinfo {year} {2005})}\BibitemShut {NoStop}%
	\bibitem [{\citenamefont {Bayandin}\ \emph {et~al.}(2006)\citenamefont
		{Bayandin}, \citenamefont {Lesovik},\ and\ \citenamefont
		{Martin}}]{Bayandin06}%
	\BibitemOpen
	\bibfield  {author} {\bibinfo {author} {\bibfnamefont {K.~V.}\ \bibnamefont
			{Bayandin}}, \bibinfo {author} {\bibfnamefont {G.~B.}\ \bibnamefont
			{Lesovik}},\ and\ \bibinfo {author} {\bibfnamefont {T.}~\bibnamefont
			{Martin}},\ }\bibfield  {title} {\bibinfo {title} {Energy entanglement in
			normal metal--superconducting forks},\ }\href
	{https://doi.org/10.1103/PhysRevB.74.085326} {\bibfield  {journal} {\bibinfo
			{journal} {Phys. Rev. B}\ }\textbf {\bibinfo {volume} {74}},\ \bibinfo
		{pages} {085326} (\bibinfo {year} {2006})}\BibitemShut {NoStop}%
	\bibitem [{\citenamefont {Chevallier}\ \emph {et~al.}(2011)\citenamefont
		{Chevallier}, \citenamefont {Rech}, \citenamefont {Jonckheere},\ and\
		\citenamefont {Martin}}]{Chevallier2011}%
	\BibitemOpen
	\bibfield  {author} {\bibinfo {author} {\bibfnamefont {D.}~\bibnamefont
			{Chevallier}}, \bibinfo {author} {\bibfnamefont {J.}~\bibnamefont {Rech}},
		\bibinfo {author} {\bibfnamefont {T.}~\bibnamefont {Jonckheere}},\ and\
		\bibinfo {author} {\bibfnamefont {T.}~\bibnamefont {Martin}},\ }\bibfield
	{title} {\bibinfo {title} {Current and noise correlations in a double-dot
			cooper-pair beam splitter},\ }\href
	{https://doi.org/10.1103/PhysRevB.83.125421} {\bibfield  {journal} {\bibinfo
			{journal} {Phys. Rev. B}\ }\textbf {\bibinfo {volume} {83}},\ \bibinfo
		{pages} {125421} (\bibinfo {year} {2011})}\BibitemShut {NoStop}%
	\bibitem [{\citenamefont {Rech}\ \emph {et~al.}(2012)\citenamefont {Rech},
		\citenamefont {Chevallier}, \citenamefont {Jonckheere},\ and\ \citenamefont
		{Martin}}]{Rech2012}%
	\BibitemOpen
	\bibfield  {author} {\bibinfo {author} {\bibfnamefont {J.}~\bibnamefont
			{Rech}}, \bibinfo {author} {\bibfnamefont {D.}~\bibnamefont {Chevallier}},
		\bibinfo {author} {\bibfnamefont {T.}~\bibnamefont {Jonckheere}},\ and\
		\bibinfo {author} {\bibfnamefont {T.}~\bibnamefont {Martin}},\ }\bibfield
	{title} {\bibinfo {title} {Current correlations in an interacting cooper-pair
			beam splitter},\ }\href {https://doi.org/10.1103/PhysRevB.85.035419}
	{\bibfield  {journal} {\bibinfo  {journal} {Phys. Rev. B}\ }\textbf {\bibinfo
			{volume} {85}},\ \bibinfo {pages} {035419} (\bibinfo {year}
		{2012})}\BibitemShut {NoStop}%
	\bibitem [{\citenamefont {Jacquet}\ \emph {et~al.}(2020)\citenamefont
		{Jacquet}, \citenamefont {Popoff}, \citenamefont {Imura}, \citenamefont
		{Rech}, \citenamefont {Jonckheere}, \citenamefont {Raymond}, \citenamefont
		{Zazunov},\ and\ \citenamefont {Martin}}]{Jacquet2020}%
	\BibitemOpen
	\bibfield  {author} {\bibinfo {author} {\bibfnamefont {R.}~\bibnamefont
			{Jacquet}}, \bibinfo {author} {\bibfnamefont {A.}~\bibnamefont {Popoff}},
		\bibinfo {author} {\bibfnamefont {K.-I.}\ \bibnamefont {Imura}}, \bibinfo
		{author} {\bibfnamefont {J.}~\bibnamefont {Rech}}, \bibinfo {author}
		{\bibfnamefont {T.}~\bibnamefont {Jonckheere}}, \bibinfo {author}
		{\bibfnamefont {L.}~\bibnamefont {Raymond}}, \bibinfo {author} {\bibfnamefont
			{A.}~\bibnamefont {Zazunov}},\ and\ \bibinfo {author} {\bibfnamefont
			{T.}~\bibnamefont {Martin}},\ }\bibfield  {title} {\bibinfo {title} {Theory
			of nonequilibrium noise in general multiterminal superconducting hybrid
			devices: Application to multiple cooper pair resonances},\ }\href
	{https://doi.org/10.1103/PhysRevB.102.064510} {\bibfield  {journal} {\bibinfo
			{journal} {Phys. Rev. B}\ }\textbf {\bibinfo {volume} {102}},\ \bibinfo
		{pages} {064510} (\bibinfo {year} {2020})}\BibitemShut {NoStop}%
	\bibitem [{\citenamefont {Benito}\ and\ \citenamefont
		{Burkard}(2020)}]{Benito2022}%
	\BibitemOpen
	\bibfield  {author} {\bibinfo {author} {\bibfnamefont {M.}~\bibnamefont
			{Benito}}\ and\ \bibinfo {author} {\bibfnamefont {G.}~\bibnamefont
			{Burkard}},\ }\bibfield  {title} {\bibinfo {title} {{Hybrid
				superconductor-semiconductor systems for quantum technology}},\ }\href
	{https://doi.org/10.1063/5.0004777} {\bibfield  {journal} {\bibinfo
			{journal} {Applied Physics Letters}\ }\textbf {\bibinfo {volume} {116}},\
		\bibinfo {pages} {190502} (\bibinfo {year} {2020})}\BibitemShut {NoStop}%
	\bibitem [{\citenamefont {Bernevig}\ \emph {et~al.}(2006)\citenamefont
		{Bernevig}, \citenamefont {Hughes},\ and\ \citenamefont
		{Zhang}}]{Bernevig2006}%
	\BibitemOpen
	\bibfield  {author} {\bibinfo {author} {\bibfnamefont {B.~A.}\ \bibnamefont
			{Bernevig}}, \bibinfo {author} {\bibfnamefont {T.~L.}\ \bibnamefont
			{Hughes}},\ and\ \bibinfo {author} {\bibfnamefont {S.-C.}\ \bibnamefont
			{Zhang}},\ }\bibfield  {title} {\bibinfo {title} {Quantum spin hall effect
			and topological phase transition in hgte quantum wells},\ }\href
	{https://doi.org/10.1126/science.1133734} {\bibfield  {journal} {\bibinfo
			{journal} {Science}\ }\textbf {\bibinfo {volume} {314}},\ \bibinfo {pages}
		{1757} (\bibinfo {year} {2006})}\BibitemShut {NoStop}%
	\bibitem [{\citenamefont {König}\ \emph {et~al.}(2007)\citenamefont {König},
		\citenamefont {Wiedmann}, \citenamefont {Brüne}, \citenamefont {Roth},
		\citenamefont {Buhmann}, \citenamefont {Molenkamp}, \citenamefont {Qi},\ and\
		\citenamefont {Zhang}}]{Konig2007}%
	\BibitemOpen
	\bibfield  {author} {\bibinfo {author} {\bibfnamefont {M.}~\bibnamefont
			{König}}, \bibinfo {author} {\bibfnamefont {S.}~\bibnamefont {Wiedmann}},
		\bibinfo {author} {\bibfnamefont {C.}~\bibnamefont {Brüne}}, \bibinfo
		{author} {\bibfnamefont {A.}~\bibnamefont {Roth}}, \bibinfo {author}
		{\bibfnamefont {H.}~\bibnamefont {Buhmann}}, \bibinfo {author} {\bibfnamefont
			{L.~W.}\ \bibnamefont {Molenkamp}}, \bibinfo {author} {\bibfnamefont {X.-L.}\
			\bibnamefont {Qi}},\ and\ \bibinfo {author} {\bibfnamefont {S.-C.}\
			\bibnamefont {Zhang}},\ }\bibfield  {title} {\bibinfo {title} {Quantum spin
			hall insulator state in hgte quantum wells},\ }\href
	{https://doi.org/10.1126/science.1148047} {\bibfield  {journal} {\bibinfo
			{journal} {Science}\ }\textbf {\bibinfo {volume} {318}},\ \bibinfo {pages}
		{766} (\bibinfo {year} {2007})}\BibitemShut {NoStop}%
	\bibitem [{\citenamefont {Qi}\ and\ \citenamefont {Zhang}(2011)}]{Qi2011}%
	\BibitemOpen
	\bibfield  {author} {\bibinfo {author} {\bibfnamefont {X.-L.}\ \bibnamefont
			{Qi}}\ and\ \bibinfo {author} {\bibfnamefont {S.-C.}\ \bibnamefont {Zhang}},\
	}\bibfield  {title} {\bibinfo {title} {Topological insulators and
			superconductors},\ }\href {https://doi.org/10.1103/RevModPhys.83.1057}
	{\bibfield  {journal} {\bibinfo  {journal} {Rev. Mod. Phys.}\ }\textbf
		{\bibinfo {volume} {83}},\ \bibinfo {pages} {1057} (\bibinfo {year}
		{2011})}\BibitemShut {NoStop}%
	\bibitem [{\citenamefont {Kane}\ and\ \citenamefont
		{Mele}(2005{\natexlab{a}})}]{Kane2005a}%
	\BibitemOpen
	\bibfield  {author} {\bibinfo {author} {\bibfnamefont {C.~L.}\ \bibnamefont
			{Kane}}\ and\ \bibinfo {author} {\bibfnamefont {E.~J.}\ \bibnamefont
			{Mele}},\ }\bibfield  {title} {\bibinfo {title} {Quantum spin hall effect in
			graphene},\ }\href {https://doi.org/10.1103/PhysRevLett.95.226801} {\bibfield
		{journal} {\bibinfo  {journal} {Phys. Rev. Lett.}\ }\textbf {\bibinfo
			{volume} {95}},\ \bibinfo {pages} {226801} (\bibinfo {year}
		{2005}{\natexlab{a}})}\BibitemShut {NoStop}%
	\bibitem [{\citenamefont {Kane}\ and\ \citenamefont
		{Mele}(2005{\natexlab{b}})}]{Kane2005b}%
	\BibitemOpen
	\bibfield  {author} {\bibinfo {author} {\bibfnamefont {C.~L.}\ \bibnamefont
			{Kane}}\ and\ \bibinfo {author} {\bibfnamefont {E.~J.}\ \bibnamefont
			{Mele}},\ }\bibfield  {title} {\bibinfo {title} {${Z}_{2}$ topological order
			and the quantum spin hall effect},\ }\href
	{https://doi.org/10.1103/PhysRevLett.95.146802} {\bibfield  {journal}
		{\bibinfo  {journal} {Phys. Rev. Lett.}\ }\textbf {\bibinfo {volume} {95}},\
		\bibinfo {pages} {146802} (\bibinfo {year} {2005}{\natexlab{b}})}\BibitemShut
	{NoStop}%
	\bibitem [{\citenamefont {{Dolcetto}}\ \emph {et~al.}(2016)\citenamefont
		{{Dolcetto}}, \citenamefont {{Sassetti}},\ and\ \citenamefont
		{{Schmidt}}}]{Dolcetto2016}%
	\BibitemOpen
	\bibfield  {author} {\bibinfo {author} {\bibfnamefont {G.}~\bibnamefont
			{{Dolcetto}}}, \bibinfo {author} {\bibfnamefont {M.}~\bibnamefont
			{{Sassetti}}},\ and\ \bibinfo {author} {\bibfnamefont {T.~L.}\ \bibnamefont
			{{Schmidt}}},\ }\bibfield  {title} {\bibinfo {title} {{Edge physics in
				two-dimensional topological insulators}},\ }\href
	{https://doi.org/10.1393/ncr/i2016-10121-7} {\bibfield  {journal} {\bibinfo
			{journal} {Nuovo Cimento Rivista Serie}\ }\textbf {\bibinfo {volume} {39}},\
		\bibinfo {pages} {113} (\bibinfo {year} {2016})},\ \Eprint
	{https://arxiv.org/abs/1511.06141} {arXiv:1511.06141 [cond-mat.mes-hall]}
	\BibitemShut {NoStop}%
	\bibitem [{\citenamefont {Martin}(2005)}]{Martin2005}%
	\BibitemOpen
	\bibfield  {author} {\bibinfo {author} {\bibfnamefont {T.}~\bibnamefont
			{Martin}},\ }\bibfield  {title} {\bibinfo {title} {Noise in mesoscopic
			physics},\ }in\ \href@noop {} {\emph {\bibinfo {booktitle} {Nanophysics:
				Coherence and Transport}}},\ \bibinfo {series and number} {Les Houches,
		Session LXXXI},\ \bibinfo {editor} {edited by\ \bibinfo {editor}
		{\bibfnamefont {H.}~\bibnamefont {Bouchiat}}, \bibinfo {editor}
		{\bibfnamefont {Y.}~\bibnamefont {Gefen}}, \bibinfo {editor} {\bibfnamefont
			{S.}~\bibnamefont {Gu{\'e}ron}}, \bibinfo {editor} {\bibfnamefont
			{G.}~\bibnamefont {Montambaux}},\ and\ \bibinfo {editor} {\bibfnamefont
			{J.}~\bibnamefont {Dalibard}}}\ (\bibinfo  {publisher} {Elsevier},\ \bibinfo
	{year} {2005})\ p.\ \bibinfo {pages} {283}\BibitemShut {NoStop}%
	\bibitem [{\citenamefont {Schrieffer}(1983)}]{Schrieffer1983}%
	\BibitemOpen
	\bibfield  {author} {\bibinfo {author} {\bibfnamefont {J.~R.}\ \bibnamefont
			{Schrieffer}},\ }\href@noop {} {\emph {\bibinfo {title} {Theory of
				superconductivity}}}\ (\bibinfo  {publisher} {CRC press},\ \bibinfo {year}
	{1983})\BibitemShut {NoStop}%
	\bibitem [{\citenamefont {Vannucci}\ \emph {et~al.}(2015)\citenamefont
		{Vannucci}, \citenamefont {Ronetti}, \citenamefont {Dolcetto}, \citenamefont
		{Carrega},\ and\ \citenamefont {Sassetti}}]{Vannucci2015}%
	\BibitemOpen
	\bibfield  {author} {\bibinfo {author} {\bibfnamefont {L.}~\bibnamefont
			{Vannucci}}, \bibinfo {author} {\bibfnamefont {F.}~\bibnamefont {Ronetti}},
		\bibinfo {author} {\bibfnamefont {G.}~\bibnamefont {Dolcetto}}, \bibinfo
		{author} {\bibfnamefont {M.}~\bibnamefont {Carrega}},\ and\ \bibinfo {author}
		{\bibfnamefont {M.}~\bibnamefont {Sassetti}},\ }\bibfield  {title} {\bibinfo
		{title} {Interference-induced thermoelectric switching and heat rectification
			in quantum hall junctions},\ }\href
	{https://doi.org/10.1103/PhysRevB.92.075446} {\bibfield  {journal} {\bibinfo
			{journal} {Phys. Rev. B}\ }\textbf {\bibinfo {volume} {92}},\ \bibinfo
		{pages} {075446} (\bibinfo {year} {2015})}\BibitemShut {NoStop}%
	\bibitem [{\citenamefont {Ronetti}\ \emph {et~al.}(2016)\citenamefont
		{Ronetti}, \citenamefont {Vannucci}, \citenamefont {Dolcetto}, \citenamefont
		{Carrega},\ and\ \citenamefont {Sassetti}}]{Ronetti2016}%
	\BibitemOpen
	\bibfield  {author} {\bibinfo {author} {\bibfnamefont {F.}~\bibnamefont
			{Ronetti}}, \bibinfo {author} {\bibfnamefont {L.}~\bibnamefont {Vannucci}},
		\bibinfo {author} {\bibfnamefont {G.}~\bibnamefont {Dolcetto}}, \bibinfo
		{author} {\bibfnamefont {M.}~\bibnamefont {Carrega}},\ and\ \bibinfo {author}
		{\bibfnamefont {M.}~\bibnamefont {Sassetti}},\ }\bibfield  {title} {\bibinfo
		{title} {Spin-thermoelectric transport induced by interactions and spin-flip
			processes in two-dimensional topological insulators},\ }\href
	{https://doi.org/10.1103/PhysRevB.93.165414} {\bibfield  {journal} {\bibinfo
			{journal} {Phys. Rev. B}\ }\textbf {\bibinfo {volume} {93}},\ \bibinfo
		{pages} {165414} (\bibinfo {year} {2016})}\BibitemShut {NoStop}%
	\bibitem [{\citenamefont {Ronetti}\ \emph {et~al.}(2017)\citenamefont
		{Ronetti}, \citenamefont {Carrega}, \citenamefont {Ferraro}, \citenamefont
		{Rech}, \citenamefont {Jonckheere}, \citenamefont {Martin},\ and\
		\citenamefont {Sassetti}}]{Ronetti2017}%
	\BibitemOpen
	\bibfield  {author} {\bibinfo {author} {\bibfnamefont {F.}~\bibnamefont
			{Ronetti}}, \bibinfo {author} {\bibfnamefont {M.}~\bibnamefont {Carrega}},
		\bibinfo {author} {\bibfnamefont {D.}~\bibnamefont {Ferraro}}, \bibinfo
		{author} {\bibfnamefont {J.}~\bibnamefont {Rech}}, \bibinfo {author}
		{\bibfnamefont {T.}~\bibnamefont {Jonckheere}}, \bibinfo {author}
		{\bibfnamefont {T.}~\bibnamefont {Martin}},\ and\ \bibinfo {author}
		{\bibfnamefont {M.}~\bibnamefont {Sassetti}},\ }\bibfield  {title} {\bibinfo
		{title} {Polarized heat current generated by quantum pumping in
			two-dimensional topological insulators},\ }\href
	{https://doi.org/10.1103/PhysRevB.95.115412} {\bibfield  {journal} {\bibinfo
			{journal} {Phys. Rev. B}\ }\textbf {\bibinfo {volume} {95}},\ \bibinfo
		{pages} {115412} (\bibinfo {year} {2017})}\BibitemShut {NoStop}%
	\bibitem [{\citenamefont {Rarity}\ and\ \citenamefont
		{Tapster}(1990)}]{Rarity90}%
	\BibitemOpen
	\bibfield  {author} {\bibinfo {author} {\bibfnamefont {J.~G.}\ \bibnamefont
			{Rarity}}\ and\ \bibinfo {author} {\bibfnamefont {P.~R.}\ \bibnamefont
			{Tapster}},\ }\bibfield  {title} {\bibinfo {title} {Experimental violation of
			bell's inequality based on phase and momentum},\ }\href
	{https://doi.org/10.1103/PhysRevLett.64.2495} {\bibfield  {journal} {\bibinfo
			{journal} {Phys. Rev. Lett.}\ }\textbf {\bibinfo {volume} {64}},\ \bibinfo
		{pages} {2495} (\bibinfo {year} {1990})}\BibitemShut {NoStop}%
	\bibitem [{\citenamefont {{Lesovik}}\ and\ \citenamefont
		{{Sadovskyy}}(2011)}]{Lesovik11}%
	\BibitemOpen
	\bibfield  {author} {\bibinfo {author} {\bibfnamefont {G.~B.}\ \bibnamefont
			{{Lesovik}}}\ and\ \bibinfo {author} {\bibfnamefont {I.~A.}\ \bibnamefont
			{{Sadovskyy}}},\ }\bibfield  {title} {\bibinfo {title} {{Scattering matrix
				approach to the description of quantum electron transport}},\ }\href
	{https://doi.org/10.3367/UFNe.0181.201110b.1041} {\bibfield  {journal}
		{\bibinfo  {journal} {Physics Uspekhi}\ }\textbf {\bibinfo {volume} {54}},\
		\bibinfo {pages} {1007} (\bibinfo {year} {2011})},\ \Eprint
	{https://arxiv.org/abs/1408.1966} {arXiv:1408.1966 [cond-mat.mes-hall]}
	\BibitemShut {NoStop}%
	\bibitem [{\citenamefont {Burset}\ \emph {et~al.}(2023)\citenamefont {Burset},
		\citenamefont {Roussel}, \citenamefont {Moskalets},\ and\ \citenamefont
		{Flindt}}]{Burset2023}%
	\BibitemOpen
	\bibfield  {author} {\bibinfo {author} {\bibfnamefont {P.}~\bibnamefont
			{Burset}}, \bibinfo {author} {\bibfnamefont {B.}~\bibnamefont {Roussel}},
		\bibinfo {author} {\bibfnamefont {M.}~\bibnamefont {Moskalets}},\ and\
		\bibinfo {author} {\bibfnamefont {C.}~\bibnamefont {Flindt}},\ }\href
	{https://arxiv.org/abs/2312.13145} {\bibinfo {title} {Tunable
			andreev-conversion of single-electron charge pulses}} (\bibinfo {year}
	{2023}),\ \Eprint {https://arxiv.org/abs/2312.13145} {arXiv:2312.13145
		[cond-mat.mes-hall]} \BibitemShut {NoStop}%
	\bibitem [{\citenamefont {Amico}\ \emph {et~al.}(2008)\citenamefont {Amico},
		\citenamefont {Fazio}, \citenamefont {Osterloh},\ and\ \citenamefont
		{Vedral}}]{Amico08}%
	\BibitemOpen
	\bibfield  {author} {\bibinfo {author} {\bibfnamefont {L.}~\bibnamefont
			{Amico}}, \bibinfo {author} {\bibfnamefont {R.}~\bibnamefont {Fazio}},
		\bibinfo {author} {\bibfnamefont {A.}~\bibnamefont {Osterloh}},\ and\
		\bibinfo {author} {\bibfnamefont {V.}~\bibnamefont {Vedral}},\ }\bibfield
	{title} {\bibinfo {title} {Entanglement in many-body systems},\ }\href
	{https://doi.org/10.1103/RevModPhys.80.517} {\bibfield  {journal} {\bibinfo
			{journal} {Rev. Mod. Phys.}\ }\textbf {\bibinfo {volume} {80}},\ \bibinfo
		{pages} {517} (\bibinfo {year} {2008})}\BibitemShut {NoStop}%
	\bibitem [{\citenamefont {Bennett}\ \emph {et~al.}(1993)\citenamefont
		{Bennett}, \citenamefont {Brassard}, \citenamefont {Cr\'epeau}, \citenamefont
		{Jozsa}, \citenamefont {Peres},\ and\ \citenamefont
		{Wootters}}]{Bennett1993}%
	\BibitemOpen
	\bibfield  {author} {\bibinfo {author} {\bibfnamefont {C.~H.}\ \bibnamefont
			{Bennett}}, \bibinfo {author} {\bibfnamefont {G.}~\bibnamefont {Brassard}},
		\bibinfo {author} {\bibfnamefont {C.}~\bibnamefont {Cr\'epeau}}, \bibinfo
		{author} {\bibfnamefont {R.}~\bibnamefont {Jozsa}}, \bibinfo {author}
		{\bibfnamefont {A.}~\bibnamefont {Peres}},\ and\ \bibinfo {author}
		{\bibfnamefont {W.~K.}\ \bibnamefont {Wootters}},\ }\bibfield  {title}
	{\bibinfo {title} {Teleporting an unknown quantum state via dual classical
			and einstein-podolsky-rosen channels},\ }\href
	{https://doi.org/10.1103/PhysRevLett.70.1895} {\bibfield  {journal} {\bibinfo
			{journal} {Phys. Rev. Lett.}\ }\textbf {\bibinfo {volume} {70}},\ \bibinfo
		{pages} {1895} (\bibinfo {year} {1993})}\BibitemShut {NoStop}%
	\bibitem [{\citenamefont {Long}\ and\ \citenamefont {Liu}(2002)}]{Long2002}%
	\BibitemOpen
	\bibfield  {author} {\bibinfo {author} {\bibfnamefont {G.~L.}\ \bibnamefont
			{Long}}\ and\ \bibinfo {author} {\bibfnamefont {X.~S.}\ \bibnamefont {Liu}},\
	}\bibfield  {title} {\bibinfo {title} {Theoretically efficient high-capacity
			quantum-key-distribution scheme},\ }\href
	{https://doi.org/10.1103/PhysRevA.65.032302} {\bibfield  {journal} {\bibinfo
			{journal} {Phys. Rev. A}\ }\textbf {\bibinfo {volume} {65}},\ \bibinfo
		{pages} {032302} (\bibinfo {year} {2002})}\BibitemShut {NoStop}%
	\bibitem [{\citenamefont {Ekert}(1991)}]{Ekert1991}%
	\BibitemOpen
	\bibfield  {author} {\bibinfo {author} {\bibfnamefont {A.~K.}\ \bibnamefont
			{Ekert}},\ }\bibfield  {title} {\bibinfo {title} {Quantum cryptography based
			on bell's theorem},\ }\href {https://doi.org/10.1103/PhysRevLett.67.661}
	{\bibfield  {journal} {\bibinfo  {journal} {Phys. Rev. Lett.}\ }\textbf
		{\bibinfo {volume} {67}},\ \bibinfo {pages} {661} (\bibinfo {year}
		{1991})}\BibitemShut {NoStop}%
	\bibitem [{\citenamefont {Bennett}\ \emph {et~al.}(1992)\citenamefont
		{Bennett}, \citenamefont {Brassard},\ and\ \citenamefont
		{Mermin}}]{Bennett1992}%
	\BibitemOpen
	\bibfield  {author} {\bibinfo {author} {\bibfnamefont {C.~H.}\ \bibnamefont
			{Bennett}}, \bibinfo {author} {\bibfnamefont {G.}~\bibnamefont {Brassard}},\
		and\ \bibinfo {author} {\bibfnamefont {N.~D.}\ \bibnamefont {Mermin}},\
	}\bibfield  {title} {\bibinfo {title} {Quantum cryptography without bell's
			theorem},\ }\href {https://doi.org/10.1103/PhysRevLett.68.557} {\bibfield
		{journal} {\bibinfo  {journal} {Phys. Rev. Lett.}\ }\textbf {\bibinfo
			{volume} {68}},\ \bibinfo {pages} {557} (\bibinfo {year} {1992})}\BibitemShut
	{NoStop}%
	\bibitem [{\citenamefont {Pan}\ \emph {et~al.}(2020)\citenamefont {Pan},
		\citenamefont {Li}, \citenamefont {Ruan}, \citenamefont {Ng},\ and\
		\citenamefont {Hanzo}}]{Pan2020}%
	\BibitemOpen
	\bibfield  {author} {\bibinfo {author} {\bibfnamefont {D.}~\bibnamefont
			{Pan}}, \bibinfo {author} {\bibfnamefont {K.}~\bibnamefont {Li}}, \bibinfo
		{author} {\bibfnamefont {D.}~\bibnamefont {Ruan}}, \bibinfo {author}
		{\bibfnamefont {S.~X.}\ \bibnamefont {Ng}},\ and\ \bibinfo {author}
		{\bibfnamefont {L.}~\bibnamefont {Hanzo}},\ }\bibfield  {title} {\bibinfo
		{title} {Single-photon-memory two-step quantum secure direct communication
			relying on einstein-podolsky-rosen pairs},\ }\href
	{https://doi.org/10.1109/ACCESS.2020.3006136} {\bibfield  {journal} {\bibinfo
			{journal} {IEEE Access}\ }\textbf {\bibinfo {volume} {8}},\ \bibinfo {pages}
		{121146} (\bibinfo {year} {2020})}\BibitemShut {NoStop}%
	\bibitem [{\citenamefont {Bennett}\ and\ \citenamefont
		{Wiesner}(1992)}]{Bennett1992b}%
	\BibitemOpen
	\bibfield  {author} {\bibinfo {author} {\bibfnamefont {C.~H.}\ \bibnamefont
			{Bennett}}\ and\ \bibinfo {author} {\bibfnamefont {S.~J.}\ \bibnamefont
			{Wiesner}},\ }\bibfield  {title} {\bibinfo {title} {Communication via one-
			and two-particle operators on einstein-podolsky-rosen states},\ }\href
	{https://doi.org/10.1103/PhysRevLett.69.2881} {\bibfield  {journal} {\bibinfo
			{journal} {Phys. Rev. Lett.}\ }\textbf {\bibinfo {volume} {69}},\ \bibinfo
		{pages} {2881} (\bibinfo {year} {1992})}\BibitemShut {NoStop}%
\end{thebibliography}
\end{document}